# The potential impact of fiscal rule changes on digital products trade[+]

Sonja Mitikj[1] and Viktor Stojkoski[2,*]

[1]Faculty of Computer Science and Engineering, University Ss. Cyril and Methodius in Skopje, North Macedonia
[2]Faculty of Economics – Skopje, University Ss. Cyril and Methodius in Skopje, North Macedonia

The WTO moratorium on customs duties on electronic transmissions lapsed in March 2026, allowing governments to impose tariffs on cross-border digital products, alongside other instruments such as digital services taxes (DSTs). Yet, how tariffs and DSTs compare in the revenue they raise and their effect on trade in digital products has not yet been assessed. Here we combine bilateral trade in 31 digital product categories in 2016–2024 with DST rates and tariffs on physical analogues, estimate gravity models and simulate alternative fiscal regimes. We find that digital trade is lower in destinations with a DST, but within destinations it declines only temporarily after adoption. In conditional scenarios for 2024 based on the event study estimate, adding tariff equivalents to DSTs raises revenue from US$2.7 to 6.9 billion and lowers trade by about 1.4%, and replacing DSTs with tariff equivalents raises US$4.3 billion. We also find that tariff-equivalent payments are attributed mainly to the United States and DST payments to subsidiary locations such as Ireland. In sum, these results suggest that tariffs on digital products could raise more revenue than DSTs at a modest cost in trade.

---

[+] We gratefully acknowledge financial and institutional support from the Hinrich Foundation. An initial version of this manuscript was published as a white paper on their website https://www.hinrichfoundation.com/research/wp/tech-digital-trade/fiscal-rule-changes-and-digital-products-trade?utm_source=linkedin&utm_medium=social-organic&utm_campaign=wp-mitikj-skojkoski-fiscal-rule-changes-digital-product-trade&utm_content=20260319-static. This article extends the white paper with an event study, additional robustness checks and a re-estimation with independent trade data.
[*] Corresponding author, vstojkoski@eccf.ukim.edu.mk

# 1. Introduction

A decade ago, buying a foreign book often meant ordering a physical copy, which was shipped and cleared customs. Today the same transaction can be paid for online, and the product can be downloaded in the form of an e-book within seconds, without any shipment crossing a border. Many products and transactions that once required physical delivery or physical presence, from digitized goods (books, music, software, and video games) to productized digital outputs (online advertising) and platform intermediation fees (short-term rental or travel booking commissions), are now traded across borders through payments and electronic delivery (Stojkoski et al., 2024).

Recent research has indicated that trade in these "*digital products*" has expanded far faster than most traditional cross-border flows (Stojkoski et al., 2024). Yet the fiscal framework has not kept pace with this transformation. This has happened mainly because the rules governing cross-border taxation are largely built around physical shipment and customs borders, while digital delivery shifts many transactions outside the traditional tariff channel and into domestic tax systems (Andrenelli and López González, 2019, 2023; Hanappi et al., 2023). The WTO moratorium on customs duties on electronic transmissions reinforced this shift by sustaining a largely duty-free border environment for digital deliveries from 1998 until it lapsed at the end of March 2026 (World Trade Organization, 2024; Hebous et al., 2026). This meant that, until March 2026, a cross-border download of software or an e-book did not face the customs duty that would apply if the same item arrived as a physical shipment. Since the moratorium lapsed, governments may impose such duties, but none has done so yet, and some WTO members have committed not to impose them among themselves (Hebous et al., 2026). Hence, fiscal treatment still depends primarily on value-added taxes (VATs) or goods and service taxes (GSTs) (Hanappi et al., 2024). However, countries differ widely in how they define the place of taxation, whether non-resident suppliers must register, and whether platforms are required to collect and remit the tax (Caragher, 2025; Hebous et al., 2026), while uncertainty over what precisely falls under "electronic transmissions" further contributes to heterogeneous treatment across jurisdictions (Andrenelli and López González, 2023).

In this paper, we ask how trade in digital products and the fiscal revenue it generates would change under different fiscal rules. We first estimate gravity models, in which bilateral trade depends on the size of the two economies and the costs of trading between them (Anderson and van Wincoop, 2003; Head and Mayer, 2014), and include the DST rate as an additional cost on cross-border digital sales. Because only a small number of destinations adopted a DST (Asquith, n.d.; Enache, 2025), this coefficient mainly reflects differences between taxing and non-taxing destinations, so we interpret it as an association and test it with stricter fixed effects, clustered inference and an event study around DST adoption (Wooldridge, 2023). We then apply DST rates and tariff equivalents to observed bilateral trade in digital products in 2016–2024, which gives a descriptive accounting benchmark of the revenue at stake. Finally, we use the estimated model to simulate two policy changes, the introduction of customs duties on electronic transmissions, which became possible when the WTO moratorium lapsed in 2026, and the combination of these duties with existing DSTs. The simulations use the short-run response of trade to DST adoption estimated in the event study, and we read them as scenarios conditional on this coefficient and not as predictions of what a change in fiscal rules would cause. We use them to describe which economies would see the largest changes in trade and fiscal revenue.

We find that digital product trade is lower in destinations with a DST. A one-percentage-point higher DST rate is associated with about 16% lower trade, and this association persists when we control for origin-by-category-by-year effects. However, this result comes from comparing destinations that tax with those that do not. Destination or country-pair fixed effects absorb this comparison, and once they are included the estimates are close to zero. Countries that adopted a DST thus imported fewer digital products than other countries, but their imports did not fall persistently after they introduced the tax. To explore the dynamics of this effect, we conduct an event study around DST adoption. We find that trade declines by 8–10% shortly after adoption (see also Hebous et al., 2026), but the decline is no longer present two years later. DSTs appear to have at most a short-lived effect on trade over the two years after adoption that the data cover. We then conduct a scenario analysis based on the event study estimate. We find that in 2024 (the last year with available data), adding tariff equivalents to DSTs would reduce digital trade by about US$9 billion, or 1.4% of the trade covered by the regression sample, and raise fiscal revenue from about US$2.7 billion to US$6.9 billion. Replacing DSTs with tariff equivalents would raise about US$4.3 billion. With the larger coefficient from the comparison across destinations, the decline in trade reaches US$49 billion (7.6%) and the revenue of the tariff-inclusive regime falls to US$4.0 billion.

Importantly, we find that the two instruments differ in the countries to which payments are attributed and in the countries that receive the revenue. We attribute each payment to the country where the supplying firm is located, which is an accounting allocation and does not show who ultimately bears the tax, since firms may pass part of it on to consumers through higher prices (Muddasani and Langenmayr, 2025). Under tariff-based regimes, fiscal receipts concentrate in middle-income economies such as Brazil, India and Mexico, which apply tariff equivalents well above the sample average, while payments are attributed to supplier origins such as the USA and China under the subsidiary rule. In contrast, DST liabilities are more closely tied to the location of subsidiaries and market-facing entities (Tørsløv, Wier and Zucman, 2023), concentrating payments in economies such as Ireland and Luxembourg, while revenues accrue primarily to large destination markets including Canada, France and Italy. In the combined regime, the geography of attributed payments and receipts lies between these two patterns, producing a mixed allocation across supplier origins and destination markets.

Overall, these conditional scenarios describe changes in the fiscal treatment of digital trade operating less through large aggregate trade contractions and more through a systematic reallocation of trade and fiscal revenues across countries.

# 2. Fiscal rules in the digital era

The rapid expansion of digitally delivered cross-border goods and services has challenged tax and customs rules designed for physical shipments. A key component of the international trade framework is the WTO Work Programme on Electronic Commerce, under which members maintained, from 1998, the practice of not imposing customs duties on electronic transmissions (ICC - International Chamber of Commerce, 2024; “WTO | Electronic Commerce Work Programme,” n.d.). The moratorium was last extended at the 13th Ministerial Conference in 2024 and lapsed at the end of March 2026, after the 14th Ministerial Conference reached no agreement. In practical terms, the moratorium implied a largely duty-free border environment for digital deliveries throughout the period we study. A software product or media title that would be dutiable if imported on a physical carrier typically does not face a customs duty when delivered as a download or stream. Although the lapse of the moratorium removed the WTO-wide commitment not to impose customs duties on electronic transmissions, some members have committed to maintaining duty-free treatment among themselves (Hebous et al., 2026).

With the tariff channel constrained, the fiscal treatment of cross-border digital transactions has increasingly shifted toward considerations of both multilateral coordination and unilateral tax measures. The multinational structure of leading digital firms has prompted policy initiatives at the Organisation for Economic Co-operation and Development (OECD), the United Nations, and the European Union. The most prominent effort is the OECD Inclusive Framework's Pillar One, which by June 2024, led negotiations between 147 countries ("Multilateral Convention to Implement Amount A of Pillar One," n.d.). The framework seeks to reallocate taxing rights between market jurisdictions for certain large multinational enterprises, reflecting the gap between physical presence and value creation in digitalized business models.

In the absence of political consensus, however, unilateral instruments have continued to expand in two directions.

First, by 2024, many countries had extended their VAT/GST coverage to digital products and services. VAT/GST is a broad consumption tax that typically applies to imports and domestic sales on the same basis and often coexists with tariffs in physical goods trade. Its extension to cross-border digital supplies has thus largely addressed the taxation of final consumption in market jurisdictions. Unlike customs duties, however, VAT/GST does not generally differentiate between domestic and foreign suppliers or among foreign countries of origin. This distinction motivates the focus of our main analysis on DSTs and tariff equivalents, while treating VAT/GST as part of the background fiscal environment, and using it only for additional robustness checks.

Second, countries have introduced DSTs and gross-basis withholding taxes to collect revenue from digital activity linked to their markets (Caragher, 2025). These measures are closer to the tariff instruments considered here because they impose charges on selected digital activities or cross-border payments, whereas VAT generally taxes domestic and imported consumption on the same basis. The comparison is imperfect, since DSTs can also apply to domestic suppliers and do not follow an origin-specific customs schedule. By 2024, DSTs and other gross-basis taxes on the digital services of non-resident suppliers had been introduced in 17 of the jurisdictions we code (Asquith, n.d.; Enache, 2025; "Taxation of the Digitalized Economy," n.d.). A DST is generally owed by the business providing the covered service, while withholding tax is deducted from the payment by the customer and remitted to the tax authority. DST liability depends on whether the business or group exceeds specified revenue thresholds. Once those thresholds are met, the tax is usually calculated as a percentage of revenues from covered activities attributable to domestic users, allowing for any exemptions or reliefs.

DSTs are generally levied on specific activities, and countries differ in which activities they tax, the rates they charge and the thresholds they set (Asquith, n.d.; Enache, 2025; "Taxation of the Digitalized Economy," n.d.). The United Kingdom, for example, introduced a 2% DST in April 2020 on revenues linked to UK users of search engines, social media platforms and online marketplaces. Groups are subject to the tax when annual revenues from these activities exceed £500 million worldwide and £25 million in the United Kingdom. The standard calculation applies the 2% rate to UK revenues above a £25 million annual allowance, although businesses can elect an alternative calculation for activities with low operating margins (HM Revenue & Customs, 2020). By contrast, France's DST, introduced in 2019, charges 3% on covered digital intermediation and targeted-advertising revenues attributable to France, with revenue thresholds of €750 million worldwide and €25 million in France. The French domestic threshold determines whether a business is subject to the tax. Unlike the UK allowance, it is not deducted from taxable revenues (Direction générale des Finances publiques, 2019).

How do these taxes affect digital trade, and how much revenue can they raise?

Existing policy analysis provides only a partial answer to this question. Hanappi et al. (2023, 2024), for example, argue that broad-based VAT is preferable to tariffs for digitized products on efficiency and revenue grounds. Their quantitative comparison, however, focuses on VAT and tariffs and calculates revenue potential without allowing trade flows to respond to taxation. Thus, it leaves open how DSTs compare with tariff-equivalent charges when trade adjusts, and how the associated payments and fiscal receipts are distributed across countries.

Similarly, Benz (2017) discusses how non-tariff policy regimes in services can be translated into tariff-equivalent trade costs using gravity models and shows that implied magnitudes can be large and highly sensitive to demand elasticities and sectoral heterogeneity. This result is directly relevant for digital-era fiscal rules, where the key wedges often operate through tax-inclusive prices, compliance costs, enforcement probability, and platform liability, and not through a posted tariff schedule. Yet, until recently, empirical progress has been constrained by a basic measurement limitation, because standard trade statistics do not directly observe many digitally delivered flows at the bilateral and product-segment level needed to evaluate counterfactual fiscal regimes.

New data construction efforts have begun to fill this gap. Bilateral services datasets such as BiTS (Li et al., 2025), which builds on BaTIS, provide evidence of how trade patterns change as services become less dependent on physical proximity. Research on digitally deliverable services also documents persistent international differences shaped by countries' earlier development paths (Di et al., 2025). These studies, however, work with broad service categories that do not separately identify many digital products. Stojkoski et al. (2024) provide more detailed estimates by constructing annual bilateral flows for 31 digital product categories from corporate revenue data. These include digitally delivered goods, digital advertising and intermediation fees such as accommodation and travel-booking commissions. The category detail allows covered digital activities to be matched with tax measures and physical product analogues, providing a basis for comparing DSTs with product-specific tariff equivalents.

Alongside these improvements in measurement, recent research has examined how digital taxation changes trade and how firms pass the cost on to their customers. Using bilateral services trade from the OECD–WTO BaTIS database, Hebous et al. (2026) find that DST adoption reduces imports of digital relative to non-digital services. Their broader comparison of tax instruments favours destination-based taxation over narrower measures that create greater distortions. The response to a DST also depends on how its cost is distributed along the supply chain. In this line,

Muddasani and Langenmayr (2025) find that Amazon increased its fees by about half the DST rate and that sellers passed most of the additional cost on to consumers. Thus, the business legally responsible for paying the tax need not bear most of its cost. Customs duties on electronic transmissions also involve a trade-off between revenue and trade. Indeed, Andrenelli and López González (2019, 2023) estimate that the revenue forgone under the WTO moratorium amounts to about 0.1% of total government revenue on average, while imposing duties would reduce trade, with particularly adverse effects for low-income countries and small firms.

Beyond revenue and trade responses, digital taxation raises questions about where taxing rights should lie and how digital activity is attributed across countries. Devereux and Vella (2018) and Becker and Englisch (2019) examine the role of market jurisdictions and user participation in allocating taxing rights. Tørsløv, Wier and Zucman (2023) show how profit shifting separates the location of reported profits from underlying economic activity. Hence, attributing a tax payment to a country does not establish where the activity takes place or who ultimately bears the tax.

Our contribution is twofold. First, we complement Hanappi et al. (2023, 2024) and Andrenelli and López González (2019, 2023) by comparing DSTs and tariff-equivalent charges on the same bilateral digital product flows, accounting for both potential revenues and conditional trade adjustments. Unlike the broad services categories used by Hebous et al. (2026), the estimates of Stojkoski et al. (2024) distinguish 31 digital product categories, allowing us to match DST coverage and product-specific tariff equivalents to a common trade base. We use this detail to compare the retention of DSTs, the addition of tariff-equivalent charges and the replacement of DSTs with those charges. Combining revenue calculations with gravity-based scenarios allows us to examine how changes in trade alter the revenue comparison across these regimes. The scenarios remain conditional on the estimated DST–trade association and the assumption that both instruments generate the same response to an equivalent fiscal charge.

Second, we complement the discussion of taxing rights and the location of digital activity in Devereux and Vella (2018) and Becker and Englisch (2019) by tracing the geographic allocation of payments and receipts under each regime. Bilateral flows attributed to supplying subsidiaries and destination markets allow us to identify which economies collect the implied revenues and which supplier economies are associated with the payments. In light of the separation between recorded multinational locations and underlying activity documented by Tørsløv, Wier and Zucman (2023), we interpret these results as accounting allocations. Unlike the pass-through analyses of Bibler, Teltser and Tremblay (2021) and Muddasani and Langenmayr (2025), our exercise does not establish who ultimately bears the tax. Instead, it shows how the choice between DSTs and tariff-equivalent charges redistributes fiscal payments and receipts across countries, alongside changes in aggregate trade and revenue.

The rest of the paper is organized as follows. The next section describes the data, the construction of the fiscal measures and the empirical strategy. We then examine the evidence on the relationship between DSTs and digital product trade, present the revenue calculations with trade held fixed, and compare the conditional scenarios for trade, revenue and their geographic allocation. The final section discusses the policy implications and the limitations of the analysis.

# 3. Data and methods

We quantify how alternative fiscal regimes could reshape these flows by combining a new bilateral dataset on digital product trade with Poisson Pseudo Maximum Likelihood (PPML) gravity estimates that map fiscal wedges into trade costs. We focus on two policy margins, the introduction of customs duties on electronic transmissions, which the lapse of the WTO moratorium in 2026 has made possible, and the role of destination-based digital services taxes (which currently are present in dozens of countries).

## 3.1. Data

We use data on digital product transactions taken from Stojkoski et al. (2024). This dataset provides annual estimates for bilateral trade in digital products across 189 countries, 31 digital sectors, and spans from 2016 to 2024. The dataset is constructed using a bottom-up pipeline that combines corporate revenue data for large digital firms with observed and model-imputed consumption patterns (Stojkoski et al., 2024).

In this dataset, digital products are defined through firm revenue models, covering digitally delivered goods, productized digital outputs, and platform intermediation fees. This is different from standard datasets on "digitally delivered trade" that bundle together heterogeneous activities such as online tutoring or other professional services delivered remotely. The dataset thus focuses on digital products for which producing an additional unit has a negligible marginal cost.

An advantage of this dataset is that it has two schemes to attribute trade to an origin country based on different revenue assignments. First is the subsidiary assignment, which attributes revenues to the fiscal residence of the subsidiary serving a destination market. The second is the headquarters assignment, which links all revenues directly to the country where the parent company is located. For example, a French sale served by an Irish subsidiary of a US parent is recorded as Ireland–France under subsidiary attribution and USA–France under headquarters attribution. In the main analysis, we use subsidiary attribution because it identifies the location of the supplying entity. The Supplementary Information reports regressions, event studies, and simulations using the headquarters attribution.

To examine how fiscal and trade policy instruments shape digital product trade, we complement the transaction data with a harmonized dataset of tax and border measures affecting digital transactions**.** This dataset covers (1) direct-tax measures (DSTs and withholding taxes) and (2) tariff-like measures, proxied via physical analogs, that vary by product and origin country. The 31 digital categories describe firm-revenue sectors, while HS6 codes classify only their physical tariff analogues.

We collect destination-sector DST and withholding tax rates by combining three sources, namely Avalara's Digital Services Tax Global Tracker, (Asquith, n.d.) Cristina Enache's DST in Europe, 2025 report (Enache, 2025), and KPMG's Taxation of the Digitalized Economy report ("Taxation of the Digitalized Economy," n.d.). Since these measures cover activities instead of following a product-by-product tariff schedule, we match the activities reported in the policy sources to the digital product categories in the trade dataset. We then assign the corresponding rates by destination, category, and year. Category-level differences thus reflect differences in the activities covered and the applicable rates. Our DST data cover DSTs together with other taxes levied on the gross revenue that non-resident suppliers earn from digital services, such as equalisation levies and withholding taxes, and they do not include levies on streaming services that fund domestic audiovisual or cultural production. Supplementary Table S1 reports the jurisdictions and rates included in our data.

Moreover, we follow Hanappi et al.(2024) and construct tariff-based counterfactuals by mapping each digital product category to its most similar dutiable physical HS6 product (reported in Supplementary Tables S2–S3). Although digital products and physical goods are not perfect substitutes, many digital products directly replicate content or functions that were historically traded as goods. These analogues therefore provide a benchmark for the duties that comparable digital products might face.

We construct these estimates in three steps. First, we manually create a concordance between physical and digital products by matching each digital product category to one or multiple HS6 categories that have the most similar use case and consumption form when delivered digitally (e.g., a streamed movie or a software download is matched with CDs). When multiple HS6 codes plausibly correspond to a single digital category, we retain all candidates and treat them as a set, without forcing a one-to-one match.

Second, we collect preferential and "most-favored-nation" (MFN) applied tariff rates data for these candidates from the UN Trade and Development TRAINS Database available at the World Integrated Trade Solution ("World Integrated Trade Solution (WITS) | Data on Export, Import, Tariff, NTM," n.d.) and the WTO Tariff and Trade Data (World Trade Organization, n.d.). Finally, we use the concordance to assign each digital product the average effective applied tariff rate of its corresponding physical HS6 products, where the effective applied tariff is defined as the minimum of the preferential and MFN applied rates. We interpret this average effective applied rate as the tariff wedge that would apply if an economically comparable transaction were delivered digitally. These rates are benchmarks, since the lapse of the moratorium does not automatically establish the same schedules, exemptions or enforcement rules for digital supplies.

We merge the digital product trade, DST, and tariff data with standard gravity controls and indexes capturing other dimensions of digital trade restrictiveness to isolate the effect of tariff wedge. We take bilateral geography and historical ties (distance, contiguity, common language, and colonial links) from CEPII's GeoDist dataset (Mayer and Zignago, 2011) and macroeconomic size variables, including gross domestic product (GDP), are sourced from the World Bank's World Development Indicators (World Bank International Economics Department Development Data Group, 1978). We also include the Digital Services Trade Restrictiveness Index (Digital STRI) of the importing country to capture non-tax policy frictions affecting cross-border digital transactions. This index is taken from the OECD (Ferencz, 2019) and combines comparable, policy-level information on barriers across five areas (infrastructure and connectivity, electronic transactions, payment systems, intellectual property rights, and other regulatory barriers) into a single composite measure that ranges from 0 (least restrictive) to 1 (most restrictive). Digital STRI measures non-tax regulatory barriers and is not a measure of digital services taxes.

Figure 1 provides a descriptive overview of the adoption and magnitude of DSTs and compares them to physical tariff-based analogs across income groups and over time. Figure 1(a) shows that DST adoption is a recent and uneven phenomenon, concentrated among high-income and lower-middle-income economies. The number of DST-adopting countries increases sharply after 2018. Interestingly, no low-income country adopts a DST over the sample period. By contrast, Figure 1(b) illustrates the near-universal application of tariffs on physical goods, with broad and stable coverage across income groups and years, highlighting a structural asymmetry between digital and physical border taxation.

Figure 1(c) compares average DST rates and tariff-equivalent rates across income groups in 2024 for countries that have adopted DSTs. While average DST rates are relatively similar across income groups, the corresponding tariff-equivalent rates display substantial heterogeneity and are highest for lower-middle-income economies. Figure 1(d) traces the evolution of these rates over time. DST rates remain relatively stable or decline modestly, whereas tariff-equivalent rates fluctuate more. These fluctuations mostly follow the changing set of countries with a DST, since each year's average covers only the destinations and categories with a positive DST rate in that year. In 2016 and 2017 this set contains only India's equalisation levy, which covers one category with a tariff equivalent of 3.4%. Pakistan and Uruguay enter in 2018. Zimbabwe enters in 2019 with 30 categories and an average tariff equivalent of 17.5%, and the average rises to 10.6%. The European DSTs that enter from 2020 have tariff equivalents close to zero, and the average falls again. The tariff equivalents of each destination change little over time, so the 2019 peak does not indicate an increase in tariffs.

These descriptive statistics suggest that DSTs are narrowly adopted and relatively homogeneous in rate design, while tariff-based wedges are widespread, heterogeneous, and often quantitatively larger. Thus, we should expect a different geography of fiscal allocations and trade adjustment across the two instruments.

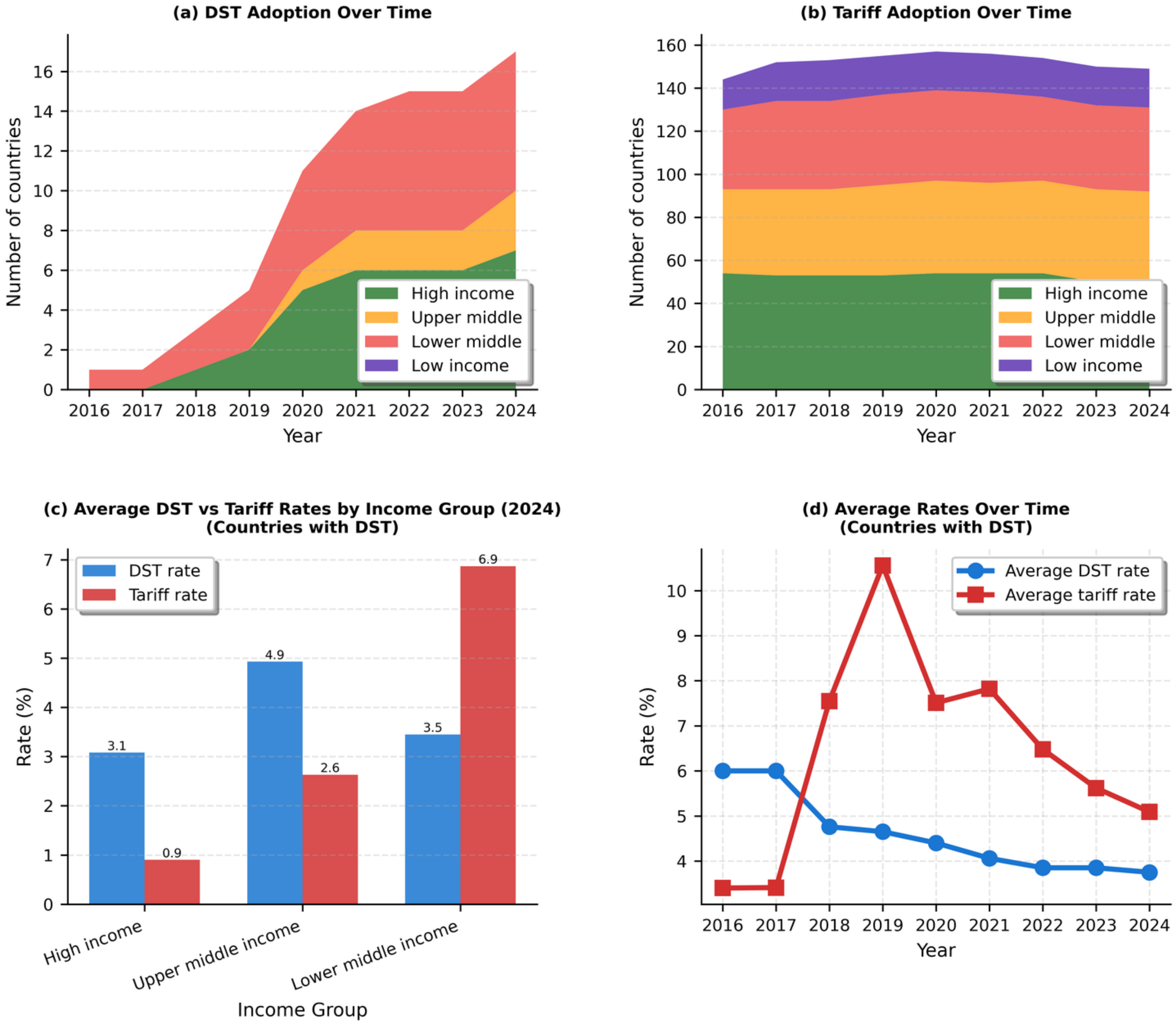

**Fig. 1 Adoption and rates of DSTs compared to physical goods tariff-based analogs.** Figure 1(a) Stacked area chart showing the number of countries adopting a DST over time by World Bank income group. Figure 1(b) Stacked area chart showing the number of countries applying tariffs on physical goods over time by income group. Figure 1(c) Bar chart comparing average DST rates and tariff-equivalent rates across income groups in 2024 for countries with DSTs. Figure 1(d) Line chart showing average DST rates and tariff-equivalent rates over time for countries with DSTs. Panels (c) and (d) average observations with a positive DST rate, and not all categories in the 17 jurisdictions with a DST. Tariff equivalents for other destinations enter the global scenarios but are not included in this conditional comparison.

## 3.2. Methods

In general, the fiscal revenues collected by destination country $d$ from origin $o$, product category $p$ and year $t$ from DST and tariffs can be written as:

$$Y_{odpt} = X_{odpt}(s_{dpt} + \tau_{odpt}), \quad (1)$$

where $X_{odpt}$ denotes the value of trade, $s_{dpt}$ is the destination/sector DST rate, and $\tau_{odpt}$ is the applied tariff rate. DST varies by destination, sector, and year, while tariffs are specific to the origin-destination-product triplet, and both rates enter as decimal shares (3% = 0.03). The resulting amounts are implied revenues, which differ from observed tax receipts and do not deduct administrative or compliance costs.

Because DSTs are already applied to digital transactions in a subset of countries, fiscal revenues from these instruments can be approximated by applying the DST rates to estimated trade values. Estimating the potential tariff revenue from digital trade is less straightforward, since digital products are not subject to customs duties.

To provide an estimate of this, we begin with a simple accounting exercise that provides an upper bound on potential tariff revenues, where we use, as tariff rates, the average tariff of the closest physical goods, as described in the data section. We assume that imposing tariffs on digital products would have no effect on observed trade volumes or patterns. While clearly unrealistic, this assumption isolates the mechanical revenue effect of tariffs and serves as a useful benchmark.

A more realistic benchmark assumes a gravity-type model that accounts for policy-induced adjustments in trade volumes and trade patterns. In this framework, bilateral trade is higher between larger economies and lower when bilateral frictions, such as distance, barriers, or taxes, increase trade costs (Anderson and van Wincoop, 2003; Head and Mayer, 2014).

To construct this benchmark, we estimate the gravity model by Poisson pseudo-maximum likelihood (PPML), the standard estimator for gravity equations (Santos Silva and Tenreyro, 2006; Yotov et al., 2016). PPML models trade flows in levels, so it keeps the observations with zero trade. It also remains consistent under the heteroskedasticity that is typical of trade data, which biases estimates from log-linear regressions (Santos Silva and Tenreyro, 2006, 2011).

The estimating equation is

$$X_{odpt} = exp\{\beta \, s_{dpt} + \gamma' Z_{odt} + F_{odpt}\} \, \varepsilon_{odpt} \quad (2)$$

where s is the DST rate of destination d on digital product category p in year t, in percentage points, and X is the value of trade in that category and year from origin o to destination d. The coefficient β is the conditional association between the DST and trade, and 100 × (exp(β) − 1) is the percentage change in trade associated with a one-percentage-point higher DST. We interpret β as a conditional association and not as a causal tax elasticity. The vector Z contains the standard gravity variables, namely log distance, contiguity, common colonizer, common official language and the log GDP of the origin and of the destination. It also includes the Digital Services Trade Restrictiveness Index (Digital STRI) of the destination, which measures regulatory barriers to digitally enabled services trade, so that the DST coefficient does not capture non-tax barriers to digital trade. The term ε is an error term with a conditional mean of one. Finally, F denotes the fixed effects, which differ across specifications.

The bilateral trade flows X of Stojkoski et al. (2024) are constructed by allocating firm revenues across countries using observed information and a gravity-type model that includes distance, GDP and common language, but no DST rates, tariffs or other policy variables. Hence, the tax measures in equation (2) do not enter the trade estimates by construction. However, the coefficients on the standard gravity variables in Z partly reflect the allocation procedure, so we do not interpret them as independent evidence of their effects on trade.

We consider five specifications that differ in their fixed effects. The first includes no fixed effects, and the second adds year and category fixed effects. With these effects, equation (2) mainly compares destinations with and without a DST, which may also differ in the size of their digital markets, their tax and enforcement capacity and their exposure to multinational platforms. To address this, the third specification replaces the year and category effects with origin-by-category-by-year effects, which absorb every shock specific to an exporter and product category in a year and nest the origin-by-year effects of structural gravity (Anderson and van Wincoop, 2003; Yotov et al., 2016). The fourth and fifth specifications add destination and country-pair fixed effects, respectively, so that the DST coefficient rests only on variation within destinations. We also report how much of the DST variation each set of fixed effects absorbs. Controls absorbed by a fixed effect, such as distance under country-pair effects, drop out of the estimation. Finally, we estimate an event study around DST adoption with the Poisson extended two-way fixed-effects estimator of Wooldridge (2023), which allows for heterogeneous effects across adoption cohorts and is consistent with the PPML gravity model. The Supplementary Information also reports estimates with the DST entered as log(1 + s) and as an adoption dummy, together with further clustering levels and other robustness checks.

We simulate trade under alternative fiscal regimes in two steps. First, we take the trade of each observation predicted by equation (2) with origin-by-category-by-year fixed effects, evaluated at the observed DST, as the baseline. Second, we change only the fiscal wedge and let trade respond with a coefficient taken from the event study, as it rests on changes in the DST within destinations around adoption. Because trade already declines before adoption in the event study, we use this coefficient as an illustrative short-run calibration and not as an identified causal response. We convert the effect in the adoption year into a coefficient per percentage point of DST, dividing by the average DST rate of the treated flows. Predicted trade in regime r is then the baseline multiplied by $\exp(\beta\ \Delta w)$, where β is this coefficient and Δw is the change in the wedge in percentage points, with all other conditions held fixed. The coefficient of equation (2) with origin-by-category-by-year effects also reflects other differences between taxing and non-taxing destinations, and we report the scenarios with it in place of β as the largest response we consider. Because these differences could move the coefficient in either direction, it is not a bound on the response of trade to the tax. The simulations thus show the changes in trade and revenue that the estimated response implies under each regime, and they should be read as conditional scenarios and not as forecasts of policy effects. The same coefficient applies to DSTs and tariff equivalents, since both raise the price of a cross-border digital transaction in proportion to its value. In the terminology of Head and Mayer (2014), the simulations are thus a Partial Trade Impact exercise, in which multilateral resistance, output, expenditure and income do not adjust, and firm relocation, platform repricing and substitution between digital and physical products are not modelled.

We consider three fiscal regimes for digital products, namely (i) the observed status quo with DST (where applicable) but no tariffs, (ii) a tariff-inclusive scenario that adds a tariff-equivalent rate on top of DST, and (iii) a no-DST scenario that removes DST while the tariff-equivalent rate is introduced.

The baseline (DST, no tariffs) corresponds to the prevailing treatment of digital trade in the sample period. This regime is directly observed in the data and is not simulated. It reflects the international framework of the sample period, in which the WTO moratorium on customs duties on electronic transmissions applied. Formally, the fiscal wedge entering the gravity equation in the baseline is:

$$\text{wedge}_{odpt} = s_{dpt}. \quad (3)$$

The tariff-inclusive regime (DST + tariffs) considers a scenario in which digital products are treated analogously to their physical counterparts for customs purposes. In addition to existing DST rates, digital transactions are assumed to face an ad valorem tariff equivalent to that applied to its closest physical analog, as determined by our digital-physical concordance. This regime represents a departure from the baseline by effectively lifting the moratorium on customs duties for digital products. The corresponding fiscal wedge is:

$$\text{wedge}_{odpt} = s_{dpt} + \tau_{odpt}. \quad (4)$$

Finally, the no-DST regime (only tariffs) considers a scenario in which destination-based digital services taxes are removed, while digital products are assumed to face only tariff-equivalent border taxation. By excluding DST while allowing tariffs to apply, this scenario isolates the extent to which border-based taxation could substitute for digital services taxes in generating fiscal revenues from digital trade. The fiscal wedge in this case is:

$$\text{wedge}_{odpt} = \tau_{odpt}. \quad (5)$$

With the wedge entering linearly, the exact proportional response of a flow to a change from wedge $w^0$ to $w^r$ (in percentage points) is $\exp\{\beta\,(w^r - w^0)\} - 1$, holding other regressors fixed.

# 4. Results

## 4.1. Empirical evidence on the fiscal-wedge coefficient

We first assess the empirical evidence on the fiscal-wedge coefficient. To this end, we estimate the five specifications of equation (2) described in Section 3 (Table 1) and an event study around DST adoption (Fig. 2). In the estimation sample, 14 of the 68 destinations introduce a DST, while the others remain in the sample with a zero rate. Since the DST is set at the destination level, we report p-values with standard errors clustered by destination and from a wild-cluster score bootstrap by destination, in addition to the standard errors clustered by origin–destination pair that are usual in gravity estimation.

Table 1 reports the DST coefficient under five sets of fixed effects, all with the gravity controls (log distance, contiguity, common colonizer, common official language, log GDP of origin and destination, and the Digital STRI of the destination). Without fixed effects (column 1) the DST is not associated with trade (0.023, SE 0.063). Once year and product-category effects are included (column 2), a one-percentage-point higher DST is associated with 12% lower trade (coefficient −0.132, p = 0.070). Column (3) replaces these with origin-by-category-by-year effects, which absorb every shock specific to an exporter and product category in a year and nest the origin-by-year effects of structural gravity. The coefficient is then −0.174 (SE 0.055, p = 0.002), which corresponds to 16% lower trade per percentage point, or about 51% at the average DST of taxed observations (4.1%). The negative association across destinations is thus not driven by exporter- or product-specific shocks. Columns (4) and (5) add destination and country-pair effects. The coefficient then falls to 0.003 (SE 0.042) and −0.017 (SE 0.044). These within-destination estimates are not statistically different from zero, and their 95% confidence intervals (−8% to +9% and −10% to +7% per percentage point) exclude the point estimates of columns (2) and (3).

**Table 1.** DST rate and digital product trade under alternative fixed effects (PPML, subsidiary attribution, 2016–2024)

| | (1) | (2) | (3) | (4) | (5) |
|---|---|---|---|---|---|
| DST rate (pp) | 0.023 | −0.132* | −0.174*** | 0.003 | −0.017 |
| | (0.063) | (0.073) | (0.055) | (0.042) | (0.044) |
| Log distance | −0.471*** | −0.490*** | −0.405*** | −0.705*** | |
| | (0.105) | (0.104) | (0.089) | (0.052) | |
| Contiguity | −0.446 | −0.478 | −0.210 | −0.320* | |
| | (0.377) | (0.367) | (0.199) | (0.177) | |
| Common colonizer | −0.667 | −0.544 | 0.601 | 0.711* | |
| | (0.484) | (0.469) | (0.570) | (0.426) | |
| Common official language | 1.003*** | 1.137*** | 0.371 | 0.087 | |
| | (0.297) | (0.295) | (0.230) | (0.205) | |
| Log GDP, origin | 0.443*** | 0.426*** | | | |
| | (0.071) | (0.078) | | | |
| Log GDP, destination | 0.783*** | 0.780*** | 0.810*** | 0.162 | 0.361* |
| | (0.076) | (0.077) | (0.054) | (0.214) | (0.191) |
| Digital STRI, destination | −1.695* | −1.282 | −1.514** | 0.330 | 0.687 |
| | (0.932) | (0.866) | (0.708) | (0.532) | (0.577) |
| Constant | −13.217*** | | | | |
| | (2.680) | | | | |
| Year and category FE | No | Yes | Yes | Yes | Yes |
| Origin × category × year FE | No | No | Yes | Yes | Yes |
| Destination FE | No | No | No | Yes | Yes |
| Country-pair FE | No | No | No | No | Yes |
| Effect of a 1 pp higher DST | +2.3% | −12.4% | −16.0% | +0.3% | −1.6% |
| p, standard errors clustered by destination | 0.737 | 0.045 | 0.007 | 0.941 | 0.632 |
| p, wild-cluster bootstrap by destination | 0.762 | 0.040 | 0.005 | 0.954 | 0.647 |
| Observations | 266,842 | 266,842 | 266,842 | 266,842 | 266,842 |
| Pseudo $R^2$ | 0.326 | 0.503 | 0.836 | 0.871 | 0.910 |

Notes: Dependent variable: bilateral digital product trade under subsidiary attribution (zeros included). The DST enters as its rate in percentage points, and the row "Effect of a 1 pp higher DST" reports 100 × (exp(b) − 1). All columns include the gravity controls, and controls absorbed by the fixed effects are not estimated (blank). Fixed effects marked Yes are included directly or nested in a finer fixed effect (origin × category × year effects contain year and category effects, and country-pair effects contain destination effects). Standard errors clustered by origin–destination pair (7,552 pairs) in parentheses. * p<0.10, ** p<0.05, *** p<0.01. The

two p-value rows give the p-value of the DST coefficient with standard errors clustered by destination (68 clusters) and from a wild-cluster score bootstrap by destination (Rademacher weights, 9,999 draws). All columns use the same observations, those that PPML keeps in every column, which excludes origins with no trade to any destination in a category and year and country pairs without any trade.

Table 2 shows where the variation that identifies the DST coefficient comes from. Year and category effects and origin-by-category-by-year effects absorb only 3% of the variance of the DST rate, so columns (2) and (3) mainly compare destinations with and without a DST. Destination effects absorb 52% of the variance, which is the entire difference between taxing and non-taxing destinations, and country-pair effects absorb the same share, as the DST does not vary by origin. Columns (4) and (5) thus rely only on changes in the DST rate within the 14 adopting destinations over time, which mostly come from the adoption of the tax itself. Destination-by-year effects, which structural gravity would add to absorb multilateral resistance, would absorb 81% of the variance and leave only differences in DST coverage across categories within a destination and year, and destination-by-category effects would raise the share to 93%. Hence, the two sets of estimates answer different questions. Columns (2) and (3) show that trade is lower in destinations that tax digital services, which may reflect the DST itself or other characteristics of these destinations. Columns (4) and (5) show that, on average, trade does not fall after a destination adopts a DST. As most DSTs were adopted in 2020 or later, this comparison covers only a few years after adoption.

**Table 2.** Share of the variation in the DST rate absorbed by alternative fixed effects

| **Fixed effects** | **Variance absorbed (%)** | **Variation left to identify the DST coefficient** |
|---|---|---|
| Year + category (column 2) | 2.8 | Differences between taxing and non-taxing destinations, and over time |
| Origin × category × year (column 3) | 3.0 | As in column (2), net of exporter–product shocks |
| Origin × category × year + destination (column 4) | 52.4 | Changes in the rate over time within the 14 adopting destinations |
| Origin × category × year + country pair (column 5) | 52.4 | As in column (4), as the DST has no origin dimension |
| Destination × year | 81.4 | Differences in DST coverage across categories within a destination and year |
| Destination × year + destination × category | 92.8 | Changes in coverage across categories within a destination over time |

Notes: Share of the variance of the DST rate (in percentage points) in the estimation panel (2,086,220 observations) explained by each set of fixed effects (exact least-squares projection). The first four rows correspond to columns (2)–(5) of Table 1. Fourteen destinations have a positive DST in at least one year.

The within-destination estimates average the response over all years after adoption and assume that it is the same for every adopter. As a result, they can hide a response that occurs only in the first years after adoption or that differs between early and late adopters, and they do not show whether trade was already changing before adoption. To examine the timing of the response, we estimate an event study around DST adoption. We define the treated unit as an origin–destination–category flow whose destination introduces a DST in that category, and event time 0 as the first year with a positive DST rate. Since most DSTs start between 2019 and 2021, we use a balanced window from two years before to two years after adoption and the cohorts observed over the whole window (adoption in 2018–2021, covering 12 destinations and 22,341 treated flows). We use the extended two-way fixed-effects Poisson estimator of Wooldridge (2023). Like the gravity regressions, it is estimated by PPML and measures the effect as a proportional change. It also estimates a separate effect for every adoption cohort and event year, which makes it robust to heterogeneous treatment effects under staggered adoption. In this model, trade is regressed on adoption-cohort-by-event-year indicators with origin-by-category-by-year and destination-by-category fixed effects, and destinations that never adopt a DST serve as controls. Indicators for the year two before adoption capture pre-trends, so all effects are measured relative to the year before adoption, and standard errors are clustered by destination. The Supplementary Information gives the full econometric setup.

Fig. 2 shows the Poisson estimates. Relative to the year before adoption, trade is 7.7% lower in the adoption year (95% interval −11% to −4%, $p < 0.001$) and 9.7% lower one year later (−17% to −2%, $p = 0.020$), and back at its pre-adoption level two years after adoption (−1.0%, $p = 0.82$). The average over the first two years is −8.7% ($p = 0.001$). Two years before adoption, however, trade is 10.4% above its level in the year before adoption ($p < 0.001$), so trade already declines before adoption. This decline has two possible readings. It may continue a downward trend in the trade of adopting destinations, and the decline after adoption would then overstate the response to the DST. It may also anticipate the tax, as most DSTs were announced one to two years before they took effect. The year before adoption would then already be affected, the decline after adoption would understate the response, and trade two years after adoption would still be about 10% lower than two years before adoption. When event time starts at the announcement of the DST, trade two years before the announcement is 7.7% above the year before ($p = 0.07$), so the earlier decline becomes smaller and only weakly significant. This fits the anticipation reading, but it is not causal evidence for it, and our data do not allow us to separate the two readings. The treated flows face an average DST of 4.0%, so the decline in the adoption year corresponds to about 2.0% lower trade per percentage point of DST. Column (3) of Table 1 implies 16.0% lower trade per percentage point, which for a DST of 4.0% would mean about 50% lower trade. The response of trade to the adoption of a DST is about one seventh of the difference in trade between taxing and non-taxing destinations.

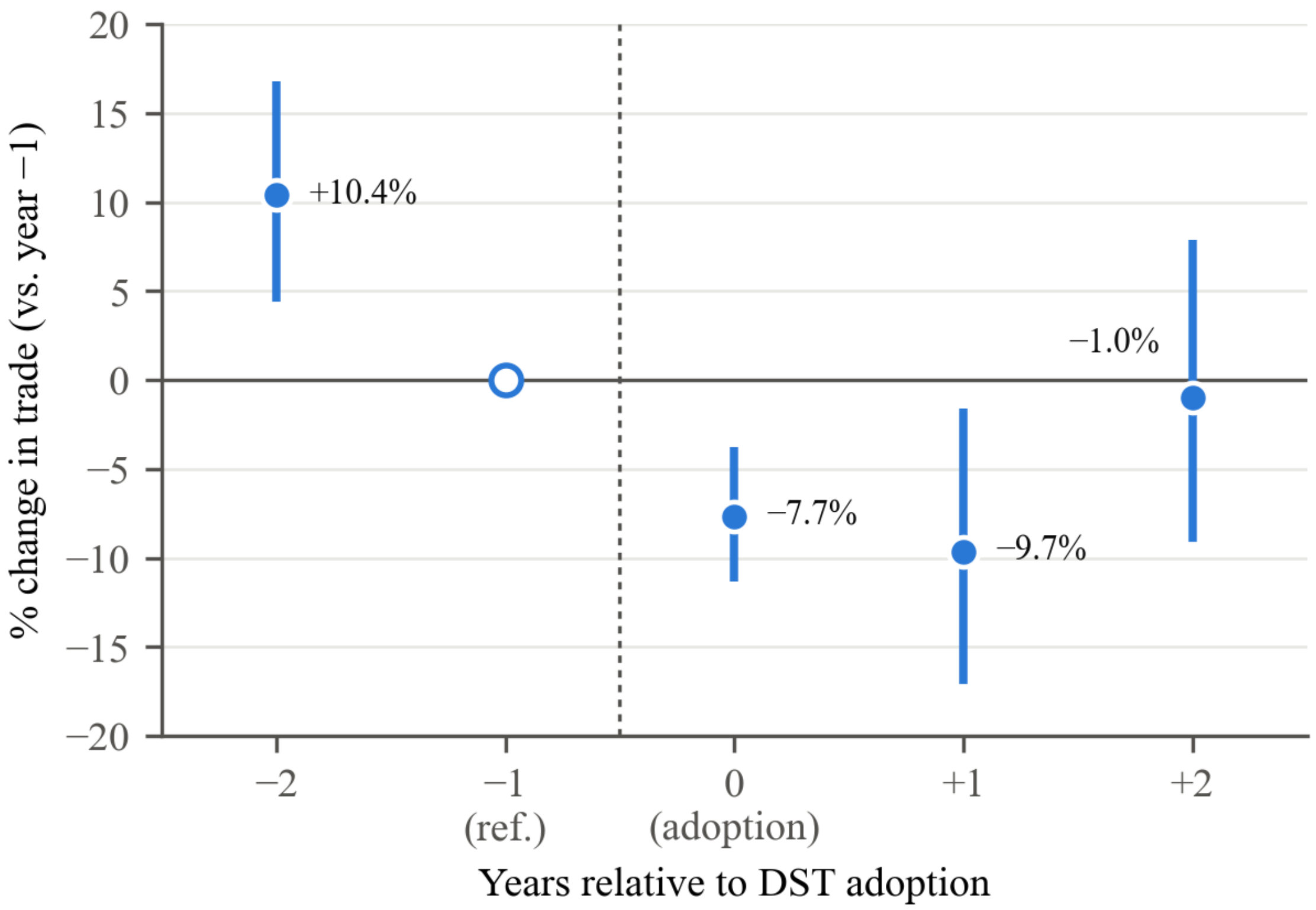

**Fig. 2 Percent change in digital product trade around DST adoption, relative to the year before adoption.** Poisson extended two-way fixed-effects estimator of Wooldridge (2023), 100 × (exp(b) − 1), with origin-by-category-by-year and destination-by-category fixed effects and never-treated controls. Subsidiary attribution, adoption cohorts 2018–2021 (12 destinations, 22,341 treated origin–destination–category flows) observed from two years before to two years after adoption, 95% confidence intervals, standard errors clustered by destination. The hollow marker is the reference year.

We investigate the robustness of our econometric results in nine ways, each reported in detail in the Supplementary Information. First, the results could depend on how the rate enters equation (2), so we also enter the DST as log(1 + rate) and as an adoption dummy. Both forms give the same pattern. The association across destinations is negative and statistically significant with origin-by-category-by-year effects, and the estimates within destinations are close to zero (Supplementary Tables S4–S5). Second, the data could place the supplying firm in the wrong country, so we attribute trade to the headquarters instead of the subsidiary. The results for the DST are the same (Supplementary Tables S6–S8). Third, we estimate the same regressions for VAT on digital services and for the DST plus VAT. VAT is the main tax on cross-border digital supplies, and unlike the DST it applies equally to domestic and foreign suppliers, so the comparison shows whether the results are specific to the DST. The association across destinations is much smaller for VAT, about 2% lower trade per percentage point. It appears only for subsidiary-attributed trade and is zero within destinations (Supplementary Tables S9–S20). Fourth, PPML drops different observations in different columns, so we also estimate each column of Table 1 on its own sample. Columns (3)–(5) are unchanged and columns (1) and

(2) move only slightly (Supplementary Tables S21–S38). Fifth, the pattern could depend on the fixed effects of Table 1 or on the level of clustering, so we use seven alternative fixed-effect structures and four sets of standard errors. The association is negative and statistically significant only in structures that compare destinations, while in structures that use variation within country pairs or within destinations it is positive and not statistically significant at the 5% level with clustered standard errors (Supplementary Table S39). Sixth, to see which controls drive the cross-sectional estimate, we add the gravity controls one group at a time. The association across destinations appears once destination GDP is controlled for, whereas the estimate within country pairs is not significant with any set of controls (Supplementary Table S40). Seventh, our bilateral flows are constructed from corporate revenue data, so we re-estimate Table 1 with bilateral trade from an independent source, the OECD–WTO Balanced Trade in Services dataset (BaTIS, Fortanier et al., 2017). BaTIS reports bilateral services trade by sector, which we convert into digitally delivered trade as Stojkoski et al. (2024) do. The pattern is the same as in Table 1, with a negative association across destinations of −0.154 and −0.167 per percentage point in columns (2) and (3) and none within destinations (Supplementary Table S41). Eighth, we vary the event study estimator. With both years before adoption as the baseline, the Poisson event study gives declines of 11.7% and 13.6% in the first two years after adoption, and a pooled short-run effect remains significant with a wild-cluster bootstrap ($p = 0.028$, Supplementary Table S42). The estimates are the same when later adopters serve as not-yet-treated controls, and averaging the cohorts by their counterfactual trade, as an imputation estimator does, gives larger declines of 15.0% and 17.3% and a decline of 9.6% two years after adoption, so the decline lasts longer in the largest DST markets. When event time starts at the year in which the DST was announced, for the ten destinations announced in 2018–2020, the estimate two years before the event is smaller and significant only at the 10% level, and the largest decline, 12.2%, comes a year after the announcement, when most of these DSTs took effect (Supplementary Table S42). We also estimate the event study with the heterogeneity-robust estimators of Callaway and Sant'Anna (2021), de Chaisemartin and D'Haultfœuille (2024) and Borusyak, Jaravel and Spiess (2024). These measure the effect on trade in levels instead of as a proportional change and use different control groups. They give similar declines in the first two years after adoption, and their confidence intervals include the Poisson estimates, but only the declines in the adoption year from the first two estimators are statistically significant, at the 10% level, and the other p-values are 0.19 or larger (Supplementary Table S43). Measured in US$, the effect depends on a few very large flows, as 1% of the positive flows account for 59% of trade, and only 12 destinations adopt a DST within the window, which widens the confidence intervals of these estimators. For the other outcomes and tax measures, the Poisson estimator gives a smaller decline under headquarters attribution and, after the extension of VAT to digital services, a decline of 5–10% in the adoption year only, which the linear estimators do not show and which rests on only seven destinations without VAT on digital services (Supplementary Tables S44–S48). With the BaTIS data, the Poisson event study gives a smaller decline than in Fig. 2, of about 5% one and two years after adoption (Supplementary Table S49). Ninth, jurisdictions differ in which taxes on digital services they levy, so we repeat Table 1 and the event study with statutory DSTs only, following the definition of Hebous et al. (2026), and with the levies on streaming services added to our main definition. Both definitions give the same pattern, with coefficients of −0.180 and −0.173 in column (3), estimates close to zero within destinations and declines of 7–12% in the adoption year (Supplementary Table S50).

Altogether, the difference between taxing and non-taxing destinations is an important feature of digital trade, as trade is lower in destinations with a DST even after controlling for origin-by-category-by-year shocks. The Digital STRI shows the same pattern in Table 1, with lower trade in destinations with higher regulatory barriers when destinations are compared (column 3) and no association within destinations (columns 4 and 5). The comparison across destinations could thus capture persistent differences in the policy environment for digital trade, of which the DST is one part. Within destinations, the adoption of a DST is followed by lower trade only in the short run, about 8% in the adoption year and 10% in the year after, and the estimators in levels give similar declines with wider confidence intervals. We use this short-run response in the simulations that follow, with the estimate for the adoption year, −0.020 per percentage point, and use the larger coefficient of column (3) as the largest response we consider.

## 4.2. Accounting benchmark

We begin our accounting exercises with a benchmark in which DST rates and tariff equivalents are applied to observed digital trade, with trade flows unchanged. We compute it on two trade bases. Figures 3 and 4 use all estimated cross-border trade in digital products. The regression sample, which keeps the 112 origins and 68 destinations for which the gravity controls and the Digital STRI are available, covers a smaller part of this trade, and the scenarios of Section 4.3 also use it. We compare the resulting revenue with the tariff revenue collected on the physical counterparts of digital products.

Figure 3 reports global trade aggregates and the fiscal revenues implied by the accounting benchmark. Figure 3(a) compares the total estimated digital product trade to the value of trade in the closest HS6 physical counterparts. Digital trade grows rapidly over 2016–2024, rising from roughly US$360 billion in 2016 to about US$1.05 trillion in 2024, while the physical counterparts increase more slowly and remain slightly below digital products at the end of the sample at around US$840 billion. This means that the gap between digital and physical narrows markedly, which motivates treating fiscal rules for digital deliveries as quantitatively relevant for aggregate trade and not only for a small niche segment.

Figure 3(b) translates these observed trade values into implied fiscal payments under the mechanical application of two wedges. Applying tariff equivalents to digital trade, interpreted as an accounting exercise of removal of the WTO moratorium on customs duties on electronic transmissions, generates the largest implied revenue stream and rises steadily from about US$4.1 billion in 2016 to about US$9.3 billion in 2024. Implied DST payments are smaller but increase sharply after 2018, reaching about US$3.5 billion by 2024, consistent with the gradual expansion of DST adoption and coverage over the period. For comparison, implied tariff revenues on the physical counterparts are relatively stable at roughly US$2.1 billion to US$2.6 billion, ending around US$2.6 billion in 2024. On the trade of the regression sample, US$652 billion in 2024, the same benchmark gives US$4.6 billion of tariff equivalents and US$2.9 billion of DST.

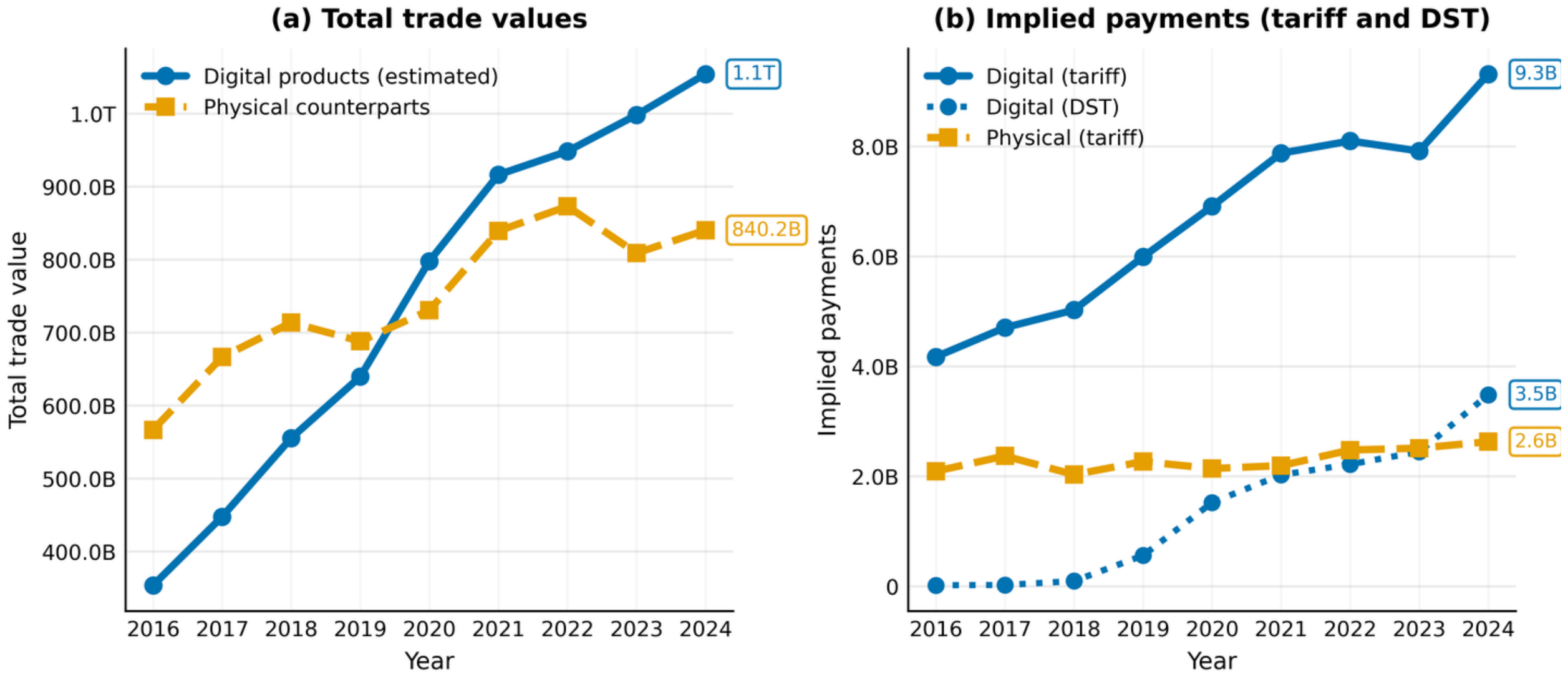

**Fig. 3 Global digital trade and implied fiscal revenues under an accounting benchmark.** (a) Total estimated global trade in digital products compared to trade in the closest HS6 physical counterparts, 2016–2024. (b) Implied fiscal revenues obtained by mechanically applying tariff-equivalent rates and DST to estimated digital trade values, and tariffs to physical counterparts.

Figure 4 illustrates the hypothetical geographic distribution of fiscal revenues and payments under the accounting benchmark, highlighting how the accounting allocation of digital taxation differs across instruments. Figure 4(a) shows the distribution of DST revenues across receiving countries. Revenues are highly concentrated in a small set of markets, with Canada, France, Italy and the United Kingdom among the largest recipients. Indeed, this result reflects the fact that unilateral DSTs are predominantly implemented in Europe, with Canada joining in 2024. Figure 4(b) reports tariff-equivalent revenues collected on digital products under a hypothetical removal of the WTO moratorium. In contrast to DST, revenues are concentrated in a different set of countries, with Brazil, China, and Argentina among the largest recipients (Supplementary Table S51). Current applied rates on physical goods are a benchmark. Governments that impose duties after the lapse of the moratorium might instead set new schedules, exemptions or enforcement rules.

Figure 4(c) depicts the distribution of DST payments across countries. Payments are concentrated in economies hosting digital subsidiaries and market-facing entities, with Ireland, the United States and Luxembourg accounting for a substantial share of total DST liabilities. Indeed, other research has also shown that DST liabilities fall disproportionately on a small number of multinational firms, largely headquartered in or operating through the United States and low-tax European hubs (Enache, 2025). Finally, Figure 4(d) shows tariff-equivalent payments on digital products. Here, implied payments are concentrated in supplier origins such as the United States under the same subsidiary-attribution rule, which does not establish headquarters incidence. Compared to DST, tariff-equivalent payments are more strongly concentrated on the supply side of digital trade.

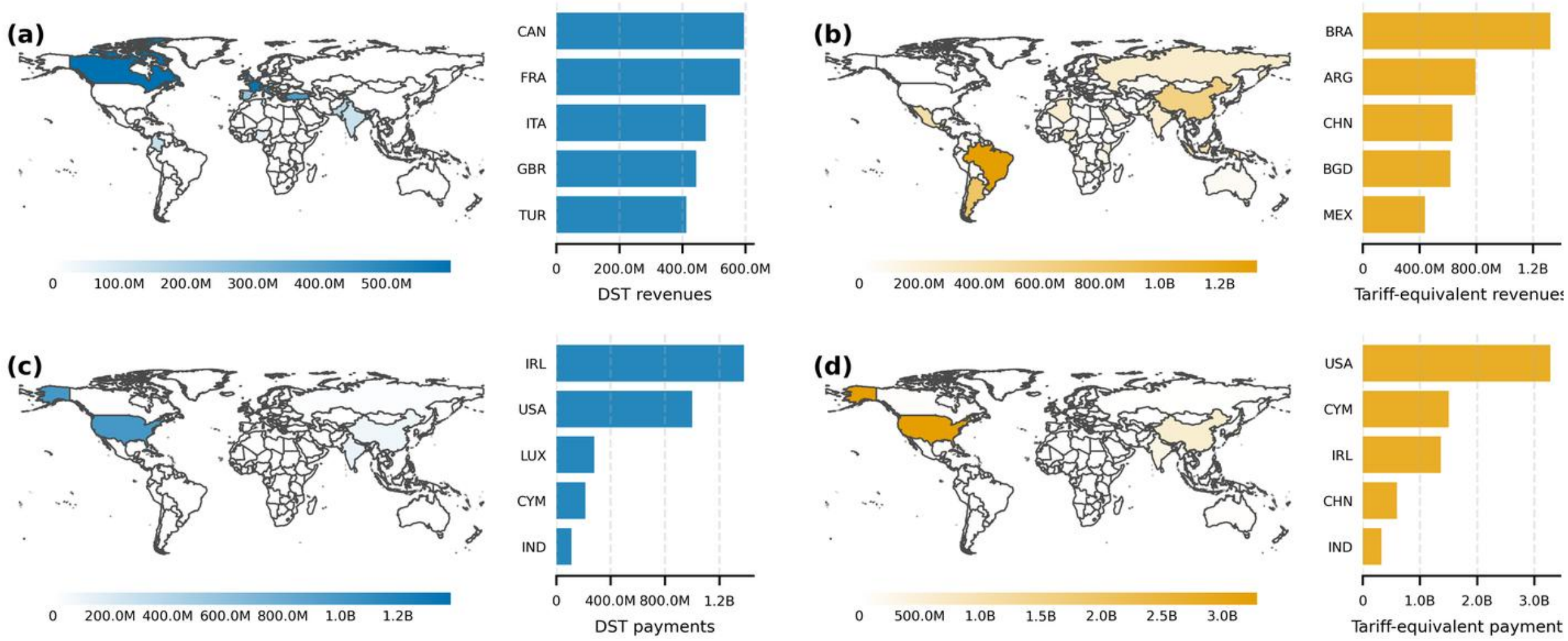

**Fig. 4 Geographic distribution of DST and tariff-equivalent revenues and payments (2024).** (a) DST revenues by destination country. (b) Tariff-equivalent revenues by destination country. (c) DST payments by origin country. (d) Tariff-equivalent payments by origin country. Payments use subsidiary attribution for both instruments and do not measure economic incidence.

The Supplementary Information extends the accounting benchmark in three directions. First, VAT is the main tax on cross-border digital supplies, so we compare the two instruments with it. Destination VAT rates applied to the trade of the regression sample would raise up to about US$101 billion in 2024, against US$4.6 billion of tariff equivalents and US$2.9 billion of DST on the same base, so DSTs and tariff equivalents are small compared with VAT (Supplementary Table S52). Second, some digital categories have only weak physical counterparts, so we change the HS6 analogues that define the tariff equivalents. Excluding the weakest analogues leaves the 2024 tariff-equivalent revenue on the regression sample practically unchanged (US$4.5 billion against US$4.6 billion), while using the lowest or the highest tariff among the analogues of each category gives US$1.2 billion or US$10.8 billion, respectively (Supplementary Tables S53–S54). Third, the geography of payments could depend on the attribution rule, so we attribute trade to the headquarters instead of the subsidiary. The recipients are practically unchanged, while payments attributed to subsidiary locations such as Ireland move to the countries where the parent companies are located, mainly the United States (Supplementary Table S55 and Figure S1).

The accounting benchmark shows that both the revenue totals and their distribution across countries depend on the instrument. While the global aggregates highlight which revenue streams are quantitatively most important, the disaggregated results show that different fiscal instruments matter for different countries, both on the revenue-collecting and revenue-paying sides. At the same time, these results are subject to an important limitation. By construction, the accounting benchmark applies fiscal wedges mechanically to observed trade values and thus abstracts from any adjustment in trade volumes or trade patterns. As a result, the implied revenues, particularly those associated with tariff-equivalent instruments, should be interpreted as upper bounds and not as predictions of realized revenues under alternative policy regimes. Moreover, the accounting approach cannot capture substitution across destinations or changes in sourcing and market structure that would likely accompany changes in fiscal treatment.

This limitation motivates the scenario analysis that follows.

### 4.3. Scenario analysis

We now turn to the scenario analysis, which allows digital trade flows to respond to changes in fiscal wedges. The figures use the event study estimate for the adoption year, −0.020 per percentage point. A one-percentage-point higher wedge thus lowers the trade of an observation by 2.0%, and wedges of 5%, 7.6%, 10% and 15% lower it by 9.5%, 14.1%, 18.2% and 26.0% relative to a zero wedge. The scenarios below should be read as conditional on the coefficient and not as predicted causal effects.

We use this estimate to evaluate three fiscal regimes, namely (i) the baseline regime that is currently in place with DST and without tariffs, (ii) a tariff-inclusive regime that adds tariff-equivalent rates to DST, and (iii) a no-DST regime that removes DST while introducing tariff-equivalent rates. These are the main options now that the WTO moratorium has lapsed, and the no-DST regime describes countries that withdraw their DSTs under a multilateral agreement on the taxation of the digital economy.

Figure 5 summarizes the implications for the covered regression sample of the fiscal regimes implied by the scenario analysis. Figure 5(a) reports the change in digital product trade in the covered sample relative to the baseline regime with DST and no tariffs. Introducing tariff-equivalent rates on top of DST leads to a persistent decline in global digital trade, reaching about US$9.3 billion in 2024, or 1.4% of the covered trade base. Removing DST while introducing tariff-equivalent rates lowers trade by about US$3.8 billion.

Figure 5(b) shows the corresponding implied fiscal revenues under each regime. Under the baseline regime, revenues generated by DST increase steadily over time, reaching approximately US$2.7 billion in 2024. This is less than the US$3.5 billion of Figure 3, which applies the rates to all estimated cross-border trade, and also slightly less than the US$2.9 billion that the accounting benchmark gives on the regression sample, as the scenario baseline distributes the trade of this sample across flows according to the gravity model. Adding tariff-equivalent rates on top of DST yields the largest revenue gains, with total revenues rising to about US$6.9 billion in 2024. The no-DST regime raises about US$4.3 billion, also more than the DST baseline, as the small trade response leaves most of the taxed base in place. Thus, the conditional scenarios illustrate a trade-off between trade volumes and fiscal revenues across regimes. Tariff-inclusive regimes raise more revenue but at the cost of larger trade reductions, while the DST-only baseline preserves higher trade volumes with more limited revenue generation. The combined regime raises the most revenue and produces the largest trade contraction.

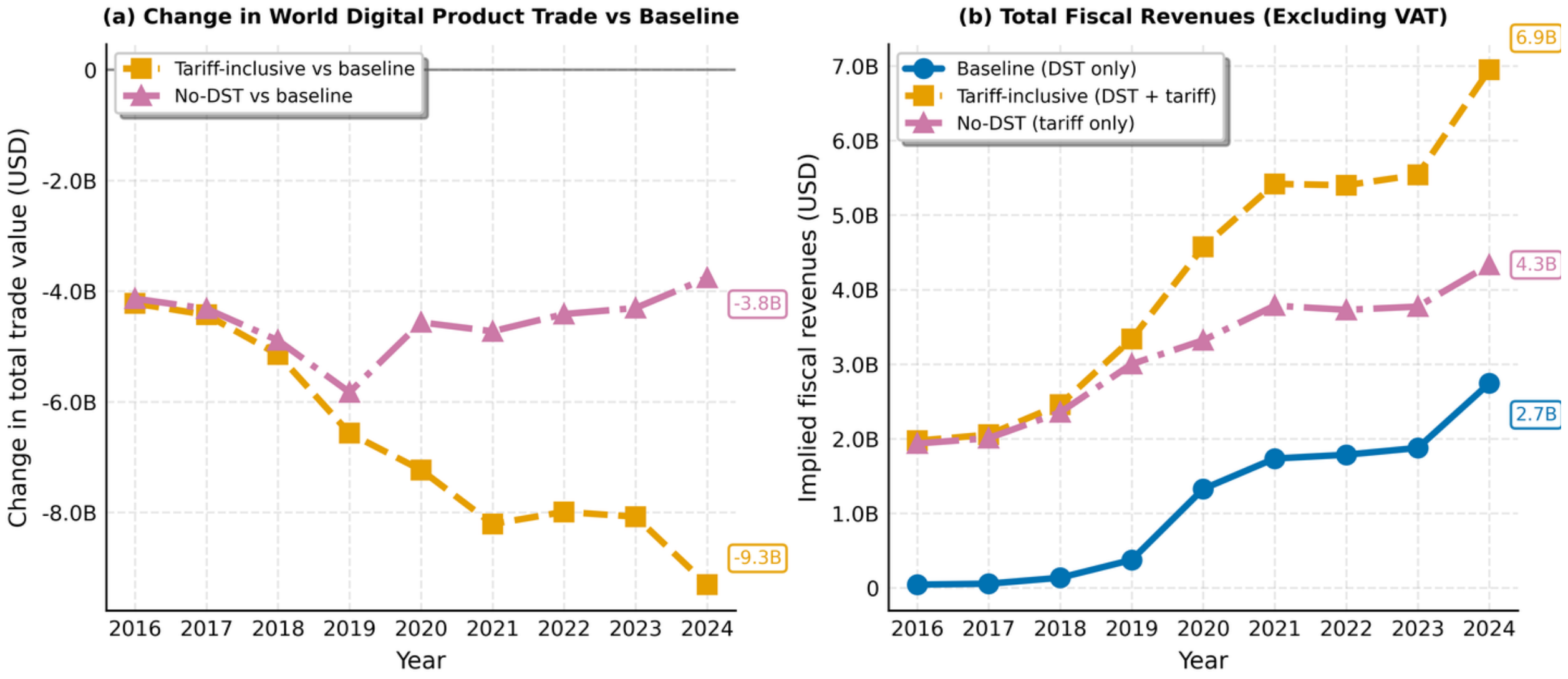

**Fig. 5 Scenario estimates for global trade and fiscal revenue effects under alternative fiscal regimes.** (a) Change in trade in the covered regression sample relative to the baseline regime with DST and no tariffs. (b) Total implied fiscal revenues excluding VAT under the three regimes: the baseline regime with DST only, a tariff-inclusive regime that adds tariff-equivalent rates to DST, and a no-DST regime with tariff-equivalent rates only. Values are shown in current US dollars.

Figure 6 summarizes the geographic allocation of fiscal payments, revenues, and trade adjustments implied by the scenarios under the two alternative policy regimes compared to the baseline scenario. Figures 6(a), (c), and (e) report outcomes under the no-DST regime, while Figures 6(b), (d), and (f) present the corresponding results under the tariff-inclusive regime (DST plus tariff-equivalent rates).

Figures 6(a) and (b) show the distribution of payments and the countries to which most payments are attributed. Under the no-DST regime (Figure 6a), payments are attributed to supplier origins under the subsidiary rule, with the United States accounting for the largest contribution, followed by Ireland and China. When tariff-equivalent rates are layered on top of DST (Figure 6b), the United States remains on top, but its relative share decreases at the expense of Ireland. China, the United Kingdom, and Japan also appear among the major contributors.

Figures 6(c) and (d) report the corresponding revenue recipients. In the no-DST regime (Figure 6 c), revenues accrue according to both the value of digital imports and the tariff-equivalent rates applied to their product-origin mix, with Brazil, India, Mexico, Argentina and Russia receiving the largest inflows. Under the tariff-inclusive regime (Figure 6d), the set of top recipients shifts toward a mix of large markets and established DST jurisdictions. Brazil and India remain the largest revenue recipients, followed by Canada, the United Kingdom and France, indicating a broader and more heterogeneous distribution of fiscal gains once tariffs are introduced alongside DST. The top recipients are thus a mix of two groups. Some, such as Brazil, India and Argentina, rank highly because of their high tariff rates. Brazil is only ninth among destinations by digital imports, but its average tariff equivalent of 6.5% is about nine times the sample average of 0.7%, and India and Argentina have small imports but average rates of 6.1% and 7.1% (Supplementary Table S51). Others, such as Canada, the United Kingdom and France, rank highly because they combine large digital imports with their existing DSTs.

Finally, Figures 6(e) and (f) show the countries experiencing the largest export declines. Under the no-DST regime (Figure 6 e), export losses are heavily concentrated in the United States, with smaller declines in China, Ireland, the United Kingdom and Japan. Under the tariff-inclusive regime (Figure 6 f), export contractions deepen and broaden. The United States again faces the largest decline, followed by Ireland, China, the United Kingdom, and Japan.

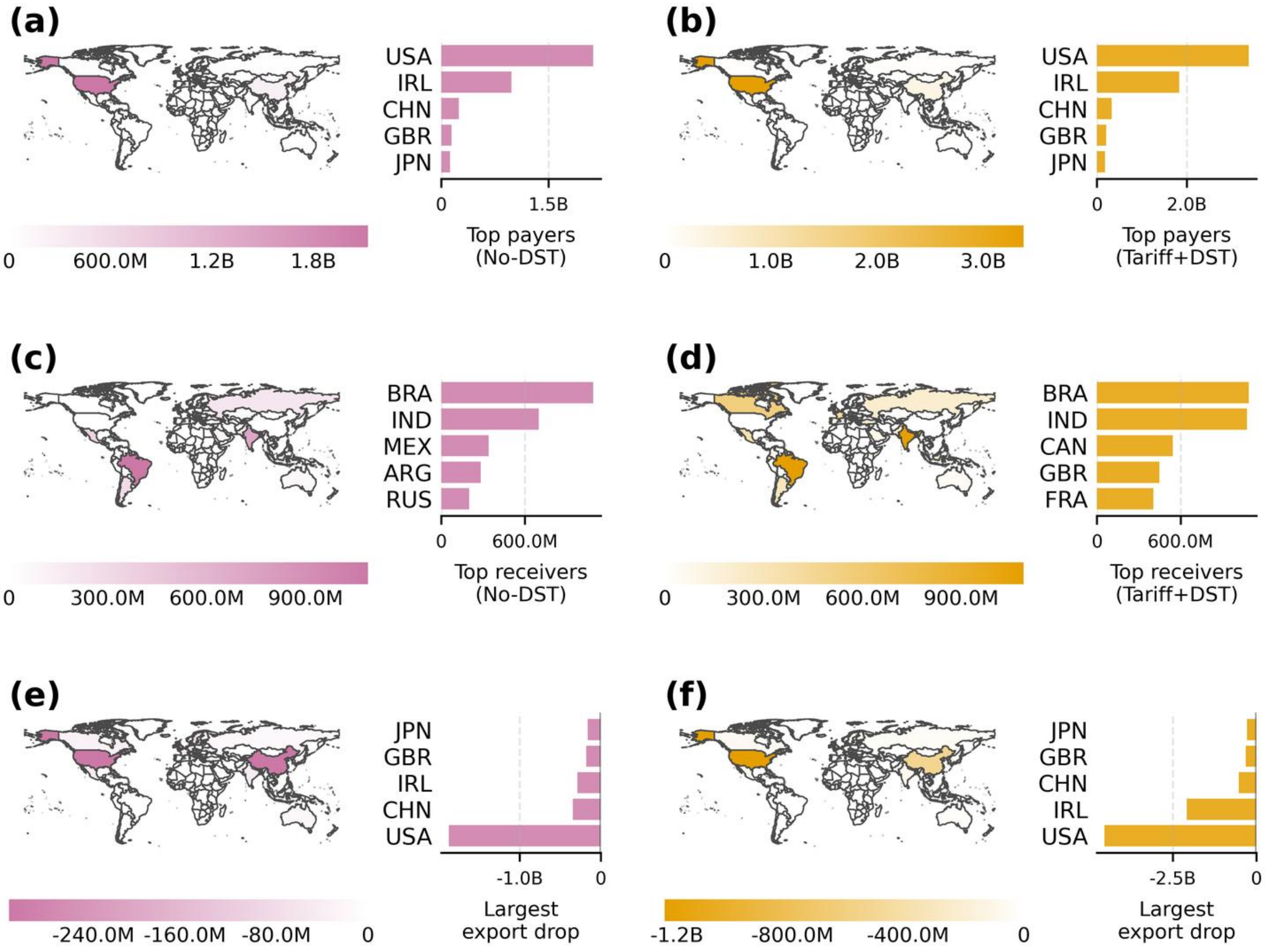

**Fig. 6 Scenario estimates of the geographic allocation of fiscal payments, revenues, and export losses under alternative regimes.** (a) Payments and the countries with the largest attributed payments under the no-DST regime. (b) Payments and the countries with the largest attributed payments under the tariff-inclusive regime (DST plus tariff-equivalent rates). (c) Revenues and top receiving countries under the no-DST regime. (d) Revenues and top receiving countries under the tariff-inclusive regime. (e) Export losses relative to the DST baseline and largest export declines under the no-DST regime. (f) Export losses and largest export declines under the tariff-inclusive regime. Payments are attributed to subsidiary-based origins, and the panels do not identify economic incidence.

Five additional analyses in the Supplementary Information show how sensitive the scenarios are to their main assumptions. First, the response of trade is the main assumption of the scenarios, so we repeat them with the other event study estimates and with the larger coefficient of column (3) of Table 1. The tariff-inclusive regime raises the most revenue for every coefficient. Its trade contraction in 2024 is about US$9 billion in the figures and US$49 billion (7.6%) with the coefficient of column (3), where a 5% wedge lowers the trade of an observation by 58%. With this coefficient the tariff-only regime raises less than the DST baseline (US$1.8 billion against US$2.7 billion), while the tariff-inclusive regime still raises the most (US$4.0 billion, Supplementary Table S56). Second, we add two further scenarios, a DST of 3% in every destination and the replacement of trade in the HS6 analogues by digital supply. These show the revenue at stake under broader policy changes. A DST of 3% in every destination would raise US$18.9 billion, almost seven times the current DST receipts, and lower trade by US$33 billion (Supplementary Table S57). Third, we repeat the geographic results under headquarters attribution, as in Section 4.2. The recipients are practically unchanged, while payments attributed to subsidiary locations such as Ireland move to the countries where the parent companies are located, mainly the United States (Supplementary Figures S2 and S3). Fourth, we give the tariff equivalents their own coefficient, estimated from goods trade in the HS6 analogues (−0.095 per percentage point) or taken from the product-level trade elasticities of Fontagné et al. (2022), as tariffs and DSTs may affect trade differently. Both imply a larger response than the event study, and the tariff-inclusive regime then raises US$4.9–5.0 billion and the tariff-only regime US$2.4–2.5 billion, slightly less than the DST baseline (Supplementary Section S4.7). The tariff-inclusive regime also raises the most revenue with every set of HS6 analogues of Section 4.2 (Supplementary Tables S53–S54). Finally, to see whether the scenario results depend on our trade data, we repeat them with bilateral digitally delivered services trade from BaTIS. The amounts are larger, because the BaTIS sectors also contain services such as banking, telecommunications and consulting, with US$9.3 billion for the DST baseline, US$20.4 billion for the tariff-inclusive regime and US$11.4 billion for the tariff-only regime, but the same ranking of the regimes and the same direction of the trade changes (Supplementary Table S61).

Overall, the scenarios show that fiscal wedges could reshape digital trade in a systematic but geographically concentrated way, with a small set of countries accounting for a disproportionate share of the implied payments, receipts, and export losses across regimes.

## 4.4. Scenarios only within DST-adopting destinations

Because DST adoption is limited to a relatively small set of destinations, the baseline regime is heterogeneous across countries, and the global tariff comparisons may partly reflect the fact that DST jurisdictions are not a random subset of markets but are typically large importers with high digital trade intensity. To isolate the implications for the jurisdictions where DST is operative, we complement the global results with a restrictive analysis that focuses only on countries that have introduced DST. In this restricted sample, the baseline is more comparable across destinations, allowing us to evaluate how adding tariff-equivalent rates changes trade and fiscal allocations within the DST-adopting group. The restriction uses destinations with a positive DST rate at least once in the sample and retains their other categories and years. It is not a restriction to positive-rate observations only.

Figure 7 shows trade and implied revenue when the analysis is restricted to DST-adopting destinations. Compared with the global results in Figure 5, the regimes differ more sharply within this group. In Figure 7(a), digital trade within DST destinations is higher when the DST is removed and only the tariff equivalents apply, whereas adding the tariff equivalents on top of the DST lowers it slightly in every year. Figure 7(b) shows that the tariff-inclusive regime again raises the most revenue, now followed by the DST baseline and then the no-DST regime. The higher trade under the no-DST regime fits the global decline in Figure 5, as the two results cover different destinations. In DST destinations the tariff equivalent replaces the DST, and trade rises where it is lower than the DST rate, while in the other destinations it is a new tax and trade falls. In sum, trade rises by about US$3.4 billion in DST destinations (US$3.0 billion in Figure 7, which rescales predicted trade within the group) and falls by US$7.2 billion elsewhere, which gives the global decline of US$3.8 billion (Supplementary Table S62).

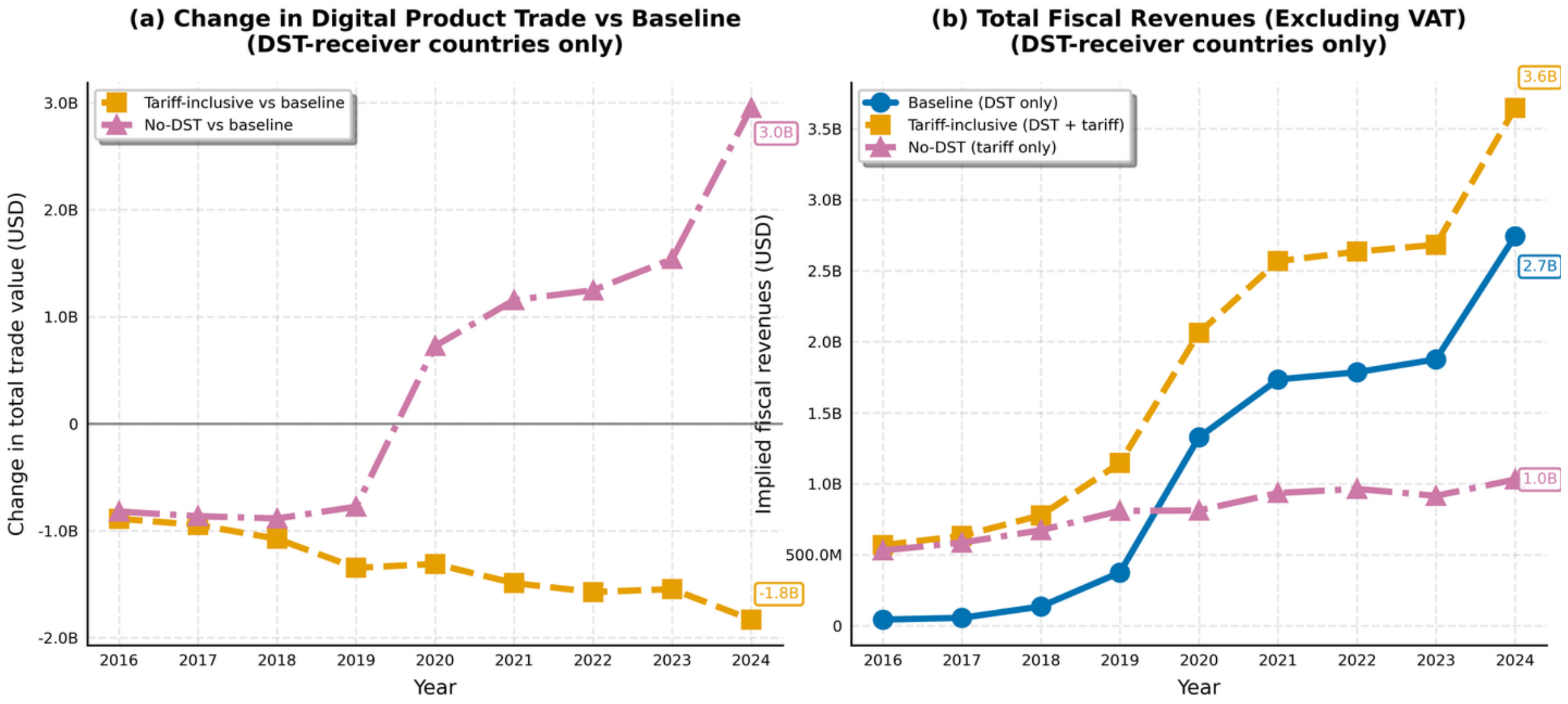

**Fig. 7 Scenario trade and fiscal revenue effects within DST-adopting destinations.** (a) Change in total digital product trade relative to the DST baseline, restricted to countries that levy DST. (b) Total implied fiscal revenues excluding VAT under alternative fiscal regimes within the same set of destinations.

Figure 8 shows the geographic allocation of the scenarios when the sample is restricted to DST-adopting destinations. Restricting the analysis to this set of countries changes the allocation results in two concrete ways relative to the global results in Figure 6. First, the countries with the largest attributed payments and receipts remain broadly the same, but the sample restriction compresses the scale. Under the no-DST regime, Figure 8(a) shows that the largest attributed payments are still in subsidiary-based supplier origins, led by the United States, Ireland and China, while Figure 8(c) shows that revenues accrue to middle-income economies such as India and Pakistan. Under the tariff-inclusive regime, Figure 8(b) again shows that Ireland and the United States dominate payments, and Figure 8(d) shows that revenues tilt toward the larger DST jurisdictions, with India, Canada and the United Kingdom at the top. In other words, attributed payments continue to be dominated by supplier locations and receipts by destination markets, just as in the global analysis. India and Pakistan are part of this group because we code India's equalisation levy and Pakistan's tax on gross payments as DSTs (Supplementary Table S1).

Second, within DST-adopting destinations, trade rises under the no-DST regime and falls under the tariff-inclusive regime. In contrast to the global results where export changes are mostly negative, Figure 8(e) shows that removing DST is associated with the largest export increases for the main supplier economies, with Ireland and the United States experiencing the strongest gains, followed by Japan, Korea and Singapore. Figure 8(f) shows the opposite under the tariff-inclusive regime. Export changes are negative and concentrated on the same supplier economies, with the largest declines borne by the United States and Ireland, followed by China and the United Kingdom. The key difference is thus not the ranking of countries by attributed payments and receipts, which remains stable, but the direction of the export response. Within DST jurisdictions, replacing the DST with a lower tariff-equivalent wedge can expand exports from the main suppliers, while layering tariffs on top of DST raises the overall wedge and generates export contractions for those same suppliers.

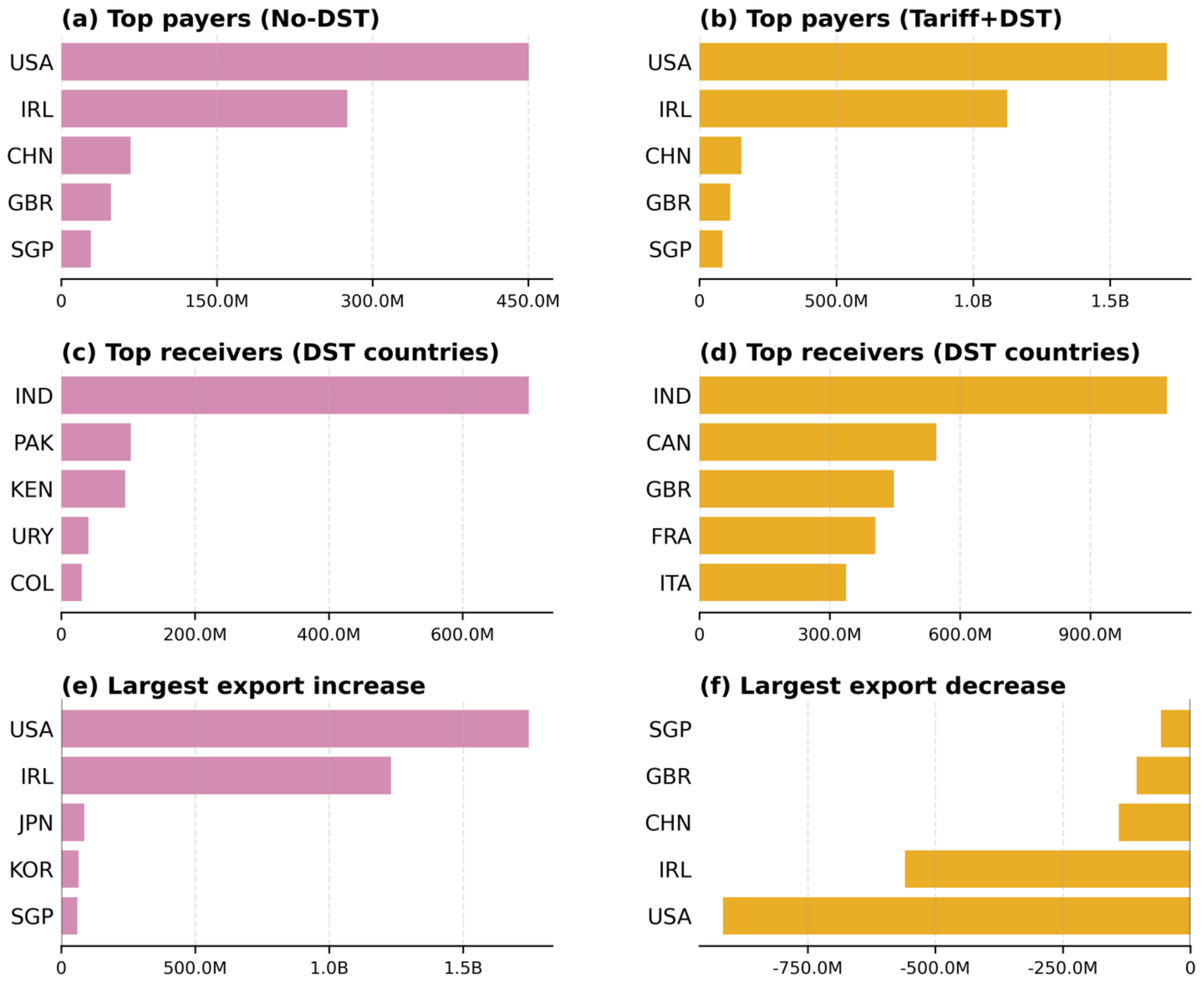

**Fig.8 Geographic allocation and trade adjustments within DST-adopting destinations.** (a) Countries with the largest attributed payments under the no-DST regime. (b) Countries with the largest attributed payments under the tariff-inclusive regime (DST plus tariff-equivalent rates). (c) Top revenue-receiving countries under the no-DST regime. (d) Top revenue-receiving countries under the tariff-inclusive regime. (e) Countries with the largest increase in digital exports relative to the DST baseline. (f) Countries with the largest decline in digital exports under the tariff-inclusive regime, relative to the DST baseline. Payment allocations use subsidiary origins, which are not estimates of economic incidence.

## 5. Conclusion

In this paper, we investigated how digital services taxes are associated with cross-border trade in digital products, and we used scenarios to describe trade and fiscal revenue if digital products also faced the tariffs applied to their physical analogues. We found a statistically significant negative association between digital product trade and DSTs across destinations, robust to origin-by-category-by-year effects, and estimates close to zero within destinations. An event study with a heterogeneity-robust Poisson estimator showed a statistically significant decline of about 8–10% in the first two years after adoption, after which trade returned to its pre-adoption level, with similar declines but wider confidence intervals from estimators in levels. This short-run decline corroborates Hebous et al. (2026), who find that DST adoption reduces imports of digital relative to non-digital services, and extends it to bilateral trade by digital product, where we do not find a decline beyond the first two years after adoption.

Based on the event study estimate, the scenarios implied small aggregate changes in trade. Introducing tariff-equivalent charges on digital products lowered trade by about 1.4% and raised the implied fiscal revenue about two and a half times, and replacing DSTs with tariff equivalents also raised more revenue than DSTs alone. These results depend on the response of trade that the scenarios assume. With the larger coefficient from the comparison across destinations, the trade contraction became large and the tariff-only regime raised less than DSTs. The tariff-only regime also raised less than DSTs when the tariff equivalents had their own coefficient, estimated from goods trade or taken from product-level elasticities. In all cases, the revenue at stake is small. In our accounting benchmark, tariff equivalents raise about US$9.3 billion in 2024 on all estimated cross-border trade and US$4.6 billion on the trade of the regression sample, of the same order as the tariff revenue potential of about US$4 billion in 2021 estimated by Hanappi et al. (2023, 2024), and small relative to public budgets, as in Andrenelli and López González (2019, 2023) and Hebous et al. (2026), who put the forgone tariff revenue at about 0.1% and 0.01–0.33% of total government revenue. VAT on the trade of the regression sample would raise up to about US$101 billion. Hanappi et al. (2023, 2024) find a smaller VAT potential of about US$10 billion in 2021, as they apply VAT only to final consumption, whereas our trade data do not separate business from consumer sales, so applying VAT to all flows gives an upper bound.

The choice of instrument also changed the countries to which payments are attributed. The largest share of DST payments was attributed to Ireland, where many large digital firms record their sales to DST countries, whereas the largest share of tariff-equivalent payments was attributed to the United States. Under headquarters attribution the recipients were unchanged, while payments attributed to subsidiary locations such as Ireland moved to the countries where the parent companies are located, mainly the United States and China (Supplementary Table S55). DST receipts accrue to the market countries that adopt them, whereas tariff-equivalent receipts accrue mostly to countries with high tariffs on the physical analogues, such as Brazil and Argentina. The two instruments would thus allocate taxing rights over digital trade to different countries, which extends the discussion of the taxing rights of market jurisdictions (Devereux and Vella, 2018; Becker and Englisch, 2019). When tariffs are layered on top of DST, the allocation combined both patterns, with larger trade contractions and a mixed geography of attributed payments and receipts. The policy choice concerns both revenue and trade openness and the allocation of fiscal payments across countries.

Our analysis has several limitations. First, we do not model the administrative, compliance and enforcement costs of applying tariffs or extending border taxation to digital products. These costs may be substantial, given the difficulty of valuing and classifying electronically delivered transactions (Andrenelli and López González, 2023). The revenue amounts are thus implied revenues conditional on existing institutional arrangements, before these costs.

Second, the geographic results describe where payments are recorded, and they depend on how trade is attributed to countries. Profit shifting separates the location of reported profits from underlying economic activity (Tørsløv, Wier and Zucman, 2023), and our headquarters attribution moves payments from subsidiary locations such as Ireland to the countries where the parent companies are located. Neither rule shows who ultimately bears the tax, which depends on pass-through to sellers and consumers (Bibler, Teltser and Tremblay, 2021; Muddasani and Langenmayr, 2025). Digital rules of origin would also require a choice among headquarters, billing entity, server and intellectual-property locations, with valuation, verification and enforcement costs that are not deducted here.

Third, the scenarios are a partial calculation. In the terminology of Head and Mayer (2014), they measure a Partial Trade Impact, which omits the adjustments of multilateral resistance, prices and incomes (Anderson and van Wincoop, 2003), as well as firm relocation, changes in platform pricing and longer-run changes in market structure. A structural general-equilibrium model of digital trade, for example with exact hat algebra (Dekle, Eaton and Kortum, 2008), is needed to assess these adjustments and is an important direction for future research.

Lastly, the association between the DST and digital trade was identified from differences between destinations and was not found within destinations, and the event study pointed to a temporary decline in trade after adoption. DSTs are recent, so with data for 2016–2024 adopters can be followed for only two years, and only 12 destinations adopt early enough to be observed over that window. Whether there is a lasting response cannot yet be assessed, and the scenarios thus rest on a short-run response of trade. The trade data are also estimated from corporate revenue data with an allocation model (Stojkoski et al., 2024) and therefore contain estimation error, which calls for further caution with the within-destination estimates. Longer panels that cover more adopters and more years after adoption would allow more precise estimates of the response of trade and its persistence.

Despite these limitations, now that the WTO moratorium has lapsed, these results could help governments understand the potential impact of DSTs and of customs duties on electronic transmissions on digital trade and fiscal revenue.

## Statements and Declarations

The authors declare no conflict of interest. This work was done with the financial and institutional assistance from the Hinrich Foundation.

# Supplementary Information for:
# The potential impact of fiscal rule changes on digital products trade

Sonja Mitikj[1] and Viktor Stojkoski[2,*]

[1]Faculty of Computer Science and Engineering, University Ss. Cyril and Methodius in Skopje, North Macedonia
[2]Faculty of Economics – Skopje, University Ss. Cyril and Methodius in Skopje, North Macedonia

* Corresponding author email address: vstojkoski@eccf.ukim.edu.mk

## Table of Contents

## S1. Data, concordance and DST rates

### S1.1 Jurisdictions with a DST in our data (Table S1)

The DST rates used in the paper come from our own compilation, which combines the Avalara Digital Services Tax Global Tracker, the Tax Foundation overview of DSTs in Europe and the KPMG summary of the taxation of the digitalized economy (Section 3 of the main text). Our DST data cover DSTs together with other taxes levied on the gross revenue that non-resident suppliers earn from digital services, such as equalisation levies, withholding taxes and taxes on a deemed share of gross revenue. Levies on streaming services that fund domestic audiovisual or cultural production are not included, because they finance a sector policy. Panel A of Supplementary Table S1 lists the 17 jurisdictions with a positive DST rate in at least one year, with the instrument, the year in which the tax was announced, the first year with a positive rate, the rates in 2024 and whether the jurisdiction is in the estimation sample. The first year with a positive rate can differ from the dates of enactment, legal effect or first payment. India's equalisation levy and Pakistan's tax on gross payments are included, which is why these countries belong to the DST-adopting destinations of Section 4.4. Panel B lists the streaming levies that the broad definition of Section S3.6 adds. Hungary has only zero rates in the sample and is therefore not among the 17 jurisdictions.

## Table S1. Jurisdictions with a DST in our data

| Jurisdiction | Instrument | Year announced | First year with a rate | Rate(s) in 2024 (%) | In estimation sample |
|---|---|---|---|---|---|
| Panel A. Main definition | | | | | |
| Austria (AUT) | DST | 2019 | 2020 | 5 | Yes |
| Canada (CAN) | DST (repealed retroactively in 2026) | 2020 | 2024 | 3 | Yes |
| Colombia (COL) | Significant economic presence tax (DST option) | 2022 | 2024 | 3 | Yes |
| Spain (ESP) | DST | 2018 | 2021 | 3 | Yes |
| France (FRA) | DST | 2018 | 2019 | 3 | Yes |
| United Kingdom (GBR) | DST | 2018 | 2020 | 2 | Yes |
| India (IND) | Equalisation levies (advertising from 2016, e-commerce from 2020) | 2016 and 2020 | 2016 | 2 and 6 | Yes |
| Italy (ITA) | DST | 2017 | 2020 | 3 | Yes |
| Kenya (KEN) | DST (repealed December 2024) | 2020 | 2021 | 1.5 | Yes |
| Nigeria (NGA) | Significant economic presence tax | 2019 | 2020 | 6 | No |
| Nepal (NPL) | DST | 2022 | 2022 | 2 | No |
| Pakistan (PAK) | Withholding tax on offshore digital services | 2018 | 2018 | 5 | Yes |
| Paraguay (PRY) | Tax on non-resident income from digital services (effective rate) | 2019 | 2021 | 4.5 | Yes |
| Tunisia (TUN) | DST | 2019 | 2020 | 3 | No |
| Turkey (TUR) | DST | 2019 | 2020 | 7.5 | Yes |
| Uruguay (URY) | Tax on non-resident income from digital services (effective rate) | 2017 | 2018 | 3.6 | Yes |
| Zimbabwe (ZWE) | Tax on foreign e-commerce and satellite broadcasting income | 2018 | 2019 | 5 | Yes |
| Panel B. Levies added in the broad definition | | | | | |
| Canada (CAN) | Streaming contribution | | 2024 | 5 | Yes |
| Switzerland (CHE) | Investment obligation for streaming services | | 2024 | 4 | Yes |
| Denmark (DNK) | Cultural levy on streaming | | 2024 | 2 | Yes |
| France (FRA) | Levy on music streaming | | 2024 | 1.2 | Yes |
| Poland (POL) | Levy on video on demand | | 2020 | 1.5 | Yes |
| Portugal (PRT) | Levy on subscription video on demand | | 2021 | 1 | Yes |

Notes: Panel A lists the jurisdictions of the main definition, which covers DSTs and gross-basis taxes on the digital services of non-resident suppliers. Panel B lists the levies on streaming services that the broad definition of Table S50 adds. The statutory definition of Table S50 keeps the DSTs and equalisation levies of Panel A. Sources: Avalara Digital Services Tax Global Tracker (Asquith, n.d.), Tax Foundation (Enache, 2025) and KPMG (Taxation of the Digitalized Economy, n.d.). Year announced is the year in which the government first announced the tax or submitted it to parliament. Effective rates apply a statutory rate to the share of gross revenue deemed to be income. Nigeria and Tunisia are not in the estimation sample because the Digital STRI is not available for them, and Nepal because the CEPII gravity variables are not available.

### S1.2 Digital products and their HS6 analogues (Tables S2–S3)

The tariff equivalents of Sections 4.2–4.4 apply to each digital category the tariff of the physical goods it can replace, as described in Section 3 of the main text. Supplementary Table S2 lists the HS6 analogues of each digital product sector, and Supplementary Table S3 describes each HS6 code. The tariff equivalent of a flow is the average applied tariff of these analogues for the same origin, destination and year.

## Table S2. Correspondence between digital product sectors and HS6 product categories

| Digital product sector | HS6 product categories |
|---|---|
| Software | 852349, 852351, 852359, 852380, 852329 |
| Payment Services | 847050, 847190, 847090, 847021, 852352, 852321 |
| Operating System | 852349, 852351, 852359 |
| Online Ride-Hailing | 847090, 847190, 851762, 852352, 852321 |
| Online Gambling | 950440, 950490 |
| Mobile Games | 950450, 852349, 852351, 852359, 950430, 950440 |
| PC and Console Games | 950450, 852349, 852351, 852359, 950430, 950440 |
| Online Travel Market | 847090, 847190, 491110 |
| Online Food Ordering | 847050, 847190, 847090, 847021 |
| Online Education | 490110, 490191, 490199, 852349, 852351, 852359 |
| eBooks | 490110, 490191, 490199, 490300, 490400, 490520, 491191, 491199 |
| Video on Demand | 852349, 852351, 852359, 370610, 370690, 852329 |
| Digital Music Streaming | 852349, 852351, 852359, 852329 |
| Digital Advertising | 490110, 490199, 490210, 490290, 491110, 491191, 491199 |
| Data Licensing | 852349, 852351, 852359, 852380, 852329 |
| Web Hosting | 847150, 847149, 847170 |
| File Hosting Service | 847150, 847149, 847170 |
| Cloud Computing | 847150, 847149, 847170 |
| Mobile Applications | 852349, 852351, 852359 |
| Online Marketplace (n.e.s.) | 847050, 847190, 847090, 847021, 490110, 490199, 491110 |
| Gaming Networks | 950450 |
| Online Dating | 490110, 490210, 490290 |
| Cybersecurity | 852349, 852351, 852359, 852380, 852329 |

Notes: The Software row applies to the nine software categories of the trade data (administrative, business intelligence, collaboration, creative, customer relationship management, enterprise resource planning, office, supply chain management and other enterprise software). HS6 lines 847130, 847141 and 847149 are excluded from the tariff data, so they do not contribute to the tariff equivalents.

## Table S3. HS6 product codes and their description

| HS6 code | Description (Harmonized System, revision 2022) |
|---|---|
| 370610 | Cinematographic film; exposed and developed, whether or not incorporating sound track or consisting only of sound track, of a width of 35mm or more |
| 370690 | Cinematographic film; exposed and developed, whether or not incorporating a sound track or consisting only of sound track, of a width less than 35mm |
| 490110 | Printed matter; in single sheets, whether or not folded |
| 490191 | Printed matter; dictionaries, encyclopedias and serial instalments thereof |
| 490199 | Printed matter; books, brochures, leaflets and similar printed matter n.e.c. in item no. 4901.10 or 4901.91 |
| 490210 | Newspapers, journals and periodicals; whether or not illustrated or containing advertising material, appearing at least four times a week |
| 490290 | Newspapers, journals and periodicals; whether or not illustrated or containing advertising material, appearing less frequently than four times a week |
| 490300 | Printed matter; children's picture, drawing or coloring books |
| 490400 | Music; printed or in manuscript, whether or not bound or illustrated |
| 490520 | Maps and hydrographic or similar charts; printed in book form, including atlases, topographical plans and similar |
| 491110 | Printed matter; trade advertising material, commercial catalogues and the like |
| 491191 | Printed matter; pictures, designs and photographs, n.e.c. in item no. 4911.10 |
| 491199 | Printed matter; n.e.c. in heading no. 4911 |
| 847021 | Calculating machines; electronic, incorporating a printing device, needing an external source of power |
| 847050 | Cash registers |
| 847090 | Machines incorporating a calculating device; n.e.c. in heading no. 8470 |
| 847149 | Automatic data processing machines; presented in the form of systems, n.e.c. in item no. 8471.30 or 8471.41 |
| 847150 | Units of automatic data processing machines; processing units other than those of item no. 8471.41 or 8471.49, whether or not containing in the same housing one or two of the following types of unit: storage units, input units or output units |
| 847170 | Units of automatic data processing machines; storage units |
| 847190 | Magnetic or optical readers, machines for transcribing data onto data media in coded form and machines for processing such data, not elsewhere specified or included |
| 851762 | Communication apparatus (excluding telephone sets or base stations); machines for the reception, conversion and transmission or regeneration of voice, images or other data, including switching and routing apparatus |
| 852321 | Magnetic media; cards incorporating a magnetic stripe, whether or not recorded, excluding products of Chapter 37 |
| 852329 | Magnetic media; other than cards incorporating a magnetic stripe, whether or not recorded, excluding products of Chapter 37 |
| 852349 | Optical media; recorded, excluding products of Chapter 37 |
| 852351 | Semiconductor media; solid-state non-volatile storage devices, whether or not recorded, excluding products of Chapter 37 |
| 852352 | Semiconductor media; smart cards, whether or not recorded, excluding products of Chapter 37 |
| 852359 | Semiconductor media; other than smart cards, whether or not recorded, excluding products of Chapter 37 |
| 852380 | Media n.e.c. in heading 8523, whether or not recorded, excluding products of Chapter 37 |
| 950430 | Games; operated by coins, banknotes, bank cards, tokens or by other means of payment, other than billiard articles and accessories, and automatic bowling alley equipment |
| 950440 | Games; playing cards |
| 950450 | Games; video game consoles and machines, other than those of subheading 9504.30 |
| 950490 | Games; articles for funfair, table or parlor games, including pintables, special tables for casino games, automatic bowling alley equipment, n.e.c. in heading 9504 |

Notes: Descriptions of the Harmonized System, revision 2022.

## S2. Gravity regressions

Section 4.1 of the main text finds a negative association between the DST and digital trade across destinations and no association within destinations. This section shows that this pattern does not depend on how the tax enters the regression, on the attribution of trade, on the estimation sample, on the clustering of the standard errors or on the controls. It also examines whether the pattern is specific to the DST or holds for any tax on digital supplies. For this purpose we repeat the regressions with the VAT rate that destinations apply to cross-border digital services and with the combined DST and VAT rate. All regressions use the estimation panel and the gravity controls of Table 1 and are estimated by PPML with standard errors clustered by origin–destination pair. The five columns of each table correspond to the five sets of fixed effects of Table 1. Each tax measure enters as its rate in percentage points, as log(1 + rate) or as a dummy for a positive rate, and the VAT rate is zero before the destination extends its VAT to non-resident digital suppliers.

### S2.1 All columns on a common sample (Tables S4–S20)

PPML drops the observations of fixed-effect groups without any positive trade, so columns (3)–(5) of Table 1 use about 267,000 observations, while columns (1) and (2) could use all 2.1 million. To make the columns comparable, Table 1 of the main text estimates every column on the observations used by every column, so that differences between columns come only from the fixed effects. Supplementary Tables S4–S20 apply the same common sample to the other combinations of outcome, tax measure and functional form.

For the DST the results are those of Table 1 in every functional form and under both attributions. With origin-by-category-by-year effects the coefficient is negative and statistically significant, −0.174 per percentage point under subsidiary attribution and −0.166 under headquarters attribution, and a positive DST is associated with 48% lower trade (−0.649 on the dummy). With destination or country-pair effects the coefficient is close to zero and not statistically significant in every form.

VAT, however, behaves differently. Under subsidiary attribution the VAT rate has only a small negative association with trade across destinations in column (3), −0.018 per percentage point, or 1.8% lower trade per point against 16.0% for the DST, and the dummy for a positive VAT rate is not statistically significant. Under headquarters attribution there is no negative association, and the dummy is even positive and significant in columns (1)–(3). Within destinations the VAT coefficient is close to zero in every form and under both attributions. The combined DST and VAT rate gives practically the same results as VAT alone (−0.019 per percentage point in column 3), because VAT rates are several times higher than DST rates and dominate the combined rate. A single wedge that combined the two taxes would thus hide the association of the DST with trade, which is why the paper uses the DST alone.

These results are what the design of VAT leads one to expect. VAT applies on the same basis to domestic and foreign suppliers, so unlike a DST or a tariff it does not change the price of imported relative to domestic digital products. A large share of digital trade is supplied to businesses, for which VAT is reverse-charged and deductible and is therefore not a cost. Most destinations had extended their VAT to non-resident digital suppliers by 2024, when 91% of the observations face a positive VAT rate, and VAT rates vary little across destinations and rarely change after adoption, which leaves little variation to identify a response. The positive dummy under headquarters attribution may reflect that the few destinations without VAT on digital services are smaller markets, and not an effect of the tax.

**Table S4. Digital product trade (subsidiary attribution) and the DST, log(1 + rate): common sample**

| | (1) | (2) | (3) | (4) | (5) |
|---|---|---|---|---|---|
| log(1 + DST rate) | 2.419 | −13.501* | −17.812*** | 0.279 | −1.703 |
| | (6.526) | (7.488) | (5.687) | (4.366) | (4.570) |
| Log distance | −0.471*** | −0.491*** | −0.406*** | −0.705*** | |
| | (0.105) | (0.104) | (0.089) | (0.052) | |
| Contiguity | −0.446 | −0.478 | −0.210 | −0.320* | |
| | (0.377) | (0.367) | (0.199) | (0.177) | |
| Common colonizer | −0.668 | −0.544 | 0.602 | 0.711* | |
| | (0.484) | (0.468) | (0.569) | (0.426) | |
| Common official language | 1.003*** | 1.137*** | 0.371 | 0.087 | |
| | (0.297) | (0.295) | (0.230) | (0.205) | |
| Log GDP, origin | 0.443*** | 0.426*** | | | |
| | (0.071) | (0.078) | | | |
| Log GDP, destination | 0.782*** | 0.780*** | 0.810*** | 0.162 | 0.361* |
| | (0.076) | (0.077) | (0.054) | (0.214) | (0.191) |
| Digital STRI, destination | −1.698* | −1.282 | −1.514** | 0.330 | 0.687 |
| | (0.933) | (0.866) | (0.708) | (0.532) | (0.577) |
| Constant | −13.216*** | | | | |
| | (2.680) | | | | |
| Year and category FE | No | Yes | Yes | Yes | Yes |
| Origin × category × year FE | No | No | Yes | Yes | Yes |
| Destination FE | No | No | No | Yes | Yes |
| Country-pair FE | No | No | No | No | Yes |
| Observations | 266,842 | 266,842 | 266,842 | 266,842 | 266,842 |
| Pseudo $R^2$ | 0.326 | 0.503 | 0.837 | 0.871 | 0.910 |

Notes: PPML. Dependent variable: bilateral digital product trade under subsidiary attribution (zeros included). Tax variable: DST, log(1 + rate). All columns include the gravity controls, and controls absorbed by the fixed effects are not estimated (blank). Fixed effects marked Yes are included directly or nested in a finer fixed effect. Standard errors clustered by origin–destination pair in parentheses. * $p<0.10$, ** $p<0.05$, *** $p<0.01$.

## Table S5. Digital product trade (subsidiary attribution) and the DST, dummy (rate > 0): common sample

| | (1) | (2) | (3) | (4) | (5) |
|---|---|---|---|---|---|
| DST > 0 (dummy) | 0.149 | −0.477 | −0.649*** | −0.010 | −0.088 |
| | (0.347) | (0.319) | (0.216) | (0.175) | (0.184) |
| Log distance | −0.470*** | −0.493*** | −0.410*** | −0.705*** | |
| | (0.104) | (0.103) | (0.089) | (0.052) | |
| Contiguity | −0.442 | −0.487 | −0.195 | −0.319* | |
| | (0.380) | (0.372) | (0.197) | (0.175) | |
| Common colonizer | −0.684 | −0.543 | 0.639 | 0.711* | |
| | (0.497) | (0.464) | (0.549) | (0.426) | |
| Common official language | 0.999*** | 1.140*** | 0.377 | 0.087 | |
| | (0.298) | (0.295) | (0.231) | (0.205) | |
| Log GDP, origin | 0.443*** | 0.426*** | | | |
| | (0.072) | (0.078) | | | |
| Log GDP, destination | 0.781*** | 0.781*** | 0.812*** | 0.160 | 0.357* |
| | (0.077) | (0.077) | (0.054) | (0.215) | (0.191) |
| Digital STRI, destination | −1.753* | −1.328 | −1.549** | 0.341 | 0.690 |
| | (0.965) | (0.876) | (0.711) | (0.528) | (0.572) |
| Constant | −13.166*** | | | | |
| | (2.729) | | | | |
| Year and category FE | No | Yes | Yes | Yes | Yes |
| Origin × category × year FE | No | No | Yes | Yes | Yes |
| Destination FE | No | No | No | Yes | Yes |
| Country-pair FE | No | No | No | No | Yes |
| Observations | 266,842 | 266,842 | 266,842 | 266,842 | 266,842 |
| Pseudo $R^2$ | 0.326 | 0.503 | 0.837 | 0.871 | 0.910 |

Notes: PPML. Dependent variable: bilateral digital product trade under subsidiary attribution (zeros included). Tax variable: DST, dummy (rate > 0). All columns include the gravity controls, and controls absorbed by the fixed effects are not estimated (blank). Fixed effects marked Yes are included directly or nested in a finer fixed effect. Standard errors clustered by origin–destination pair in parentheses. * p<0.10, ** p<0.05, *** p<0.01.

## Table S6. Digital product trade (headquarters attribution) and the DST, rate in percentage points: common sample

| | (1) | (2) | (3) | (4) | (5) |
|---|---|---|---|---|---|
| DST rate (pp) | 0.033 | −0.113* | −0.166*** | −0.011 | −0.013 |
| | (0.059) | (0.064) | (0.051) | (0.014) | (0.013) |
| Log distance | −0.184 | −0.175 | −0.246* | −0.401*** | |
| | (0.130) | (0.131) | (0.140) | (0.060) | |
| Contiguity | −0.591** | −0.626** | −0.883*** | −0.823*** | |
| | (0.236) | (0.247) | (0.239) | (0.276) | |
| Common colonizer | 0.334 | 0.471 | 0.582 | 0.820* | |
| | (0.951) | (0.992) | (0.867) | (0.482) | |
| Common official language | 0.511** | 0.787*** | 0.505** | 0.662*** | |
| | (0.254) | (0.262) | (0.236) | (0.125) | |
| Log GDP, origin | 1.047*** | 1.018*** | | | |
| | (0.062) | (0.088) | | | |
| Log GDP, destination | 0.844*** | 0.819*** | 0.922*** | 0.274** | 0.352*** |
| | (0.075) | (0.083) | (0.038) | (0.124) | (0.097) |
| Digital STRI, destination | −2.990*** | −2.452*** | −2.497*** | 0.239 | 0.089 |
| | (1.052) | (0.916) | (0.830) | (0.501) | (0.325) |
| Constant | −34.210*** | | | | |
| | (3.198) | | | | |
| Year and category FE | No | Yes | Yes | Yes | Yes |
| Origin × category × year FE | No | No | Yes | Yes | Yes |
| Destination FE | No | No | No | Yes | Yes |
| Country-pair FE | No | No | No | No | Yes |
| Observations | 137,854 | 137,854 | 137,854 | 137,854 | 137,854 |
| Pseudo $R^2$ | 0.549 | 0.781 | 0.942 | 0.963 | 0.986 |

Notes: PPML. Dependent variable: bilateral digital product trade under headquarters attribution (zeros included). Tax variable: DST, rate in percentage points. All columns include the gravity controls, and controls absorbed by the fixed effects are not estimated (blank). Fixed effects marked Yes are included directly or nested in a finer fixed effect. Standard errors clustered by origin–destination pair in parentheses. * p<0.10, ** p<0.05, *** p<0.01.

## Table S7. Digital product trade (headquarters attribution) and the DST, log(1 + rate): common sample

| | (1) | (2) | (3) | (4) | (5) |
|---|---|---|---|---|---|
| log(1 + DST rate) | 3.469 | −11.537* | −16.973*** | −1.133 | −1.355 |
| | (6.152) | (6.537) | (5.190) | (1.417) | (1.310) |
| Log distance | −0.184 | −0.175 | −0.247* | −0.401*** | |
| | (0.130) | (0.131) | (0.139) | (0.060) | |
| Contiguity | −0.591** | −0.626** | −0.884*** | −0.823*** | |
| | (0.236) | (0.247) | (0.239) | (0.276) | |
| Common colonizer | 0.334 | 0.471 | 0.584 | 0.820* | |
| | (0.951) | (0.992) | (0.867) | (0.481) | |
| Common official language | 0.511** | 0.787*** | 0.505** | 0.662*** | |
| | (0.254) | (0.262) | (0.236) | (0.125) | |
| Log GDP, origin | 1.047*** | 1.018*** | | | |
| | (0.062) | (0.088) | | | |
| Log GDP, destination | 0.844*** | 0.819*** | 0.922*** | 0.274** | 0.352*** |
| | (0.075) | (0.083) | (0.038) | (0.124) | (0.097) |
| Digital STRI, destination | −2.993*** | −2.452*** | −2.497*** | 0.239 | 0.089 |
| | (1.052) | (0.915) | (0.829) | (0.501) | (0.325) |
| Constant | −34.206*** | | | | |
| | (3.198) | | | | |
| Year and category FE | No | Yes | Yes | Yes | Yes |
| Origin × category × year FE | No | No | Yes | Yes | Yes |
| Destination FE | No | No | No | Yes | Yes |
| Country-pair FE | No | No | No | No | Yes |
| Observations | 137,854 | 137,854 | 137,854 | 137,854 | 137,854 |
| Pseudo $R^2$ | 0.549 | 0.781 | 0.942 | 0.963 | 0.986 |

Notes: PPML. Dependent variable: bilateral digital product trade under headquarters attribution (zeros included). Tax variable: DST, log(1 + rate). All columns include the gravity controls, and controls absorbed by the fixed effects are not estimated (blank). Fixed effects marked Yes are included directly or nested in a finer fixed effect. Standard errors clustered by origin–destination pair in parentheses. * p<0.10, ** p<0.05, *** p<0.01.

## Table S8. Digital product trade (headquarters attribution) and the DST, dummy (rate > 0): common sample

| | (1) | (2) | (3) | (4) | (5) |
|---|---|---|---|---|---|
| DST > 0 (dummy) | 0.213 | −0.369 | −0.573*** | −0.054 | −0.064 |
| | (0.326) | (0.274) | (0.206) | (0.048) | (0.046) |
| Log distance | −0.183 | −0.177 | −0.253* | −0.400*** | |
| | (0.130) | (0.130) | (0.135) | (0.060) | |
| Contiguity | −0.588** | −0.639*** | −0.916*** | −0.823*** | |
| | (0.236) | (0.248) | (0.236) | (0.275) | |
| Common colonizer | 0.310 | 0.474 | 0.677 | 0.826* | |
| | (0.956) | (0.992) | (0.830) | (0.480) | |
| Common official language | 0.498* | 0.795*** | 0.524** | 0.662*** | |
| | (0.256) | (0.262) | (0.237) | (0.125) | |
| Log GDP, origin | 1.045*** | 1.019*** | | | |
| | (0.061) | (0.088) | | | |
| Log GDP, destination | 0.841*** | 0.820*** | 0.925*** | 0.274** | 0.353*** |
| | (0.075) | (0.084) | (0.038) | (0.125) | (0.097) |
| Digital STRI, destination | −3.096*** | −2.505*** | −2.555*** | 0.238 | 0.088 |
| | (1.045) | (0.906) | (0.819) | (0.500) | (0.328) |
| Constant | −34.054*** | | | | |
| | (3.200) | | | | |
| Year and category FE | No | Yes | Yes | Yes | Yes |
| Origin × category × year FE | No | No | Yes | Yes | Yes |
| Destination FE | No | No | No | Yes | Yes |
| Country-pair FE | No | No | No | No | Yes |
| Observations | 137,854 | 137,854 | 137,854 | 137,854 | 137,854 |
| Pseudo $R^2$ | 0.549 | 0.780 | 0.942 | 0.963 | 0.986 |

Notes: PPML. Dependent variable: bilateral digital product trade under headquarters attribution (zeros included). Tax variable: DST, dummy (rate > 0). All columns include the gravity controls, and controls absorbed by the fixed effects are not estimated (blank). Fixed effects marked Yes are included directly or nested in a finer fixed effect. Standard errors clustered by origin–destination pair in parentheses. * p<0.10, ** p<0.05, *** p<0.01.

## Table S9. Digital product trade (subsidiary attribution) and the DST + VAT, rate in percentage points: common sample

| | (1) | (2) | (3) | (4) | (5) |
|---|---|---|---|---|---|
| DST + VAT rate (pp) | −0.009 | −0.013* | −0.019*** | 0.001 | 0.002 |
| | (0.007) | (0.008) | (0.005) | (0.006) | (0.006) |
| Log distance | −0.505*** | −0.527*** | −0.478*** | −0.705*** | |
| | (0.102) | (0.102) | (0.101) | (0.052) | |
| Contiguity | −0.498 | −0.514 | −0.348* | −0.320* | |
| | (0.382) | (0.366) | (0.206) | (0.177) | |
| Common colonizer | −0.719 | −0.725 | 0.354 | 0.711* | |
| | (0.454) | (0.442) | (0.541) | (0.426) | |
| Common official language | 0.974*** | 1.082*** | 0.242 | 0.087 | |
| | (0.302) | (0.298) | (0.244) | (0.206) | |
| Log GDP, origin | 0.445*** | 0.422*** | | | |
| | (0.072) | (0.078) | | | |
| Log GDP, destination | 0.779*** | 0.770*** | 0.805*** | 0.157 | 0.357* |
| | (0.074) | (0.075) | (0.053) | (0.209) | (0.186) |
| Digital STRI, destination | −1.557* | −1.511* | −1.769** | 0.329 | 0.625 |
| | (0.874) | (0.872) | (0.773) | (0.530) | (0.565) |
| Constant | −12.722*** | | | | |
| | (2.512) | | | | |
| Year and category FE | No | Yes | Yes | Yes | Yes |
| Origin × category × year FE | No | No | Yes | Yes | Yes |
| Destination FE | No | No | No | Yes | Yes |
| Country-pair FE | No | No | No | No | Yes |
| Observations | 266,842 | 266,842 | 266,842 | 266,842 | 266,842 |
| Pseudo $R^2$ | 0.327 | 0.502 | 0.837 | 0.871 | 0.910 |

Notes: PPML. Dependent variable: bilateral digital product trade under subsidiary attribution (zeros included). Tax variable: DST + VAT, rate in percentage points. All columns include the gravity controls, and controls absorbed by the fixed effects are not estimated (blank). Fixed effects marked Yes are included directly or nested in a finer fixed effect. Standard errors clustered by origin–destination pair in parentheses. * $p<0.10$, ** $p<0.05$, *** $p<0.01$.

## Table S10. Digital product trade (subsidiary attribution) and the DST + VAT, log(1 + rate): common sample

| | (1) | (2) | (3) | (4) | (5) |
|---|---|---|---|---|---|
| log(1 + DST + VAT rate) | −0.981 | −1.443* | −2.113*** | 0.038 | 0.233 |
| | (0.826) | (0.870) | (0.604) | (0.573) | (0.584) |
| Log distance | −0.504*** | −0.526*** | −0.475*** | −0.705*** | |
| | (0.103) | (0.103) | (0.102) | (0.052) | |
| Contiguity | −0.497 | −0.514 | −0.342* | −0.320* | |
| | (0.382) | (0.367) | (0.205) | (0.177) | |
| Common colonizer | −0.715 | −0.722 | 0.360 | 0.711* | |
| | (0.454) | (0.442) | (0.543) | (0.426) | |
| Common official language | 0.975*** | 1.083*** | 0.246 | 0.087 | |
| | (0.302) | (0.298) | (0.244) | (0.206) | |
| Log GDP, origin | 0.445*** | 0.422*** | | | |
| | (0.072) | (0.078) | | | |
| Log GDP, destination | 0.779*** | 0.771*** | 0.806*** | 0.159 | 0.356* |
| | (0.074) | (0.076) | (0.054) | (0.208) | (0.186) |
| Digital STRI, destination | −1.558* | −1.511* | −1.771** | 0.333 | 0.626 |
| | (0.877) | (0.876) | (0.777) | (0.530) | (0.563) |
| Constant | −12.761*** | | | | |
| | (2.516) | | | | |
| Year and category FE | No | Yes | Yes | Yes | Yes |
| Origin × category × year FE | No | No | Yes | Yes | Yes |
| Destination FE | No | No | No | Yes | Yes |
| Country-pair FE | No | No | No | No | Yes |
| Observations | 266,842 | 266,842 | 266,842 | 266,842 | 266,842 |
| Pseudo $R^2$ | 0.327 | 0.502 | 0.836 | 0.871 | 0.910 |

Notes: PPML. Dependent variable: bilateral digital product trade under subsidiary attribution (zeros included). Tax variable: DST + VAT, log(1 + rate). All columns include the gravity controls, and controls absorbed by the fixed effects are not estimated (blank). Fixed effects marked Yes are included directly or nested in a finer fixed effect. Standard errors clustered by origin–destination pair in parentheses. * p<0.10, ** p<0.05, *** p<0.01.

## Table S11. Digital product trade (subsidiary attribution) and the DST + VAT, dummy (rate > 0): common sample

| | (1) | (2) | (3) | (4) | (5) |
|---|---|---|---|---|---|
| DST + VAT > 0 (dummy) | 0.061 | −0.110 | −0.139 | −0.046 | 0.010 |
| | (0.159) | (0.181) | (0.142) | (0.067) | (0.062) |
| Log distance | −0.467*** | −0.489*** | −0.397*** | −0.705*** | |
| | (0.109) | (0.110) | (0.099) | (0.052) | |
| Contiguity | −0.441 | −0.468 | −0.219 | −0.319* | |
| | (0.376) | (0.362) | (0.197) | (0.177) | |
| Common colonizer | −0.651 | −0.633 | 0.490 | 0.711* | |
| | (0.480) | (0.470) | (0.604) | (0.426) | |
| Common official language | 1.007*** | 1.121*** | 0.329 | 0.088 | |
| | (0.300) | (0.298) | (0.231) | (0.206) | |
| Log GDP, origin | 0.443*** | 0.423*** | | | |
| | (0.071) | (0.078) | | | |
| Log GDP, destination | 0.780*** | 0.780*** | 0.812*** | 0.194 | 0.361* |
| | (0.079) | (0.081) | (0.059) | (0.203) | (0.184) |
| Digital STRI, destination | −1.632* | −1.629* | −1.957** | 0.338 | 0.647 |
| | (0.891) | (0.897) | (0.816) | (0.522) | (0.554) |
| Constant | −13.240*** | | | | |
| | (2.652) | | | | |
| Year and category FE | No | Yes | Yes | Yes | Yes |
| Origin × category × year FE | No | No | Yes | Yes | Yes |
| Destination FE | No | No | No | Yes | Yes |
| Country-pair FE | No | No | No | No | Yes |
| Observations | 266,842 | 266,842 | 266,842 | 266,842 | 266,842 |
| Pseudo $R^2$ | 0.326 | 0.501 | 0.834 | 0.871 | 0.910 |

Notes: PPML. Dependent variable: bilateral digital product trade under subsidiary attribution (zeros included). Tax variable: DST + VAT, dummy (rate > 0). All columns include the gravity controls, and controls absorbed by the fixed effects are not estimated (blank). Fixed effects marked Yes are included directly or nested in a finer fixed effect. Standard errors clustered by origin–destination pair in parentheses. * p<0.10, ** p<0.05, *** p<0.01.

## Table S12. Digital product trade (headquarters attribution) and the DST + VAT, rate in percentage points: common sample

| | (1) | (2) | (3) | (4) | (5) |
|---|---|---|---|---|---|
| DST + VAT rate (pp) | 0.010 | 0.011 | 0.002 | 0.001 | 0.002 |
| | (0.007) | (0.009) | (0.005) | (0.002) | (0.002) |
| Log distance | −0.151 | −0.138 | −0.217 | −0.401*** | |
| | (0.128) | (0.128) | (0.151) | (0.060) | |
| Contiguity | −0.454** | −0.494** | −0.771*** | −0.822*** | |
| | (0.216) | (0.218) | (0.243) | (0.276) | |
| Common colonizer | 0.385 | 0.492 | 0.496 | 0.820* | |
| | (0.953) | (0.994) | (0.912) | (0.480) | |
| Common official language | 0.546** | 0.781*** | 0.434* | 0.661*** | |
| | (0.249) | (0.259) | (0.248) | (0.125) | |
| Log GDP, origin | 1.035*** | 1.002*** | | | |
| | (0.055) | (0.080) | | | |
| Log GDP, destination | 0.850*** | 0.819*** | 0.906*** | 0.269** | 0.345*** |
| | (0.073) | (0.082) | (0.044) | (0.124) | (0.099) |
| Digital STRI, destination | −2.950*** | −2.891*** | −3.006*** | 0.204 | 0.037 |
| | (1.052) | (1.089) | (1.051) | (0.497) | (0.321) |
| Constant | −34.471*** | | | | |
| | (3.037) | | | | |
| Year and category FE | No | Yes | Yes | Yes | Yes |
| Origin × category × year FE | No | No | Yes | Yes | Yes |
| Destination FE | No | No | No | Yes | Yes |
| Country-pair FE | No | No | No | No | Yes |
| Observations | 137,854 | 137,854 | 137,854 | 137,854 | 137,854 |
| Pseudo $R^2$ | 0.550 | 0.781 | 0.939 | 0.963 | 0.986 |

Notes: PPML. Dependent variable: bilateral digital product trade under headquarters attribution (zeros included). Tax variable: DST + VAT, rate in percentage points. All columns include the gravity controls, and controls absorbed by the fixed effects are not estimated (blank). Fixed effects marked Yes are included directly or nested in a finer fixed effect. Standard errors clustered by origin–destination pair in parentheses. * $p<0.10$, ** $p<0.05$, *** $p<0.01$.

## Table S13. Digital product trade (headquarters attribution) and the DST + VAT, log(1 + rate): common sample

| | (1) | (2) | (3) | (4) | (5) |
|---|---|---|---|---|---|
| log(1 + DST + VAT rate) | 1.179 | 1.274 | 0.203 | 0.151 | 0.221 |
| | (0.770) | (0.983) | (0.563) | (0.242) | (0.215) |
| Log distance | −0.151 | −0.138 | −0.216 | −0.401*** | |
| | (0.128) | (0.128) | (0.151) | (0.060) | |
| Contiguity | −0.452** | −0.492** | −0.767*** | −0.822*** | |
| | (0.215) | (0.218) | (0.243) | (0.276) | |
| Common colonizer | 0.384 | 0.490 | 0.497 | 0.820* | |
| | (0.953) | (0.994) | (0.913) | (0.480) | |
| Common official language | 0.544** | 0.779*** | 0.435* | 0.661*** | |
| | (0.249) | (0.259) | (0.248) | (0.125) | |
| Log GDP, origin | 1.035*** | 1.002*** | | | |
| | (0.055) | (0.080) | | | |
| Log GDP, destination | 0.849*** | 0.818*** | 0.906*** | 0.268** | 0.344*** |
| | (0.073) | (0.082) | (0.044) | (0.124) | (0.099) |
| Digital STRI, destination | −2.952*** | −2.893*** | −3.008*** | 0.205 | 0.037 |
| | (1.051) | (1.087) | (1.052) | (0.497) | (0.321) |
| Constant | −34.450*** | | | | |
| | (3.036) | | | | |
| Year and category FE | No | Yes | Yes | Yes | Yes |
| Origin × category × year FE | No | No | Yes | Yes | Yes |
| Destination FE | No | No | No | Yes | Yes |
| Country-pair FE | No | No | No | No | Yes |
| Observations | 137,854 | 137,854 | 137,854 | 137,854 | 137,854 |
| Pseudo $R^2$ | 0.550 | 0.781 | 0.939 | 0.963 | 0.986 |

Notes: PPML. Dependent variable: bilateral digital product trade under headquarters attribution (zeros included). Tax variable: DST + VAT, log(1 + rate). All columns include the gravity controls, and controls absorbed by the fixed effects are not estimated (blank). Fixed effects marked Yes are included directly or nested in a finer fixed effect. Standard errors clustered by origin–destination pair in parentheses. * $p<0.10$, ** $p<0.05$, *** $p<0.01$.

## Table S14. Digital product trade (headquarters attribution) and the DST + VAT, dummy (rate > 0): common sample

| | (1) | (2) | (3) | (4) | (5) |
|---|---|---|---|---|---|
| DST + VAT > 0 (dummy) | 0.306*** | 0.274** | 0.174** | −0.012 | 0.014 |
| | (0.112) | (0.133) | (0.086) | (0.039) | (0.041) |
| Log distance | −0.169 | −0.155 | −0.215 | −0.401*** | |
| | (0.129) | (0.132) | (0.153) | (0.059) | |
| Contiguity | −0.505** | −0.549** | −0.744*** | −0.822*** | |
| | (0.228) | (0.238) | (0.248) | (0.276) | |
| Common colonizer | 0.318 | 0.416 | 0.481 | 0.821* | |
| | (0.952) | (0.992) | (0.912) | (0.481) | |
| Common official language | 0.521** | 0.754*** | 0.430* | 0.661*** | |
| | (0.249) | (0.260) | (0.249) | (0.125) | |
| Log GDP, origin | 1.038*** | 1.009*** | | | |
| | (0.060) | (0.086) | | | |
| Log GDP, destination | 0.831*** | 0.801*** | 0.898*** | 0.282** | 0.346*** |
| | (0.076) | (0.084) | (0.045) | (0.122) | (0.095) |
| Digital STRI, destination | −2.848*** | −2.783*** | −2.962*** | 0.220 | 0.056 |
| | (1.006) | (1.041) | (1.053) | (0.496) | (0.323) |
| Constant | −33.962*** | | | | |
| | (3.132) | | | | |
| Year and category FE | No | Yes | Yes | Yes | Yes |
| Origin × category × year FE | No | No | Yes | Yes | Yes |
| Destination FE | No | No | No | Yes | Yes |
| Country-pair FE | No | No | No | No | Yes |
| Observations | 137,854 | 137,854 | 137,854 | 137,854 | 137,854 |
| Pseudo $R^2$ | 0.550 | 0.780 | 0.939 | 0.963 | 0.986 |

Notes: PPML. Dependent variable: bilateral digital product trade under headquarters attribution (zeros included). Tax variable: DST + VAT, dummy (rate > 0). All columns include the gravity controls, and controls absorbed by the fixed effects are not estimated (blank). Fixed effects marked Yes are included directly or nested in a finer fixed effect. Standard errors clustered by origin–destination pair in parentheses. * p<0.10, ** p<0.05, *** p<0.01.

**Table S15. Digital product trade (subsidiary attribution) and the VAT, rate in percentage points: common sample**

| | (1) | (2) | (3) | (4) | (5) |
|---|---|---|---|---|---|
| VAT rate (pp) | −0.009 | −0.012 | −0.018*** | 0.001 | 0.004 |
| | (0.007) | (0.008) | (0.005) | (0.004) | (0.004) |
| Log distance | −0.506*** | −0.521*** | −0.469*** | −0.705*** | |
| | (0.103) | (0.102) | (0.102) | (0.052) | |
| Contiguity | −0.500 | −0.506 | −0.334* | −0.320* | |
| | (0.381) | (0.364) | (0.203) | (0.178) | |
| Common colonizer | −0.727 | −0.724 | 0.355 | 0.711* | |
| | (0.453) | (0.444) | (0.550) | (0.426) | |
| Common official language | 0.971*** | 1.085*** | 0.244 | 0.087 | |
| | (0.302) | (0.299) | (0.245) | (0.206) | |
| Log GDP, origin | 0.445*** | 0.422*** | | | |
| | (0.072) | (0.078) | | | |
| Log GDP, destination | 0.778*** | 0.770*** | 0.805*** | 0.158 | 0.347* |
| | (0.074) | (0.076) | (0.053) | (0.207) | (0.186) |
| Digital STRI, destination | −1.579* | −1.561* | −1.839** | 0.332 | 0.618 |
| | (0.871) | (0.874) | (0.784) | (0.528) | (0.557) |
| Constant | −12.691*** | | | | |
| | (2.521) | | | | |
| Year and category FE | No | Yes | Yes | Yes | Yes |
| Origin × category × year FE | No | No | Yes | Yes | Yes |
| Destination FE | No | No | No | Yes | Yes |
| Country-pair FE | No | No | No | No | Yes |
| Observations | 266,842 | 266,842 | 266,842 | 266,842 | 266,842 |
| Pseudo $R^2$ | 0.327 | 0.502 | 0.836 | 0.871 | 0.910 |

Notes: PPML. Dependent variable: bilateral digital product trade under subsidiary attribution (zeros included). Tax variable: VAT, rate in percentage points. All columns include the gravity controls, and controls absorbed by the fixed effects are not estimated (blank). Fixed effects marked Yes are included directly or nested in a finer fixed effect. Standard errors clustered by origin–destination pair in parentheses. * p<0.10, ** p<0.05, *** p<0.01.

## Table S16. Digital product trade (subsidiary attribution) and the VAT, log(1 + rate): common sample

| | (1) | (2) | (3) | (4) | (5) |
|---|---|---|---|---|---|
| log(1 + VAT rate) | −1.015 | −1.296 | −1.948*** | 0.048 | 0.380 |
| | (0.808) | (0.876) | (0.616) | (0.477) | (0.435) |
| Log distance | −0.504*** | −0.520*** | −0.466*** | −0.705*** | |
| | (0.104) | (0.103) | (0.102) | (0.052) | |
| Contiguity | −0.498 | −0.506 | −0.330 | −0.320* | |
| | (0.381) | (0.365) | (0.203) | (0.178) | |
| Common colonizer | −0.722 | −0.721 | 0.361 | 0.711* | |
| | (0.453) | (0.445) | (0.552) | (0.426) | |
| Common official language | 0.974*** | 1.086*** | 0.248 | 0.087 | |
| | (0.303) | (0.299) | (0.245) | (0.206) | |
| Log GDP, origin | 0.445*** | 0.422*** | | | |
| | (0.072) | (0.078) | | | |
| Log GDP, destination | 0.779*** | 0.771*** | 0.806*** | 0.158 | 0.347* |
| | (0.074) | (0.076) | (0.054) | (0.206) | (0.185) |
| Digital STRI, destination | −1.579* | −1.560* | −1.839** | 0.333 | 0.619 |
| | (0.874) | (0.877) | (0.788) | (0.528) | (0.557) |
| Constant | −12.739*** | | | | |
| | (2.525) | | | | |
| Year and category FE | No | Yes | Yes | Yes | Yes |
| Origin × category × year FE | No | No | Yes | Yes | Yes |
| Destination FE | No | No | No | Yes | Yes |
| Country-pair FE | No | No | No | No | Yes |
| Observations | 266,842 | 266,842 | 266,842 | 266,842 | 266,842 |
| Pseudo $R^2$ | 0.327 | 0.502 | 0.836 | 0.871 | 0.910 |

Notes: PPML. Dependent variable: bilateral digital product trade under subsidiary attribution (zeros included). Tax variable: VAT, log(1 + rate). All columns include the gravity controls, and controls absorbed by the fixed effects are not estimated (blank). Fixed effects marked Yes are included directly or nested in a finer fixed effect. Standard errors clustered by origin–destination pair in parentheses. * p<0.10, ** p<0.05, *** p<0.01.

## Table S17. Digital product trade (subsidiary attribution) and the VAT, dummy (rate > 0): common sample

| | (1) | (2) | (3) | (4) | (5) |
|---|---|---|---|---|---|
| VAT > 0 (dummy) | 0.069 | −0.091 | −0.124 | −0.047 | 0.009 |
| | (0.159) | (0.181) | (0.143) | (0.067) | (0.062) |
| Log distance | −0.467*** | −0.487*** | −0.396*** | −0.705*** | |
| | (0.109) | (0.110) | (0.100) | (0.052) | |
| Contiguity | −0.439 | −0.466 | −0.216 | −0.319* | |
| | (0.376) | (0.362) | (0.197) | (0.177) | |
| Common colonizer | −0.648 | −0.636 | 0.488 | 0.711* | |
| | (0.480) | (0.469) | (0.604) | (0.426) | |
| Common official language | 1.008*** | 1.121*** | 0.329 | 0.088 | |
| | (0.300) | (0.298) | (0.232) | (0.206) | |
| Log GDP, origin | 0.443*** | 0.423*** | | | |
| | (0.071) | (0.078) | | | |
| Log GDP, destination | 0.780*** | 0.780*** | 0.812*** | 0.194 | 0.362** |
| | (0.080) | (0.081) | (0.059) | (0.203) | (0.184) |
| Digital STRI, destination | −1.629* | −1.635* | −1.964** | 0.339 | 0.647 |
| | (0.887) | (0.892) | (0.812) | (0.522) | (0.554) |
| Constant | −13.237*** | | | | |
| | (2.655) | | | | |
| Year and category FE | No | Yes | Yes | Yes | Yes |
| Origin × category × year FE | No | No | Yes | Yes | Yes |
| Destination FE | No | No | No | Yes | Yes |
| Country-pair FE | No | No | No | No | Yes |
| Observations | 266,842 | 266,842 | 266,842 | 266,842 | 266,842 |
| Pseudo $R^2$ | 0.326 | 0.501 | 0.833 | 0.871 | 0.910 |

Notes: PPML. Dependent variable: bilateral digital product trade under subsidiary attribution (zeros included). Tax variable: VAT, dummy (rate > 0). All columns include the gravity controls, and controls absorbed by the fixed effects are not estimated (blank). Fixed effects marked Yes are included directly or nested in a finer fixed effect. Standard errors clustered by origin–destination pair in parentheses. * p<0.10, ** p<0.05, *** p<0.01.

## Table S18. Digital product trade (headquarters attribution) and the VAT, rate in percentage points: common sample

| | (1) | (2) | (3) | (4) | (5) |
|---|---|---|---|---|---|
| VAT rate (pp) | 0.010 | 0.013 | 0.004 | 0.003 | 0.003 |
| | (0.007) | (0.009) | (0.005) | (0.002) | (0.002) |
| Log distance | −0.151 | −0.133 | −0.206 | −0.401*** | |
| | (0.128) | (0.129) | (0.154) | (0.060) | |
| Contiguity | −0.456** | −0.474** | −0.731*** | −0.822*** | |
| | (0.216) | (0.216) | (0.246) | (0.276) | |
| Common colonizer | 0.390 | 0.506 | 0.511 | 0.820* | |
| | (0.953) | (0.995) | (0.922) | (0.480) | |
| Common official language | 0.547** | 0.791*** | 0.442* | 0.661*** | |
| | (0.248) | (0.258) | (0.247) | (0.125) | |
| Log GDP, origin | 1.036*** | 1.000*** | | | |
| | (0.055) | (0.080) | | | |
| Log GDP, destination | 0.850*** | 0.821*** | 0.906*** | 0.263** | 0.338*** |
| | (0.073) | (0.081) | (0.044) | (0.123) | (0.097) |
| Digital STRI, destination | −2.918*** | −2.855*** | −3.007*** | 0.198 | 0.029 |
| | (1.050) | (1.096) | (1.066) | (0.496) | (0.322) |
| Constant | −34.494*** | | | | |
| | (3.040) | | | | |
| Year and category FE | No | Yes | Yes | Yes | Yes |
| Origin × category × year FE | No | No | Yes | Yes | Yes |
| Destination FE | No | No | No | Yes | Yes |
| Country-pair FE | No | No | No | No | Yes |
| Observations | 137,854 | 137,854 | 137,854 | 137,854 | 137,854 |
| Pseudo $R^2$ | 0.550 | 0.781 | 0.939 | 0.963 | 0.986 |

Notes: PPML. Dependent variable: bilateral digital product trade under headquarters attribution (zeros included). Tax variable: VAT, rate in percentage points. All columns include the gravity controls, and controls absorbed by the fixed effects are not estimated (blank). Fixed effects marked Yes are included directly or nested in a finer fixed effect. Standard errors clustered by origin–destination pair in parentheses. * p<0.10, ** p<0.05, *** p<0.01.

## Table S19. Digital product trade (headquarters attribution) and the VAT, log(1 + rate): common sample

| | (1) | (2) | (3) | (4) | (5) |
|---|---|---|---|---|---|
| log(1 + VAT rate) | 1.173 | 1.484 | 0.442 | 0.255 | 0.347 |
| | (0.772) | (1.005) | (0.546) | (0.260) | (0.238) |
| Log distance | −0.151 | −0.134 | −0.206 | −0.401*** | |
| | (0.128) | (0.129) | (0.154) | (0.060) | |
| Contiguity | −0.453** | −0.473** | −0.729*** | −0.822*** | |
| | (0.216) | (0.216) | (0.245) | (0.276) | |
| Common colonizer | 0.389 | 0.505 | 0.510 | 0.820* | |
| | (0.953) | (0.995) | (0.922) | (0.480) | |
| Common official language | 0.546** | 0.789*** | 0.442* | 0.661*** | |
| | (0.249) | (0.258) | (0.248) | (0.125) | |
| Log GDP, origin | 1.036*** | 1.000*** | | | |
| | (0.055) | (0.080) | | | |
| Log GDP, destination | 0.850*** | 0.820*** | 0.906*** | 0.263** | 0.338*** |
| | (0.073) | (0.081) | (0.044) | (0.123) | (0.097) |
| Digital STRI, destination | −2.920*** | −2.858*** | −3.008*** | 0.200 | 0.030 |
| | (1.049) | (1.093) | (1.065) | (0.496) | (0.322) |
| Constant | −34.472*** | | | | |
| | (3.037) | | | | |
| Year and category FE | No | Yes | Yes | Yes | Yes |
| Origin × category × year FE | No | No | Yes | Yes | Yes |
| Destination FE | No | No | No | Yes | Yes |
| Country-pair FE | No | No | No | No | Yes |
| Observations | 137,854 | 137,854 | 137,854 | 137,854 | 137,854 |
| Pseudo $R^2$ | 0.550 | 0.781 | 0.939 | 0.963 | 0.986 |

Notes: PPML. Dependent variable: bilateral digital product trade under headquarters attribution (zeros included). Tax variable: VAT, log(1 + rate). All columns include the gravity controls, and controls absorbed by the fixed effects are not estimated (blank). Fixed effects marked Yes are included directly or nested in a finer fixed effect. Standard errors clustered by origin–destination pair in parentheses. * p<0.10, ** p<0.05, *** p<0.01.

## Table S20. Digital product trade (headquarters attribution) and the VAT, dummy (rate > 0): common sample

| | (1) | (2) | (3) | (4) | (5) |
|---|---|---|---|---|---|
| VAT > 0 (dummy) | 0.312*** | 0.289** | 0.186** | −0.017 | 0.009 |
| | (0.113) | (0.133) | (0.086) | (0.039) | (0.040) |
| Log distance | −0.168 | −0.154 | −0.214 | −0.401*** | |
| | (0.129) | (0.132) | (0.153) | (0.059) | |
| Contiguity | −0.503** | −0.544** | −0.741*** | −0.821*** | |
| | (0.228) | (0.238) | (0.247) | (0.276) | |
| Common colonizer | 0.324 | 0.423 | 0.482 | 0.822* | |
| | (0.951) | (0.992) | (0.913) | (0.481) | |
| Common official language | 0.522** | 0.756*** | 0.432* | 0.661*** | |
| | (0.249) | (0.260) | (0.249) | (0.125) | |
| Log GDP, origin | 1.038*** | 1.008*** | | | |
| | (0.060) | (0.086) | | | |
| Log GDP, destination | 0.830*** | 0.800*** | 0.897*** | 0.285** | 0.349*** |
| | (0.076) | (0.084) | (0.045) | (0.122) | (0.095) |
| Digital STRI, destination | −2.832*** | −2.764*** | −2.949*** | 0.221 | 0.057 |
| | (1.005) | (1.038) | (1.052) | (0.496) | (0.324) |
| Constant | −33.935*** | | | | |
| | (3.130) | | | | |
| Year and category FE | No | Yes | Yes | Yes | Yes |
| Origin × category × year FE | No | No | Yes | Yes | Yes |
| Destination FE | No | No | No | Yes | Yes |
| Country-pair FE | No | No | No | No | Yes |
| Observations | 137,854 | 137,854 | 137,854 | 137,854 | 137,854 |
| Pseudo $R^2$ | 0.550 | 0.780 | 0.939 | 0.963 | 0.986 |

Notes: PPML. Dependent variable: bilateral digital product trade under headquarters attribution (zeros included). Tax variable: VAT, dummy (rate > 0). All columns include the gravity controls, and controls absorbed by the fixed effects are not estimated (blank). Fixed effects marked Yes are included directly or nested in a finer fixed effect. Standard errors clustered by origin–destination pair in parentheses. * p<0.10, ** p<0.05, *** p<0.01.

### S2.2 Each column on its own estimation sample (Tables S21–S38)

Supplementary Tables S21–S38 repeat the regressions with each column on its own estimation sample, so that columns (1) and (2) use all observations. Only these two columns change, and only slightly, for example from −0.132 to −0.146 for the DST rate in column (2) under subsidiary attribution, and all conclusions of Section S2.1 hold.

## Table S21. Digital product trade (subsidiary attribution) and the DST, rate in percentage points

| | (1) | (2) | (3) | (4) | (5) |
|---|---|---|---|---|---|
| DST rate (pp) | 0.021 | −0.146* | −0.174*** | 0.003 | −0.017 |
| | (0.066) | (0.080) | (0.055) | (0.042) | (0.044) |
| Log distance | −0.615*** | −0.622*** | −0.405*** | −0.705*** | |
| | (0.118) | (0.117) | (0.089) | (0.052) | |
| Contiguity | −0.948* | −0.972** | −0.210 | −0.320* | |
| | (0.485) | (0.492) | (0.199) | (0.177) | |
| Common colonizer | −0.618 | −0.494 | 0.601 | 0.711* | |
| | (0.552) | (0.545) | (0.570) | (0.426) | |
| Common official language | 1.224*** | 1.240*** | 0.371 | 0.087 | |
| | (0.338) | (0.338) | (0.230) | (0.205) | |
| Log GDP, origin | 0.847*** | 0.846*** | | | |
| | (0.064) | (0.063) | | | |
| Log GDP, destination | 0.803*** | 0.803*** | 0.810*** | 0.162 | 0.361* |
| | (0.087) | (0.087) | (0.054) | (0.214) | (0.191) |
| Digital STRI, destination | −1.552 | −1.196 | −1.514** | 0.330 | 0.687 |
| | (0.991) | (0.906) | (0.708) | (0.532) | (0.577) |
| Constant | −25.212*** | | | | |
| | (2.634) | | | | |
| Year and category FE | No | Yes | Yes | Yes | Yes |
| Origin × category × year FE | No | No | Yes | Yes | Yes |
| Destination FE | No | No | No | Yes | Yes |
| Country-pair FE | No | No | No | No | Yes |
| Observations | 2,086,220 | 2,086,220 | 267,642 | 267,642 | 267,174 |
| Pseudo $R^2$ | 0.379 | 0.549 | 0.837 | 0.871 | 0.910 |

Notes: PPML. Dependent variable: bilateral digital product trade under subsidiary attribution (zeros included). Tax variable: DST, rate in percentage points. All columns include the gravity controls, and controls absorbed by the fixed effects are not estimated (blank). Fixed effects marked Yes are included directly or nested in a finer fixed effect. Standard errors clustered by origin–destination pair in parentheses. * $p<0.10$, ** $p<0.05$, *** $p<0.01$. Each column uses its own estimation sample (PPML drops origin–category–year groups without trade).

## Table S22. Digital product trade (subsidiary attribution) and the DST, log(1 + rate)

| | (1) | (2) | (3) | (4) | (5) |
|---|---|---|---|---|---|
| log(1 + DST rate) | 2.236 | −14.996* | −17.812*** | 0.279 | −1.703 |
| | (6.833) | (8.176) | (5.687) | (4.366) | (4.570) |
| Log distance | −0.615*** | −0.622*** | −0.406*** | −0.705*** | |
| | (0.118) | (0.117) | (0.089) | (0.052) | |
| Contiguity | −0.948* | −0.972** | −0.210 | −0.320* | |
| | (0.485) | (0.493) | (0.199) | (0.177) | |
| Common colonizer | −0.618 | −0.494 | 0.602 | 0.711* | |
| | (0.552) | (0.545) | (0.569) | (0.426) | |
| Common official language | 1.224*** | 1.240*** | 0.371 | 0.087 | |
| | (0.338) | (0.338) | (0.230) | (0.205) | |
| Log GDP, origin | 0.847*** | 0.846*** | | | |
| | (0.064) | (0.063) | | | |
| Log GDP, destination | 0.803*** | 0.803*** | 0.810*** | 0.162 | 0.361* |
| | (0.087) | (0.087) | (0.054) | (0.214) | (0.191) |
| Digital STRI, destination | −1.554 | −1.196 | −1.514** | 0.330 | 0.687 |
| | (0.992) | (0.906) | (0.708) | (0.532) | (0.577) |
| Constant | −25.211*** | | | | |
| | (2.635) | | | | |
| Year and category FE | No | Yes | Yes | Yes | Yes |
| Origin × category × year FE | No | No | Yes | Yes | Yes |
| Destination FE | No | No | No | Yes | Yes |
| Country-pair FE | No | No | No | No | Yes |
| Observations | 2,086,220 | 2,086,220 | 267,642 | 267,642 | 267,174 |
| Pseudo $R^2$ | 0.379 | 0.549 | 0.837 | 0.871 | 0.910 |

Notes: PPML. Dependent variable: bilateral digital product trade under subsidiary attribution (zeros included). Tax variable: DST, log(1 + rate). All columns include the gravity controls, and controls absorbed by the fixed effects are not estimated (blank). Fixed effects marked Yes are included directly or nested in a finer fixed effect. Standard errors clustered by origin–destination pair in parentheses. * p<0.10, ** p<0.05, *** p<0.01. Each column uses its own estimation sample (PPML drops origin–category–year groups without trade).

## Table S23. Digital product trade (subsidiary attribution) and the DST, dummy (rate > 0)

| | (1) | (2) | (3) | (4) | (5) |
|---|---|---|---|---|---|
| DST > 0 (dummy) | 0.121 | −0.537 | −0.649*** | −0.010 | −0.088 |
| | (0.360) | (0.343) | (0.216) | (0.175) | (0.184) |
| Log distance | −0.615*** | −0.625*** | −0.410*** | −0.705*** | |
| | (0.118) | (0.117) | (0.089) | (0.052) | |
| Contiguity | −0.946* | −0.985** | −0.195 | −0.319* | |
| | (0.487) | (0.498) | (0.197) | (0.175) | |
| Common colonizer | −0.634 | −0.490 | 0.639 | 0.711* | |
| | (0.564) | (0.545) | (0.549) | (0.426) | |
| Common official language | 1.221*** | 1.246*** | 0.377 | 0.087 | |
| | (0.339) | (0.338) | (0.231) | (0.205) | |
| Log GDP, origin | 0.847*** | 0.847*** | | | |
| | (0.064) | (0.063) | | | |
| Log GDP, destination | 0.802*** | 0.804*** | 0.812*** | 0.160 | 0.357* |
| | (0.088) | (0.086) | (0.054) | (0.215) | (0.191) |
| Digital STRI, destination | −1.593 | −1.246 | −1.549** | 0.341 | 0.690 |
| | (1.026) | (0.919) | (0.711) | (0.528) | (0.572) |
| Constant | −25.171*** | | | | |
| | (2.692) | | | | |
| Year and category FE | No | Yes | Yes | Yes | Yes |
| Origin × category × year FE | No | No | Yes | Yes | Yes |
| Destination FE | No | No | No | Yes | Yes |
| Country-pair FE | No | No | No | No | Yes |
| Observations | 2,086,220 | 2,086,220 | 267,642 | 267,642 | 267,174 |
| Pseudo $R^2$ | 0.379 | 0.549 | 0.837 | 0.871 | 0.910 |

Notes: PPML. Dependent variable: bilateral digital product trade under subsidiary attribution (zeros included). Tax variable: DST, dummy (rate > 0). All columns include the gravity controls, and controls absorbed by the fixed effects are not estimated (blank). Fixed effects marked Yes are included directly or nested in a finer fixed effect. Standard errors clustered by origin–destination pair in parentheses. * p<0.10, ** p<0.05, *** p<0.01. Each column uses its own estimation sample (PPML drops origin–category–year groups without trade).

## Table S24. Digital product trade (headquarters attribution) and the DST, rate in percentage points

| | (1) | (2) | (3) | (4) | (5) |
|---|---|---|---|---|---|
| DST rate (pp) | 0.037 | −0.115* | −0.166*** | −0.011 | −0.013 |
| | (0.063) | (0.060) | (0.051) | (0.014) | (0.013) |
| Log distance | −0.237 | −0.242* | −0.246* | −0.401*** | |
| | (0.145) | (0.143) | (0.140) | (0.060) | |
| Contiguity | −0.706*** | −0.715*** | −0.883*** | −0.823*** | |
| | (0.270) | (0.272) | (0.239) | (0.276) | |
| Common colonizer | 0.046 | 0.116 | 0.582 | 0.820* | |
| | (1.005) | (1.012) | (0.867) | (0.481) | |
| Common official language | 0.704*** | 0.749*** | 0.505** | 0.662*** | |
| | (0.271) | (0.268) | (0.236) | (0.125) | |
| Log GDP, origin | 1.700*** | 1.708*** | | | |
| | (0.077) | (0.081) | | | |
| Log GDP, destination | 0.815*** | 0.825*** | 0.922*** | 0.274** | 0.352*** |
| | (0.086) | (0.088) | (0.038) | (0.124) | (0.097) |
| Digital STRI, destination | −2.835** | −2.362** | −2.497*** | 0.239 | 0.089 |
| | (1.102) | (0.929) | (0.830) | (0.501) | (0.325) |
| Constant | −53.319*** | | | | |
| | (3.981) | | | | |
| Year and category FE | No | Yes | Yes | Yes | Yes |
| Origin × category × year FE | No | No | Yes | Yes | Yes |
| Destination FE | No | No | No | Yes | Yes |
| Country-pair FE | No | No | No | No | Yes |
| Observations | 2,086,220 | 2,086,220 | 138,587 | 138,587 | 138,185 |
| Pseudo $R^2$ | 0.642 | 0.813 | 0.942 | 0.963 | 0.986 |

Notes: PPML. Dependent variable: bilateral digital product trade under headquarters attribution (zeros included). Tax variable: DST, rate in percentage points. All columns include the gravity controls, and controls absorbed by the fixed effects are not estimated (blank). Fixed effects marked Yes are included directly or nested in a finer fixed effect. Standard errors clustered by origin–destination pair in parentheses. * p<0.10, ** p<0.05, *** p<0.01. Each column uses its own estimation sample (PPML drops origin–category–year groups without trade).

## Table S25. Digital product trade (headquarters attribution) and the DST, log(1 + rate)

| | (1) | (2) | (3) | (4) | (5) |
|---|---|---|---|---|---|
| log(1 + DST rate) | 3.931 | −11.720* | −16.973*** | −1.133 | −1.355 |
| | (6.512) | (6.170) | (5.190) | (1.417) | (1.310) |
| Log distance | −0.237 | −0.242* | −0.247* | −0.401*** | |
| | (0.145) | (0.143) | (0.139) | (0.060) | |
| Contiguity | −0.706*** | −0.715*** | −0.884*** | −0.823*** | |
| | (0.270) | (0.272) | (0.239) | (0.276) | |
| Common colonizer | 0.045 | 0.116 | 0.584 | 0.820* | |
| | (1.005) | (1.012) | (0.867) | (0.481) | |
| Common official language | 0.704*** | 0.749*** | 0.505** | 0.662*** | |
| | (0.271) | (0.268) | (0.236) | (0.125) | |
| Log GDP, origin | 1.700*** | 1.708*** | | | |
| | (0.077) | (0.081) | | | |
| Log GDP, destination | 0.815*** | 0.825*** | 0.922*** | 0.274** | 0.352*** |
| | (0.086) | (0.088) | (0.038) | (0.124) | (0.097) |
| Digital STRI, destination | −2.838** | −2.362** | −2.497*** | 0.239 | 0.089 |
| | (1.103) | (0.929) | (0.829) | (0.501) | (0.325) |
| Constant | −53.316*** | | | | |
| | (3.981) | | | | |
| Year and category FE | No | Yes | Yes | Yes | Yes |
| Origin × category × year FE | No | No | Yes | Yes | Yes |
| Destination FE | No | No | No | Yes | Yes |
| Country-pair FE | No | No | No | No | Yes |
| Observations | 2,086,220 | 2,086,220 | 138,587 | 138,587 | 138,185 |
| Pseudo $R^2$ | 0.642 | 0.813 | 0.942 | 0.963 | 0.986 |

Notes: PPML. Dependent variable: bilateral digital product trade under headquarters attribution (zeros included). Tax variable: DST, log(1 + rate). All columns include the gravity controls, and controls absorbed by the fixed effects are not estimated (blank). Fixed effects marked Yes are included directly or nested in a finer fixed effect. Standard errors clustered by origin–destination pair in parentheses. * p<0.10, ** p<0.05, *** p<0.01. Each column uses its own estimation sample (PPML drops origin–category–year groups without trade).

## Table S26. Digital product trade (headquarters attribution) and the DST, dummy (rate > 0)

| | (1) | (2) | (3) | (4) | (5) |
|---|---|---|---|---|---|
| DST > 0 (dummy) | 0.235 | −0.377 | −0.573*** | −0.054 | −0.064 |
| | (0.347) | (0.254) | (0.206) | (0.048) | (0.046) |
| Log distance | −0.236 | −0.244* | −0.253* | −0.400*** | |
| | (0.145) | (0.142) | (0.135) | (0.060) | |
| Contiguity | −0.706*** | −0.727*** | −0.916*** | −0.823*** | |
| | (0.272) | (0.274) | (0.236) | (0.275) | |
| Common colonizer | 0.029 | 0.109 | 0.677 | 0.826* | |
| | (1.006) | (1.013) | (0.830) | (0.480) | |
| Common official language | 0.689** | 0.758*** | 0.524** | 0.662*** | |
| | (0.273) | (0.269) | (0.237) | (0.125) | |
| Log GDP, origin | 1.698*** | 1.709*** | | | |
| | (0.077) | (0.081) | | | |
| Log GDP, destination | 0.812*** | 0.826*** | 0.925*** | 0.274** | 0.353*** |
| | (0.086) | (0.088) | (0.038) | (0.125) | (0.097) |
| Digital STRI, destination | −2.953*** | −2.417*** | −2.555*** | 0.238 | 0.088 |
| | (1.103) | (0.920) | (0.819) | (0.500) | (0.328) |
| Constant | −53.150*** | | | | |
| | (3.987) | | | | |
| Year and category FE | No | Yes | Yes | Yes | Yes |
| Origin × category × year FE | No | No | Yes | Yes | Yes |
| Destination FE | No | No | No | Yes | Yes |
| Country-pair FE | No | No | No | No | Yes |
| Observations | 2,086,220 | 2,086,220 | 138,587 | 138,587 | 138,185 |
| Pseudo $R^2$ | 0.642 | 0.813 | 0.942 | 0.963 | 0.986 |

Notes: PPML. Dependent variable: bilateral digital product trade under headquarters attribution (zeros included). Tax variable: DST, dummy (rate > 0). All columns include the gravity controls, and controls absorbed by the fixed effects are not estimated (blank). Fixed effects marked Yes are included directly or nested in a finer fixed effect. Standard errors clustered by origin–destination pair in parentheses. * p<0.10, ** p<0.05, *** p<0.01. Each column uses its own estimation sample (PPML drops origin–category–year groups without trade).

## Table S27. Digital product trade (subsidiary attribution) and the DST + VAT, rate in percentage points

| | (1) | (2) | (3) | (4) | (5) |
|---|---|---|---|---|---|
| DST + VAT rate (pp) | −0.008 | −0.013 | −0.019*** | 0.001 | 0.002 |
| | (0.008) | (0.008) | (0.005) | (0.006) | (0.006) |
| Log distance | −0.643*** | −0.655*** | −0.478*** | −0.705*** | |
| | (0.113) | (0.114) | (0.101) | (0.052) | |
| Contiguity | −1.004** | −1.028** | −0.348* | −0.320* | |
| | (0.495) | (0.502) | (0.206) | (0.177) | |
| Common colonizer | −0.653 | −0.681 | 0.354 | 0.711* | |
| | (0.536) | (0.532) | (0.541) | (0.426) | |
| Common official language | 1.200*** | 1.184*** | 0.242 | 0.087 | |
| | (0.339) | (0.338) | (0.244) | (0.206) | |
| Log GDP, origin | 0.847*** | 0.842*** | | | |
| | (0.064) | (0.064) | | | |
| Log GDP, destination | 0.799*** | 0.791*** | 0.805*** | 0.157 | 0.357* |
| | (0.085) | (0.085) | (0.053) | (0.209) | (0.186) |
| Digital STRI, destination | −1.437 | −1.461 | −1.769** | 0.328 | 0.625 |
| | (0.928) | (0.911) | (0.773) | (0.530) | (0.565) |
| Constant | −24.715*** | | | | |
| | (2.432) | | | | |
| Year and category FE | No | Yes | Yes | Yes | Yes |
| Origin × category × year FE | No | No | Yes | Yes | Yes |
| Destination FE | No | No | No | Yes | Yes |
| Country-pair FE | No | No | No | No | Yes |
| Observations | 2,086,220 | 2,086,220 | 267,642 | 267,642 | 267,174 |
| Pseudo $R^2$ | 0.379 | 0.549 | 0.837 | 0.871 | 0.910 |

Notes: PPML. Dependent variable: bilateral digital product trade under subsidiary attribution (zeros included). Tax variable: DST + VAT, rate in percentage points. All columns include the gravity controls, and controls absorbed by the fixed effects are not estimated (blank). Fixed effects marked Yes are included directly or nested in a finer fixed effect. Standard errors clustered by origin–destination pair in parentheses. * p<0.10, ** p<0.05, *** p<0.01. Each column uses its own estimation sample (PPML drops origin–category–year groups without trade).

## Table S28. Digital product trade (subsidiary attribution) and the DST + VAT, log(1 + rate)

| | (1) | (2) | (3) | (4) | (5) |
|---|---|---|---|---|---|
| log(1 + DST + VAT rate) | −0.931 | −1.501 | −2.113*** | 0.038 | 0.233 |
| | (0.889) | (0.937) | (0.604) | (0.573) | (0.584) |
| Log distance | −0.642*** | −0.654*** | −0.475*** | −0.705*** | |
| | (0.114) | (0.115) | (0.102) | (0.052) | |
| Contiguity | −1.003** | −1.028** | −0.342* | −0.320* | |
| | (0.496) | (0.503) | (0.205) | (0.177) | |
| Common colonizer | −0.650 | −0.678 | 0.360 | 0.711* | |
| | (0.537) | (0.532) | (0.543) | (0.426) | |
| Common official language | 1.201*** | 1.185*** | 0.246 | 0.087 | |
| | (0.340) | (0.338) | (0.244) | (0.206) | |
| Log GDP, origin | 0.847*** | 0.842*** | | | |
| | (0.064) | (0.064) | | | |
| Log GDP, destination | 0.800*** | 0.792*** | 0.806*** | 0.159 | 0.356* |
| | (0.085) | (0.085) | (0.054) | (0.208) | (0.186) |
| Digital STRI, destination | −1.436 | −1.459 | −1.771** | 0.333 | 0.626 |
| | (0.931) | (0.914) | (0.777) | (0.530) | (0.563) |
| Constant | −24.752*** | | | | |
| | (2.437) | | | | |
| Year and category FE | No | Yes | Yes | Yes | Yes |
| Origin × category × year FE | No | No | Yes | Yes | Yes |
| Destination FE | No | No | No | Yes | Yes |
| Country-pair FE | No | No | No | No | Yes |
| Observations | 2,086,220 | 2,086,220 | 267,642 | 267,642 | 267,174 |
| Pseudo $R^2$ | 0.379 | 0.549 | 0.837 | 0.871 | 0.910 |

Notes: PPML. Dependent variable: bilateral digital product trade under subsidiary attribution (zeros included). Tax variable: DST + VAT, log(1 + rate). All columns include the gravity controls, and controls absorbed by the fixed effects are not estimated (blank). Fixed effects marked Yes are included directly or nested in a finer fixed effect. Standard errors clustered by origin–destination pair in parentheses. * $p<0.10$, ** $p<0.05$, *** $p<0.01$. Each column uses its own estimation sample (PPML drops origin–category–year groups without trade).

## Table S29. Digital product trade (subsidiary attribution) and the DST + VAT, dummy (rate > 0)

| | (1) | (2) | (3) | (4) | (5) |
|---|---|---|---|---|---|
| DST + VAT > 0 (dummy) | 0.053 | −0.130 | −0.139 | −0.046 | 0.010 |
| | (0.163) | (0.182) | (0.142) | (0.067) | (0.062) |
| Log distance | −0.612*** | −0.621*** | −0.397*** | −0.705*** | |
| | (0.121) | (0.122) | (0.099) | (0.052) | |
| Contiguity | −0.941* | −0.964** | −0.219 | −0.319* | |
| | (0.486) | (0.490) | (0.197) | (0.177) | |
| Common colonizer | −0.604 | −0.610 | 0.490 | 0.711* | |
| | (0.549) | (0.547) | (0.604) | (0.426) | |
| Common official language | 1.228*** | 1.217*** | 0.329 | 0.088 | |
| | (0.340) | (0.340) | (0.231) | (0.206) | |
| Log GDP, origin | 0.847*** | 0.843*** | | | |
| | (0.064) | (0.064) | | | |
| Log GDP, destination | 0.802*** | 0.804*** | 0.812*** | 0.194 | 0.361* |
| | (0.090) | (0.091) | (0.059) | (0.203) | (0.184) |
| Digital STRI, destination | −1.496 | −1.565* | −1.957** | 0.338 | 0.647 |
| | (0.947) | (0.942) | (0.816) | (0.522) | (0.554) |
| Constant | −25.229*** | | | | |
| | (2.611) | | | | |
| Year and category FE | No | Yes | Yes | Yes | Yes |
| Origin × category × year FE | No | No | Yes | Yes | Yes |
| Destination FE | No | No | No | Yes | Yes |
| Country-pair FE | No | No | No | No | Yes |
| Observations | 2,086,220 | 2,086,220 | 267,642 | 267,642 | 267,174 |
| Pseudo $R^2$ | 0.379 | 0.548 | 0.834 | 0.871 | 0.910 |

Notes: PPML. Dependent variable: bilateral digital product trade under subsidiary attribution (zeros included). Tax variable: DST + VAT, dummy (rate > 0). All columns include the gravity controls, and controls absorbed by the fixed effects are not estimated (blank). Fixed effects marked Yes are included directly or nested in a finer fixed effect. Standard errors clustered by origin–destination pair in parentheses. * p<0.10, ** p<0.05, *** p<0.01. Each column uses its own estimation sample (PPML drops origin–category–year groups without trade).

## Table S30. Digital product trade (headquarters attribution) and the DST + VAT, rate in percentage points

| | (1) | (2) | (3) | (4) | (5) |
|---|---|---|---|---|---|
| DST + VAT rate (pp) | 0.012 | 0.010 | 0.002 | 0.001 | 0.002 |
| | (0.009) | (0.009) | (0.005) | (0.002) | (0.002) |
| Log distance | −0.209 | −0.216 | −0.217 | −0.401*** | |
| | (0.141) | (0.141) | (0.151) | (0.060) | |
| Contiguity | −0.575** | −0.598** | −0.771*** | −0.822*** | |
| | (0.242) | (0.246) | (0.243) | (0.276) | |
| Common colonizer | 0.121 | 0.117 | 0.496 | 0.820* | |
| | (1.002) | (1.004) | (0.912) | (0.480) | |
| Common official language | 0.743*** | 0.738*** | 0.434* | 0.661*** | |
| | (0.265) | (0.265) | (0.248) | (0.125) | |
| Log GDP, origin | 1.689*** | 1.695*** | | | |
| | (0.071) | (0.075) | | | |
| Log GDP, destination | 0.823*** | 0.823*** | 0.906*** | 0.269** | 0.345*** |
| | (0.084) | (0.086) | (0.044) | (0.124) | (0.099) |
| Digital STRI, destination | −2.787** | −2.774** | −3.006*** | 0.204 | 0.037 |
| | (1.103) | (1.100) | (1.051) | (0.497) | (0.321) |
| Constant | −53.686*** | | | | |
| | (3.777) | | | | |
| Year and category FE | No | Yes | Yes | Yes | Yes |
| Origin × category × year FE | No | No | Yes | Yes | Yes |
| Destination FE | No | No | No | Yes | Yes |
| Country-pair FE | No | No | No | No | Yes |
| Observations | 2,086,220 | 2,086,220 | 138,587 | 138,587 | 138,185 |
| Pseudo $R^2$ | 0.643 | 0.813 | 0.939 | 0.963 | 0.986 |

Notes: PPML. Dependent variable: bilateral digital product trade under headquarters attribution (zeros included). Tax variable: DST + VAT, rate in percentage points. All columns include the gravity controls, and controls absorbed by the fixed effects are not estimated (blank). Fixed effects marked Yes are included directly or nested in a finer fixed effect. Standard errors clustered by origin–destination pair in parentheses. * p<0.10, ** p<0.05, *** p<0.01. Each column uses its own estimation sample (PPML drops origin–category–year groups without trade).

## Table S31. Digital product trade (headquarters attribution) and the DST + VAT, log(1 + rate)

| | (1) | (2) | (3) | (4) | (5) |
|---|---|---|---|---|---|
| log(1 + DST + VAT rate) | 1.376 | 1.142 | 0.203 | 0.151 | 0.221 |
| | (1.004) | (1.012) | (0.563) | (0.241) | (0.215) |
| Log distance | −0.210 | −0.216 | −0.216 | −0.401*** | |
| | (0.141) | (0.142) | (0.151) | (0.060) | |
| Contiguity | −0.573** | −0.596** | −0.767*** | −0.822*** | |
| | (0.242) | (0.246) | (0.243) | (0.276) | |
| Common colonizer | 0.119 | 0.117 | 0.497 | 0.820* | |
| | (1.002) | (1.004) | (0.913) | (0.480) | |
| Common official language | 0.741*** | 0.736*** | 0.435* | 0.661*** | |
| | (0.265) | (0.265) | (0.248) | (0.125) | |
| Log GDP, origin | 1.689*** | 1.695*** | | | |
| | (0.071) | (0.075) | | | |
| Log GDP, destination | 0.822*** | 0.823*** | 0.906*** | 0.268** | 0.344*** |
| | (0.084) | (0.086) | (0.044) | (0.124) | (0.099) |
| Digital STRI, destination | −2.789** | −2.776** | −3.008*** | 0.205 | 0.037 |
| | (1.101) | (1.099) | (1.051) | (0.497) | (0.321) |
| Constant | −53.658*** | | | | |
| | (3.778) | | | | |
| Year and category FE | No | Yes | Yes | Yes | Yes |
| Origin × category × year FE | No | No | Yes | Yes | Yes |
| Destination FE | No | No | No | Yes | Yes |
| Country-pair FE | No | No | No | No | Yes |
| Observations | 2,086,220 | 2,086,220 | 138,587 | 138,587 | 138,185 |
| Pseudo $R^2$ | 0.643 | 0.813 | 0.939 | 0.963 | 0.986 |

Notes: PPML. Dependent variable: bilateral digital product trade under headquarters attribution (zeros included). Tax variable: DST + VAT, log(1 + rate). All columns include the gravity controls, and controls absorbed by the fixed effects are not estimated (blank). Fixed effects marked Yes are included directly or nested in a finer fixed effect. Standard errors clustered by origin–destination pair in parentheses. * p<0.10, ** p<0.05, *** p<0.01. Each column uses its own estimation sample (PPML drops origin–category–year groups without trade).

## Table S32. Digital product trade (headquarters attribution) and the DST + VAT, dummy (rate > 0)

| | (1) | (2) | (3) | (4) | (5) |
|---|---|---|---|---|---|
| DST + VAT > 0 (dummy) | 0.277** | 0.265* | 0.174** | −0.012 | 0.014 |
| | (0.130) | (0.135) | (0.086) | (0.039) | (0.041) |
| Log distance | −0.223 | −0.226 | −0.215 | −0.401*** | |
| | (0.144) | (0.145) | (0.153) | (0.059) | |
| Contiguity | −0.636** | −0.639** | −0.744*** | −0.822*** | |
| | (0.262) | (0.262) | (0.248) | (0.276) | |
| Common colonizer | 0.048 | 0.061 | 0.481 | 0.821* | |
| | (1.002) | (1.004) | (0.912) | (0.481) | |
| Common official language | 0.713*** | 0.715*** | 0.430* | 0.661*** | |
| | (0.266) | (0.266) | (0.249) | (0.125) | |
| Log GDP, origin | 1.692*** | 1.699*** | | | |
| | (0.075) | (0.079) | | | |
| Log GDP, destination | 0.803*** | 0.806*** | 0.898*** | 0.282** | 0.346*** |
| | (0.088) | (0.089) | (0.045) | (0.122) | (0.095) |
| Digital STRI, destination | −2.695** | −2.684** | −2.962*** | 0.220 | 0.056 |
| | (1.051) | (1.059) | (1.053) | (0.496) | (0.323) |
| Constant | −53.126*** | | | | |
| | (3.916) | | | | |
| Year and category FE | No | Yes | Yes | Yes | Yes |
| Origin × category × year FE | No | No | Yes | Yes | Yes |
| Destination FE | No | No | No | Yes | Yes |
| Country-pair FE | No | No | No | No | Yes |
| Observations | 2,086,220 | 2,086,220 | 138,587 | 138,587 | 138,185 |
| Pseudo $R^2$ | 0.642 | 0.813 | 0.939 | 0.963 | 0.986 |

Notes: PPML. Dependent variable: bilateral digital product trade under headquarters attribution (zeros included). Tax variable: DST + VAT, dummy (rate > 0). All columns include the gravity controls, and controls absorbed by the fixed effects are not estimated (blank). Fixed effects marked Yes are included directly or nested in a finer fixed effect. Standard errors clustered by origin–destination pair in parentheses. * p<0.10, ** p<0.05, *** p<0.01. Each column uses its own estimation sample (PPML drops origin–category–year groups without trade).

## Table S33. Digital product trade (subsidiary attribution) and the VAT, rate in percentage points

| | (1) | (2) | (3) | (4) | (5) |
|---|---|---|---|---|---|
| VAT rate (pp) | −0.009 | −0.012 | −0.018*** | 0.001 | 0.004 |
| | (0.008) | (0.008) | (0.005) | (0.004) | (0.004) |
| Log distance | −0.644*** | −0.650*** | −0.469*** | −0.705*** | |
| | (0.114) | (0.114) | (0.102) | (0.052) | |
| Contiguity | −1.006** | −1.017** | −0.334* | −0.320* | |
| | (0.495) | (0.498) | (0.203) | (0.178) | |
| Common colonizer | −0.660 | −0.684 | 0.355 | 0.711* | |
| | (0.535) | (0.533) | (0.550) | (0.426) | |
| Common official language | 1.198*** | 1.186*** | 0.244 | 0.087 | |
| | (0.340) | (0.338) | (0.245) | (0.206) | |
| Log GDP, origin | 0.847*** | 0.841*** | | | |
| | (0.064) | (0.064) | | | |
| Log GDP, destination | 0.798*** | 0.791*** | 0.805*** | 0.158 | 0.347* |
| | (0.085) | (0.085) | (0.053) | (0.207) | (0.186) |
| Digital STRI, destination | −1.457 | −1.510* | −1.839** | 0.332 | 0.618 |
| | (0.926) | (0.915) | (0.784) | (0.528) | (0.557) |
| Constant | −24.686*** | | | | |
| | (2.441) | | | | |
| Year and category FE | No | Yes | Yes | Yes | Yes |
| Origin × category × year FE | No | No | Yes | Yes | Yes |
| Destination FE | No | No | No | Yes | Yes |
| Country-pair FE | No | No | No | No | Yes |
| Observations | 2,086,220 | 2,086,220 | 267,642 | 267,642 | 267,174 |
| Pseudo $R^2$ | 0.379 | 0.548 | 0.836 | 0.871 | 0.910 |

Notes: PPML. Dependent variable: bilateral digital product trade under subsidiary attribution (zeros included). Tax variable: VAT, rate in percentage points. All columns include the gravity controls, and controls absorbed by the fixed effects are not estimated (blank). Fixed effects marked Yes are included directly or nested in a finer fixed effect. Standard errors clustered by origin–destination pair in parentheses. * p<0.10, ** p<0.05, *** p<0.01. Each column uses its own estimation sample (PPML drops origin–category–year groups without trade).

## Table S34. Digital product trade (subsidiary attribution) and the VAT, log(1 + rate)

| | (1) | (2) | (3) | (4) | (5) |
|---|---|---|---|---|---|
| log(1 + VAT rate) | −0.964 | −1.351 | −1.948*** | 0.048 | 0.380 |
| | (0.869) | (0.937) | (0.616) | (0.477) | (0.435) |
| Log distance | −0.642*** | −0.649*** | −0.466*** | −0.705*** | |
| | (0.114) | (0.115) | (0.102) | (0.052) | |
| Contiguity | −1.005** | −1.017** | −0.330 | −0.320* | |
| | (0.495) | (0.500) | (0.203) | (0.178) | |
| Common colonizer | −0.657 | −0.682 | 0.361 | 0.711* | |
| | (0.536) | (0.533) | (0.552) | (0.426) | |
| Common official language | 1.200*** | 1.187*** | 0.248 | 0.087 | |
| | (0.340) | (0.339) | (0.245) | (0.206) | |
| Log GDP, origin | 0.847*** | 0.842*** | | | |
| | (0.064) | (0.064) | | | |
| Log GDP, destination | 0.799*** | 0.792*** | 0.806*** | 0.158 | 0.347* |
| | (0.085) | (0.085) | (0.054) | (0.206) | (0.185) |
| Digital STRI, destination | −1.455 | −1.507 | −1.839** | 0.333 | 0.619 |
| | (0.928) | (0.918) | (0.788) | (0.528) | (0.557) |
| Constant | −24.730*** | | | | |
| | (2.446) | | | | |
| Year and category FE | No | Yes | Yes | Yes | Yes |
| Origin × category × year FE | No | No | Yes | Yes | Yes |
| Destination FE | No | No | No | Yes | Yes |
| Country-pair FE | No | No | No | No | Yes |
| Observations | 2,086,220 | 2,086,220 | 267,642 | 267,642 | 267,174 |
| Pseudo $R^2$ | 0.379 | 0.548 | 0.836 | 0.871 | 0.910 |

Notes: PPML. Dependent variable: bilateral digital product trade under subsidiary attribution (zeros included). Tax variable: VAT, log(1 + rate). All columns include the gravity controls, and controls absorbed by the fixed effects are not estimated (blank). Fixed effects marked Yes are included directly or nested in a finer fixed effect. Standard errors clustered by origin–destination pair in parentheses. * p<0.10, ** p<0.05, *** p<0.01. Each column uses its own estimation sample (PPML drops origin–category–year groups without trade).

## Table S35. Digital product trade (subsidiary attribution) and the VAT, dummy (rate > 0)

| | (1) | (2) | (3) | (4) | (5) |
|---|---|---|---|---|---|
| VAT > 0 (dummy) | 0.059 | −0.112 | −0.124 | −0.047 | 0.009 |
| | (0.163) | (0.181) | (0.143) | (0.067) | (0.062) |
| Log distance | −0.611*** | −0.620*** | −0.396*** | −0.705*** | |
| | (0.121) | (0.122) | (0.100) | (0.052) | |
| Contiguity | −0.940* | −0.962** | −0.216 | −0.319* | |
| | (0.486) | (0.490) | (0.197) | (0.177) | |
| Common colonizer | −0.603 | −0.614 | 0.488 | 0.711* | |
| | (0.549) | (0.546) | (0.604) | (0.426) | |
| Common official language | 1.229*** | 1.218*** | 0.329 | 0.088 | |
| | (0.340) | (0.340) | (0.232) | (0.206) | |
| Log GDP, origin | 0.847*** | 0.843*** | | | |
| | (0.064) | (0.064) | | | |
| Log GDP, destination | 0.801*** | 0.803*** | 0.812*** | 0.194 | 0.362** |
| | (0.090) | (0.091) | (0.059) | (0.203) | (0.184) |
| Digital STRI, destination | −1.493 | −1.572* | −1.964** | 0.339 | 0.647 |
| | (0.944) | (0.938) | (0.812) | (0.522) | (0.554) |
| Constant | −25.226*** | | | | |
| | (2.613) | | | | |
| Year and category FE | No | Yes | Yes | Yes | Yes |
| Origin × category × year FE | No | No | Yes | Yes | Yes |
| Destination FE | No | No | No | Yes | Yes |
| Country-pair FE | No | No | No | No | Yes |
| Observations | 2,086,220 | 2,086,220 | 267,642 | 267,642 | 267,174 |
| Pseudo $R^2$ | 0.379 | 0.548 | 0.834 | 0.871 | 0.910 |

Notes: PPML. Dependent variable: bilateral digital product trade under subsidiary attribution (zeros included). Tax variable: VAT, dummy (rate > 0). All columns include the gravity controls, and controls absorbed by the fixed effects are not estimated (blank). Fixed effects marked Yes are included directly or nested in a finer fixed effect. Standard errors clustered by origin–destination pair in parentheses. * p<0.10, ** p<0.05, *** p<0.01. Each column uses its own estimation sample (PPML drops origin–category–year groups without trade).

**Table S36. Digital product trade (headquarters attribution) and the VAT, rate in percentage points**

| | (1) | (2) | (3) | (4) | (5) |
|---|---|---|---|---|---|
| VAT rate (pp) | 0.012 | 0.012 | 0.004 | 0.003 | 0.003 |
| | (0.009) | (0.009) | (0.005) | (0.002) | (0.002) |
| Log distance | −0.210 | −0.212 | −0.206 | −0.401*** | |
| | (0.141) | (0.142) | (0.154) | (0.060) | |
| Contiguity | −0.576** | −0.579** | −0.731*** | −0.822*** | |
| | (0.242) | (0.244) | (0.246) | (0.276) | |
| Common colonizer | 0.126 | 0.133 | 0.511 | 0.820* | |
| | (1.002) | (1.004) | (0.922) | (0.480) | |
| Common official language | 0.744*** | 0.746*** | 0.442* | 0.661*** | |
| | (0.264) | (0.264) | (0.247) | (0.125) | |
| Log GDP, origin | 1.690*** | 1.695*** | | | |
| | (0.071) | (0.074) | | | |
| Log GDP, destination | 0.824*** | 0.825*** | 0.906*** | 0.263** | 0.338*** |
| | (0.084) | (0.086) | (0.044) | (0.123) | (0.097) |
| Digital STRI, destination | −2.748** | −2.741** | −3.007*** | 0.198 | 0.029 |
| | (1.102) | (1.108) | (1.066) | (0.496) | (0.322) |
| Constant | −53.711*** | | | | |
| | (3.782) | | | | |
| Year and category FE | No | Yes | Yes | Yes | Yes |
| Origin × category × year FE | No | No | Yes | Yes | Yes |
| Destination FE | No | No | No | Yes | Yes |
| Country-pair FE | No | No | No | No | Yes |
| Observations | 2,086,220 | 2,086,220 | 138,587 | 138,587 | 138,185 |
| Pseudo $R^2$ | 0.643 | 0.813 | 0.939 | 0.963 | 0.986 |

Notes: PPML. Dependent variable: bilateral digital product trade under headquarters attribution (zeros included). Tax variable: VAT, rate in percentage points. All columns include the gravity controls, and controls absorbed by the fixed effects are not estimated (blank). Fixed effects marked Yes are included directly or nested in a finer fixed effect. Standard errors clustered by origin–destination pair in parentheses. * p<0.10, ** p<0.05, *** p<0.01. Each column uses its own estimation sample (PPML drops origin–category–year groups without trade).

## Table S37. Digital product trade (headquarters attribution) and the VAT, log(1 + rate)

| | (1) | (2) | (3) | (4) | (5) |
|---|---|---|---|---|---|
| log(1 + VAT rate) | 1.370 | 1.345 | 0.442 | 0.255 | 0.347 |
| | (1.008) | (1.034) | (0.546) | (0.260) | (0.238) |
| Log distance | −0.210 | −0.212 | −0.206 | −0.401*** | |
| | (0.141) | (0.142) | (0.154) | (0.060) | |
| Contiguity | −0.574** | −0.578** | −0.729*** | −0.822*** | |
| | (0.242) | (0.244) | (0.245) | (0.276) | |
| Common colonizer | 0.125 | 0.132 | 0.510 | 0.820* | |
| | (1.002) | (1.004) | (0.922) | (0.480) | |
| Common official language | 0.743*** | 0.745*** | 0.442* | 0.661*** | |
| | (0.265) | (0.264) | (0.248) | (0.125) | |
| Log GDP, origin | 1.690*** | 1.695*** | | | |
| | (0.071) | (0.074) | | | |
| Log GDP, destination | 0.823*** | 0.825*** | 0.906*** | 0.263** | 0.338*** |
| | (0.084) | (0.085) | (0.044) | (0.123) | (0.097) |
| Digital STRI, destination | −2.751** | −2.744** | −3.008*** | 0.200 | 0.030 |
| | (1.100) | (1.106) | (1.065) | (0.496) | (0.322) |
| Constant | −53.683*** | | | | |
| | (3.781) | | | | |
| Year and category FE | No | Yes | Yes | Yes | Yes |
| Origin × category × year FE | No | No | Yes | Yes | Yes |
| Destination FE | No | No | No | Yes | Yes |
| Country-pair FE | No | No | No | No | Yes |
| Observations | 2,086,220 | 2,086,220 | 138,587 | 138,587 | 138,185 |
| Pseudo $R^2$ | 0.643 | 0.813 | 0.939 | 0.963 | 0.986 |

Notes: PPML. Dependent variable: bilateral digital product trade under headquarters attribution (zeros included). Tax variable: VAT, log(1 + rate). All columns include the gravity controls, and controls absorbed by the fixed effects are not estimated (blank). Fixed effects marked Yes are included directly or nested in a finer fixed effect. Standard errors clustered by origin–destination pair in parentheses. * p<0.10, ** p<0.05, *** p<0.01. Each column uses its own estimation sample (PPML drops origin–category–year groups without trade).

**Table S38. Digital product trade (headquarters attribution) and the VAT, dummy (rate > 0)**

| | (1) | (2) | (3) | (4) | (5) |
|---|---|---|---|---|---|
| VAT > 0 (dummy) | 0.284** | 0.280** | 0.186** | −0.017 | 0.009 |
| | (0.131) | (0.136) | (0.086) | (0.039) | (0.040) |
| Log distance | −0.222 | −0.225 | −0.214 | −0.401*** | |
| | (0.144) | (0.144) | (0.153) | (0.059) | |
| Contiguity | −0.633** | −0.634** | −0.741*** | −0.821*** | |
| | (0.262) | (0.262) | (0.247) | (0.276) | |
| Common colonizer | 0.055 | 0.070 | 0.482 | 0.822* | |
| | (1.002) | (1.004) | (0.913) | (0.481) | |
| Common official language | 0.714*** | 0.717*** | 0.432* | 0.661*** | |
| | (0.267) | (0.266) | (0.249) | (0.125) | |
| Log GDP, origin | 1.692*** | 1.699*** | | | |
| | (0.075) | (0.079) | | | |
| Log GDP, destination | 0.802*** | 0.805*** | 0.897*** | 0.285** | 0.349*** |
| | (0.088) | (0.089) | (0.045) | (0.122) | (0.095) |
| Digital STRI, destination | −2.681** | −2.666** | −2.949*** | 0.221 | 0.057 |
| | (1.049) | (1.056) | (1.052) | (0.496) | (0.324) |
| Constant | −53.104*** | | | | |
| | (3.914) | | | | |
| Year and category FE | No | Yes | Yes | Yes | Yes |
| Origin × category × year FE | No | No | Yes | Yes | Yes |
| Destination FE | No | No | No | Yes | Yes |
| Country-pair FE | No | No | No | No | Yes |
| Observations | 2,086,220 | 2,086,220 | 138,587 | 138,587 | 138,185 |
| Pseudo $R^2$ | 0.642 | 0.813 | 0.939 | 0.963 | 0.986 |

Notes: PPML. Dependent variable: bilateral digital product trade under headquarters attribution (zeros included). Tax variable: VAT, dummy (rate > 0). All columns include the gravity controls, and controls absorbed by the fixed effects are not estimated (blank). Fixed effects marked Yes are included directly or nested in a finer fixed effect. Standard errors clustered by origin–destination pair in parentheses. * p<0.10, ** p<0.05, *** p<0.01. Each column uses its own estimation sample (PPML drops origin–category–year groups without trade).

### S2.3 Alternative fixed-effect structures and clustering (Table S39)

Table 1 uses five sets of fixed effects and clusters the standard errors by country pair. Supplementary Table S39 shows that the conclusions do not depend on these choices. It reports seven further fixed-effect structures, from year and category effects (model A) to exporter-by-importer-by-product effects with origin-by-year effects (model G), each without and with the gravity controls, and the p-values under four types of standard errors, heteroskedasticity-robust and clustered by country pair, by country pair and category, and by destination. The DST rate enters in percentage points, as in Table 1. The coefficient is negative only in models A and B, which compare destinations with and without a DST, and it is statistically significant there with the gravity controls (−0.146 and −0.158, or 13.6% and 14.6% lower trade per percentage point). Models C–G, which compare flows within country pairs or within destinations, give small positive coefficients (0.027 to 0.043) that are not statistically significant at the 5% level under any clustering. The association between the DST and digital trade thus comes from differences between destinations, as in Table 1.

**Table S39. DST coefficient (rate in percentage points) under alternative fixed effects and standard errors (PPML, subsidiary attribution)**

| Model and fixed effects | Controls | Coef. | Effect of 1 pp | N | p: HC1 | p: pair | p: pair × cat. | p: dest. |
|---|---|---|---|---|---|---|---|---|
| A. year + category | none | −0.088 | −8.4% | 2,086,220 | 0.000 | 0.146 | 0.077 | 0.275 |
| A. year + category | gravity | −0.146 | −13.6% | 2,086,220 | 0.000 | 0.066 | 0.025 | 0.026 |
| B. origin × year + category | none | −0.090 | −8.6% | 1,430,859 | 0.000 | 0.105 | 0.044 | 0.265 |
| B. origin × year + category | gravity | −0.158 | −14.6% | 1,430,859 | 0.000 | 0.003 | 0.001 | 0.013 |
| C. pair + origin × year + category | none | 0.032 | +3.2% | 1,416,352 | 0.126 | 0.504 | 0.385 | 0.349 |
| C. pair + origin × year + category | gravity | 0.032 | +3.2% | 1,416,352 | 0.127 | 0.509 | 0.390 | 0.360 |
| D. origin × year + destination × category | none | 0.040 | +4.1% | 1,429,179 | 0.276 | 0.146 | 0.122 | 0.157 |
| D. origin × year + destination × category | gravity | 0.040 | +4.1% | 1,429,179 | 0.151 | 0.105 | 0.101 | 0.086 |
| E. pair + origin × year + destination × category | none | 0.039 | +3.9% | 1,414,690 | 0.145 | 0.120 | 0.110 | 0.137 |
| E. pair + origin × year + destination × category | gravity | 0.039 | +4.0% | 1,414,690 | 0.140 | 0.113 | 0.106 | 0.114 |
| F. exporter × importer × product + year | none | 0.043 | +4.4% | 192,269 | 0.042 | 0.164 | 0.116 | 0.132 |
| F. exporter × importer × product + year | gravity | 0.027 | +2.7% | 192,269 | 0.209 | 0.356 | 0.319 | 0.347 |
| G. exporter × importer × product + origin × year | none | 0.029 | +2.9% | 184,474 | 0.127 | 0.236 | 0.247 | 0.305 |
| G. exporter × importer × product + origin × year | gravity | 0.029 | +2.9% | 184,474 | 0.125 | 0.227 | 0.242 | 0.278 |

Notes: p-values with heteroskedasticity-robust (HC1) standard errors and with clustering by country pair (7,552 clusters), by origin–destination–category (234,112) and by destination (68). Controls: none, or the gravity controls (absorbed controls dropped).

### S2.4 Sensitivity to controls (Table S40)

The association across destinations could reflect differences in economic size between taxing and non-taxing destinations. Supplementary Table S40 therefore adds the gravity controls one group at a time in model B, which compares destinations, and in model C, which compares flows within country pairs. In model B the coefficient is −0.090 without controls, −0.148 with destination GDP alone and −0.021 with the other gravity controls but without destination GDP. Destinations with a DST are on average larger economies, which import more digital products, so the negative association appears once destinations of similar size are compared. In model C the coefficient is positive and not statistically significant with every set of controls.

**Table S40. Sensitivity of the DST coefficient (rate in percentage points) to controls (PPML, subsidiary attribution)**

| Model | Controls | Coef. | Effect of 1 pp | p: HC1 | p: pair | p: pair × category | p: destination |
|---|---|---|---|---|---|---|---|
| B | no controls | −0.090 | −8.6% | 0.000 | 0.105 | 0.044 | 0.265 |
| B | destination GDP only | −0.148 | −13.7% | 0.000 | 0.014 | 0.003 | 0.075 |
| B | gravity without destination GDP | −0.021 | −2.1% | 0.331 | 0.707 | 0.654 | 0.760 |
| B | gravity (incl. destination GDP) | −0.158 | −14.6% | 0.000 | 0.003 | 0.001 | 0.013 |
| C | no controls | 0.032 | +3.2% | 0.126 | 0.504 | 0.385 | 0.349 |
| C | destination GDP only | 0.033 | +3.4% | 0.111 | 0.487 | 0.366 | 0.338 |
| C | gravity without destination GDP | 0.030 | +3.0% | 0.151 | 0.533 | 0.416 | 0.380 |
| C | gravity (incl. destination GDP) | 0.032 | +3.2% | 0.127 | 0.509 | 0.390 | 0.360 |

Notes: Same sample and conventions as Table S39.

### S2.5 Table 1 with bilateral digitally delivered services from BaTIS (Table S41)

Our bilateral flows are estimated from corporate revenue data (Stojkoski et al., 2024), so the results of Table 1 could reflect how these flows are constructed. The OECD–WTO Balanced Trade in Services dataset (BaTIS, Fortanier et al., 2017) reports bilateral services trade by EBOPS 2010 sector, so Table 1 can be re-estimated on an independent source. We construct bilateral digitally delivered trade as Stojkoski et al. (2024) do for their comparison with BaTIS. We keep the digitally deliverable sectors except insurance and pension services, to which none of our categories belongs, namely financial services (SG), charges for the use of intellectual property (SH), telecommunications, computer and information services (SI), other business services (SJ) and personal, cultural and recreational services (SK). Each balanced bilateral flow of the December 2025 release is multiplied by the digitally delivered share of its exporter, sector and year, which is the WTO estimate of Mode 1 exports (World Trade Organization, 2026) divided by the exporter's BaTIS exports to the world. In 30% of exporter–sector–years the WTO estimate exceeds the BaTIS exports, and the share is set to one. The panel has the 112 origins, 68 destinations, years and gravity controls of Table 1, which gives 336,505 observations, 7% of them zero, and US$3.0 trillion of trade in 2024. The DST rate of a destination–sector–year is the mean DST rate of the categories assigned to the sector. Software, operating systems, cybersecurity, cloud computing, hosting, mobile applications and data licensing belong to SI, video, music, games, gambling, education, e-books and dating to SK, advertising and online intermediation to SJ, and payment services to SG. SH contains none of our categories and has a rate of zero.

Table S41 re-estimates the five columns of Table 1 with the EBOPS sector in place of the category. The coefficient on the DST rate is −0.154 in column (2) and −0.167 in column (3), against −0.132 and −0.174 in Table 1, and close to zero within destinations, 0.017 in column (4) and 0.001 in column (5). The BaTIS data thus show the same negative association across destinations and the same absence of an association within destinations. The results are practically the same for digitally deliverable trade, without the delivered shares, and without SH.

The BaTIS sectors are broader than our categories. Other business services also contain consulting, research and technical services, and the DST rate of a sector averages taxed and untaxed categories. The BaTIS flows are themselves partly estimated, as missing and asymmetric bilateral flows are completed and balanced with gravity-type models. Despite these differences, BaTIS shows the same association across destinations and the same absence of an association within destinations as Table 1.

**Table S41. Bilateral digitally delivered services trade from BaTIS and the DST: Table 1 re-estimated (PPML, DST rate in pp)**

| | (1) | (2) | (3) | (4) | (5) |
|---|---|---|---|---|---|
| Our data (Table 1) | 0.023 | −0.132* | −0.174*** | 0.003 | −0.017 |
| | (0.063) | (0.073) | (0.055) | (0.042) | (0.044) |
| BaTIS, digitally delivered | 0.007 | −0.154*** | −0.167*** | 0.017 | 0.001 |
| | (0.030) | (0.033) | (0.026) | (0.014) | (0.011) |
| BaTIS, digitally deliverable | 0.011 | −0.157*** | −0.170*** | 0.015 | 0.002 |
| | (0.031) | (0.033) | (0.028) | (0.013) | (0.011) |
| BaTIS, digitally delivered, without SH | −0.012 | −0.169*** | −0.175*** | −0.020 | −0.024** |
| | (0.030) | (0.033) | (0.026) | (0.013) | (0.011) |
| Year and sector FE | No | Yes | No | No | No |
| Origin × sector × year FE | No | No | Yes | Yes | Yes |
| Destination FE | No | No | No | Yes | No |
| Country-pair FE | No | No | No | No | Yes |
| Observations (BaTIS, digitally delivered) | 315,582 | 315,582 | 315,582 | 315,582 | 315,582 |

Notes: PPML of bilateral trade on the DST rate in percentage points and the gravity controls of Table 1, on the observations used by every column. The rows on BaTIS use EBOPS sectors in place of the 31 categories, with the fixed effects of Table 1 defined accordingly. Digitally delivered: BaTIS flows times the WTO digitally delivered share of the exporter, sector and year. Digitally deliverable: BaTIS flows of SG, SH, SI, SJ and SK. Standard errors clustered by country pair in parentheses. * p<0.10, ** p<0.05, *** p<0.01.

## S3. Event studies

### S3.1 Design

The event studies compare the trade of the same flows before and after their destination adopts a tax. Besides the DST, we study VAT on digital services and the combined DST and VAT rate, each under subsidiary and headquarters attribution. For the DST and the combined rate, a flow is treated from the first year g in which its destination applies a positive rate to its category. For VAT, a flow is treated from the first year in which its destination applies VAT to digital services, and all categories of the destination are treated together. An adoption cohort is the group of treated units that start in the same year g, and event time is $k = t - g$. Flows that are already taxed in 2016 have no pre-adoption years and are dropped.

The data cover 2016–2024 and most DSTs start in 2019–2021. We therefore use a balanced window from two years before to two years after adoption, $k \in \{-2, -1, 0, 1, 2\}$, with the year before adoption as the reference, and estimate effects only for the cohorts observed over the whole window, 2018–2021 for the DST and 2018–2022 for VAT and the combined rate. Every event year thus rests on the same cohorts. For the DST these are the adopters of 2018 (Pakistan and Uruguay), 2019 (France and Zimbabwe), 2020 (Austria, Italy, Turkey, the United Kingdom and a second taxed category in India) and 2021 (Kenya, Paraguay and Spain). All estimators compare treated flows only with control flows of the same origin and category in the same year, which removes every origin-by-category-by-year shock, as the fixed effects of column (3) of Table 1 do.

### S3.2 Poisson estimator of the main text (Table S42)

The main event study of the paper (Fig. 2) uses the Poisson estimator of Wooldridge (2023), because, like the gravity regressions, it is estimated by PPML and measures the effect as a proportional change. Wooldridge (2023) shows that a pooled Poisson regression with cohort-by-year treatment indicators is robust to heterogeneous effects across cohorts and event years. We estimate the model by PPML with an indicator for each cohort g and event year $k \geq 0$, never-treated flows as controls, and origin-by-category-by-year and destination-by-category fixed effects. The destination-by-category effects remove permanent differences between taxing and non-taxing destinations. Event-year effects average the cohort coefficients with weights equal to the number of treated destination–category units and are reported as 100 × (exp(b) − 1). Column (1) of Supplementary Table S42 also includes indicators for $k = -2$, so that every effect is measured relative to the year before adoption, as in Fig. 2. Column (2) reports the standard form of the estimator, with both pre-adoption years as the baseline, and a pooled indicator for the adoption year and the following year, tested with a wild-cluster score bootstrap (Rademacher weights, 9,999 draws).

The Poisson estimates show a short-run decline after adoption. Relative to the year before adoption, trade is 7.7% lower in the adoption year and 9.7% lower a year later, both statistically significant, and back at its pre-adoption level two years after adoption. Trade two years before adoption is 10.4% above the reference year, so trade already declines before adoption. This may continue an earlier downward trend, which would overstate the response, or anticipate DSTs announced one to two years before they took effect, which would understate it. With both pre-adoption years as the baseline, the declines are 11.7% and 13.6%, and the pooled short-run effect is significant with the wild-cluster bootstrap ($p = 0.028$). The scenarios of the main text use the effect in the adoption year, which corresponds to −0.020 per percentage point of DST.

Columns (3) and (4) of Supplementary Table S42 vary two choices of the estimator, the control group and the weights of the adoption cohorts. Column (3) adds the later adopters (2022–2024) as controls in the years before they adopt, as de Chaisemartin and D'Haultfœuille (2024) do. The estimates are practically unchanged, 7.4% and 9.3% lower trade in the first two years after adoption, because only 37 destination–category units of later adopters enter the comparison. Column (4) changes how the effects of the four adoption cohorts are averaged. Columns (1)–(3) weight each cohort by its number of treated destination–category pairs, so that a small destination counts as much as a large one. Column (4) weights each cohort by the trade its flows would have had without the DST, so that it measures the effect on the total trade of the taxed flows. This is the estimate of an imputation estimator, the Poisson counterpart of Borusyak, Jaravel and Spiess (2024), which predicts the trade of each taxed flow without the tax from the untaxed observations and compares it with actual trade (Wooldridge, 2023). With these weights the declines are larger, 15.0% in the adoption year and 17.3% a year later, and trade is still 9.6% lower two years after adoption. The difference comes from the cohorts. The adopters of 2019 and 2020, which include France, the United Kingdom, Italy and Turkey, account for about four fifths of the counterfactual trade, and their trade falls by 7–23% in the first three years after adoption, whereas the adopters of 2021, Spain, Kenya and Paraguay, show no decline. The decline after adoption is thus larger and lasts longer in the largest DST markets.

Column (5) measures event time from the year in which the DST was announced, as firms may react to the announcement before the tax takes effect. The announcement year is the year in which the government first announced the tax or submitted it to parliament, as reported in Supplementary Table S1. Italy and Uruguay, both announced in 2017, are left out because the year two before their announcement is not observed. Trade two years before the announcement is 7.7% above the year before (p = 0.07), against 10.4% before adoption in column (1), so the earlier decline becomes smaller and only weakly significant. The largest decline, 12.2%, comes a year after the announcement, when most of these DSTs took effect. This fits the reading that part of the decline before adoption anticipates the tax, but it does not establish it.

**Table S42. Event study around DST adoption with the Poisson estimator of Wooldridge (2023): digital product trade (subsidiary attribution)**

| | (1) Relative to k = −1 | (2) Both pre-adoption years as baseline | (3) Not-yet-treated controls | (4) Imputation, trade-weighted | (5) Event time from announcement |
|---|---|---|---|---|---|
| k = −2 | 10.4***<br>[4, 17] | 0 (baseline) | 9.8***<br>[4, 16] | 0 (baseline) | 7.7*<br>[−1, 17] |
| k = 0 (adoption year) | −7.7***<br>[−11, −4] | −11.7***<br>[−16, −7] | −7.4***<br>[−11, −3] | −15.0***<br>[−21, −9] | −5.2<br>[−12, 2] |
| k = +1 | −9.7**<br>[−17, −2] | −13.6***<br>[−21, −6] | −9.3**<br>[−17, −1] | −17.3***<br>[−24, −10] | −12.2**<br>[−22, −2] |
| k = +2 | −1.0<br>[−9, 8] | −5.4<br>[−13, 3] | −0.6<br>[−9, 8] | −9.6**<br>[−18, −1] | −4.9<br>[−21, 15] |
| Short-run average (k = 0, +1) | −8.7***<br>[−13, −4] | −12.7***<br>[−17, −8] | −8.4***<br>[−13, −3] | −16.2***<br>[−22, −11] | −8.8**<br>[−17, 0] |
| Post-adoption average (k = 0 to +2) | −6.2**<br>[−12, 0] | −10.3***<br>[−16, −5] | −5.9*<br>[−12, 0] | −13.7***<br>[−20, −8] | −7.5<br>[−17, 4] |
| Pooled short-run indicator | | −15.9*** | | | |
| | | [−23, −9] | | | |
| wild-cluster bootstrap p | | 0.028 | | | |
| Observations | 250,460 | 250,460 | 254,522 | 250,460 | 249,195 |
| Cohort weights | treated units | treated units | treated units | counterfactual trade | treated units |
| Fixed effects | origin × category × year; destination × category | origin × category × year; destination × category | origin × category × year; destination × category | origin × category × year; destination × category | origin × category × year; destination × category |
| Adoption cohorts; controls | 2018–2021; never treated | 2018–2021; never treated | 2018–2021; never and not yet treated | 2018–2021; never treated | announced 2018–2020; never treated |

Notes: PPML. Effects in %, 100 × (exp(b) − 1). 95% confidence intervals in brackets. Standard errors clustered by destination. * p<0.10, ** p<0.05, *** p<0.01. Columns (1)–(3) average the cohort coefficients with weights equal to the number of treated destination–category units, column (4) with weights equal to their counterfactual trade (delta-method standard errors). Column (5) measures event time from the year in which the DST was announced and leaves out Italy and Uruguay, announced in 2017, whose year two before announcement is not observed.

### S3.3 Linear estimators

Heterogeneity-robust difference-in-differences estimators differ in the comparisons they make and in how they weight cohorts and event years, so we check the Poisson estimates with three linear estimators, those of Callaway and Sant'Anna (2021), de Chaisemartin and D'Haultfœuille (2024) and Borusyak, Jaravel and Spiess (2024). We apply them to the DST, to VAT and to the combined DST and VAT rate, under both attributions, and estimate the Poisson event study for the same six cases. The linear estimators use a balanced panel, in which the unit of observation is an origin–destination–category flow observed in every year. This gives 226,977 flows, 97% of the flows of the estimation panel. As in the gravity regressions, the zero flows are kept, and trade is zero in about 93% of the observations of this panel. The linear estimators remove permanent differences between flows by differencing from the year before adoption, the counterpart of the country-pair effects of Table 1.

Callaway and Sant'Anna (2021) and de Chaisemartin and D'Haultfœuille (2024) estimate the effect for adoption cohort g in year t as

$$\mathrm{ATT}(g,t) = \sum_{s} \frac{n_{gs}}{n_g} \left[ \left( \bar{Y}_{gs,t} - \bar{Y}_{gs,g-1} \right) - \left( \bar{Y}_{Cs,t} - \bar{Y}_{Cs,g-1} \right) \right]$$

where the bars denote the mean trade of the treated flows of cohort g and of the control flows C in stratum s, the stratum being the origin and category of the flow, and the weights are the shares of the treated flows of cohort g in each stratum. Callaway and Sant'Anna (2021) use never-treated flows as controls, and de Chaisemartin and D'Haultfœuille (2024) use flows that are not yet treated in both years, with the pre-adoption estimates as placebos. Event-time effects average these group–time effects over cohorts, with weights equal to the number of treated flows. Borusyak, Jaravel and Spiess (2024) instead estimate flow effects and stratum-by-year effects by least squares on the untreated observations only,

$$Y_{it} = \alpha_i + \lambda_{s(i),t} + \varepsilon_{it}$$

impute the untreated outcome of each treated observation as

$$\hat{Y}_{it}(0) = \hat{\alpha}_i + \hat{\lambda}_{s(i),t}$$

and average the difference between observed and imputed trade by event time.

The three estimators measure the effect on trade in levels (US$ million). The zero flows enter with their value of zero, but because the effect is measured in US$, the largest flows carry most of the weight. To compare the estimates with the proportional effects of the Poisson estimator, we report each effect as a percentage of the counterfactual trade of the treated flows,

$$\text{Effect}_k = 100 \times \frac{\text{ATT}_k}{E[Y_{it}(0) \mid \text{treated}]}$$

The average post-adoption effect is the mean over k = 0, 1, 2. Dividing it by the average tax rate of the treated flows gives the effect per percentage point, and ln(1 + effect/100) divided by the average log(1 + rate) gives the implied coefficient on log(1 + rate). Standard errors are clustered by destination, the level at which the tax is set. For Callaway–Sant'Anna and de Chaisemartin–D'Haultfœuille we use a multiplier bootstrap with 999 draws, in which each destination receives an independent exponential weight, and for Borusyak–Jaravel–Spiess a delete-one-destination jackknife. The pre-trend test is a Wald test that the estimate two years before adoption is zero.

### S3.3 Results (Tables S43–S48)

Supplementary Tables S43–S48 report the event studies for each tax measure and attribution, with the three linear estimators in columns (1)–(3) and the Poisson estimator of Section S3.2 in column (4). For the DST the linear estimators show the pattern of the Poisson estimates in Fig. 2 of the main text. Trade is 11–19% lower for subsidiary-attributed trade and 2–6% lower for headquarters-attributed trade in the adoption year and the following year, and it returns to its pre-adoption level two years after adoption. With only 12 adopting destinations, only the declines in the adoption year from Callaway–Sant'Anna and de Chaisemartin–D'Haultfœuille are statistically significant, at the 10% level, and the pre-trend tests do not reject. The Poisson estimator gives a smaller but significant decline under headquarters attribution, 4.5% in the adoption year and 7.4% a year later.

The lack of significance of most linear estimates does not contradict the Poisson estimates. In the first two years after adoption the linear estimates have the same sign as the Poisson estimates and a similar or larger size, and all their 95% confidence intervals include the Poisson estimates. Their confidence intervals are wide because they measure the effect in US$ and digital trade is

highly concentrated. The largest flow is worth US$26 billion and 1% of the positive flows account for 59% of trade, so the linear estimates depend on a small number of very large flows. The linear estimators also assume that, without the tax, treated and control flows would have changed by the same amount in US$, which is hard to reconcile with flows that differ by several orders of magnitude. The Poisson estimator, like the gravity model, assumes instead that they would have changed by the same proportion. For these reasons the main text and the scenarios use the Poisson estimates, and the linear estimators confirm the direction and the size of the decline.

For VAT and the combined DST and VAT rate the estimators disagree. The Poisson estimator shows a decline of 5–10% in the adoption year only, with no significant effect in the following years, whereas the linear estimators show no decline and trade rises gradually, though not significantly. Both comparisons rest on few controls. Of the 68 destinations, 28 had already extended their VAT to digital services by 2016 and most of the others do so during the sample, so only seven destinations, Brazil, Colombia, the Dominican Republic, Ecuador, Ethiopia, Paraguay and the Philippines, serve as never-treated controls. A decline that lasts beyond the adoption year is thus found only for the DST.

## Table S43. Event study around DST adoption: digital product trade (subsidiary attribution)

| | (1) Callaway–Sant'Anna | (2) de Chaisemartin–D'Haultfœuille | (3) Borusyak–Jaravel–Spiess | (4) Poisson (Wooldridge) |
|---|---|---|---|---|
| k = −2 | 15.93 | 15.10 | 2.50 | 10.39*** |
| | (19.90) | (18.72) | (6.72) | (3.16) |
| k = −1 (reference) | 0 | 0 | 0 | 0 |
| k = 0 (adoption year) | −12.76* | −12.33* | −18.57 | −7.65*** |
| | (7.72) | (7.45) | (14.25) | (1.92) |
| k = +1 | −10.87 | −10.61 | −16.19 | −9.65** |
| | (9.63) | (9.33) | (14.57) | (3.94) |
| k = +2 | 1.69 | 1.94 | −4.08 | −0.96 |
| | (16.23) | (16.03) | (19.00) | (4.33) |
| Average post-adoption effect (%) | −7.31 | −7.00 | −12.95 | −6.16** |
| | (10.21) | (9.96) | (15.31) | (2.99) |
| % per percentage point of the rate | −1.842 | −1.763 | −3.261 | −1.552** |
| | (2.577) | (2.512) | (3.856) | (0.754) |
| Implied coefficient on log(1 + rate) | −1.96 | −1.87 | −3.58 | −1.64** |
| | (2.97) | (2.88) | (4.46) | (0.82) |
| Pre-trend test, p (k = −2) | 0.424 | 0.420 | 0.710 | 0.001 |
| Controls | never treated | not yet treated | untreated observations | never treated |
| Treated flows | 22,341 | 22,341 | 22,341 | 22,341 |
| Flows in the panel | 226,977 | 226,977 | 226,977 | 229,894 |
| Adoption cohorts | 2018–2021 | 2018–2021 | 2018–2021 | 2018–2021 |
| Standard errors | bootstrap (999) | bootstrap (999) | jackknife | clustered |

Notes: Event = first year with a DST (destination x category). Effects in % of the treated flows' counterfactual trade. k = years relative to adoption, reference k = −1. Treated flows are compared only with control flows of the same

origin and category. Standard errors clustered by destination in parentheses (multiplier bootstrap for (1)–(2), delete-one-destination jackknife for (3), clustered standard errors for (4)). * p<0.10, ** p<0.05, *** p<0.01. Column (4) is the Poisson estimator of Fig. 2 (PPML with origin × category × year and destination × category fixed effects, never-treated controls), with effects 100 × (exp(b) − 1).

## Table S44. Event study around DST adoption: digital product trade (headquarters attribution)

| | (1) Callaway–Sant'Anna | (2) de Chaisemartin–D'Haultfœuille | (3) Borusyak–Jaravel–Spiess | (4) Poisson (Wooldridge) |
|---|---|---|---|---|
| k = −2 | 4.45 | 4.36 | 1.66 | 5.53*** |
| | (8.44) | (8.07) | (5.66) | (1.06) |
| k = −1 (reference) | 0 | 0 | 0 | 0 |
| k = 0 (adoption year) | −4.40 | −4.14 | −5.62 | −4.49*** |
| | (8.51) | (8.27) | (13.92) | (1.56) |
| k = +1 | −2.78 | −2.40 | −3.68 | −7.41** |
| | (15.59) | (15.46) | (19.18) | (3.63) |
| k = +2 | 2.73 | 3.26 | 1.97 | −2.62 |
| | (18.92) | (18.95) | (22.37) | (3.37) |
| Average post-adoption effect (%) | −1.48 | −1.09 | −2.44 | −4.86* |
| | (14.10) | (13.98) | (18.33) | (2.66) |
| % per percentage point of the rate | −0.374 | −0.275 | −0.615 | −1.225* |
| | (3.542) | (3.511) | (4.618) | (0.669) |
| Implied coefficient on log(1 + rate) | −0.39 | −0.28 | −0.64 | −1.29* |
| | (3.44) | (3.39) | (4.92) | (0.72) |
| Pre-trend test, p (k = −2) | 0.598 | 0.589 | 0.770 | 0.000 |
| Controls | never treated | not yet treated | untreated observations | never treated |
| Treated flows | 22,341 | 22,341 | 22,341 | 22,341 |
| Flows in the panel | 226,977 | 226,977 | 226,977 | 229,894 |
| Adoption cohorts | 2018–2021 | 2018–2021 | 2018–2021 | 2018–2021 |
| Standard errors | bootstrap (999) | bootstrap (999) | jackknife | clustered |

Notes: Event = first year with a DST (destination x category). Effects in % of the treated flows' counterfactual trade. k = years relative to adoption, reference k = −1. Treated flows are compared only with control flows of the same origin and category. Standard errors clustered by destination in parentheses (multiplier bootstrap for (1)–(2), delete-one-destination jackknife for (3), clustered standard errors for (4)). * p<0.10, ** p<0.05, *** p<0.01. Column (4) is the Poisson estimator of Fig. 2 (PPML with origin × category × year and destination × category fixed effects, never-treated controls), with effects 100 × (exp(b) − 1).

## Table S45. Event study around DST + VAT adoption: digital product trade (subsidiary attribution)

| | (1) Callaway–Sant'Anna | (2) de Chaisemartin–D'Haultfœuille | (3) Borusyak–Jaravel–Spiess | (4) Poisson (Wooldridge) |
|---|---|---|---|---|
| k = −2 | −16.36 | −21.56 | −10.45 | −2.91 |
| | (23.84) | (20.68) | (11.58) | (2.37) |
| k = −1 (reference) | 0 | 0 | 0 | 0 |
| k = 0 (adoption year) | −4.19 | 1.88 | 13.56 | −9.69*** |
| | (8.24) | (3.46) | (14.59) | (1.92) |
| k = +1 | 4.17 | 15.70 | 27.89 | −4.14 |
| | (16.80) | (12.57) | (27.68) | (4.59) |
| k = +2 | 32.99 | 48.09 | 58.57 | 8.30 |
| | (39.14) | (41.06) | (61.20) | (7.00) |
| Average post-adoption effect (%) | 10.99 | 21.89 | 33.34 | −2.13 |
| | (19.50) | (17.95) | (34.05) | (3.63) |
| % per percentage point of the rate | 0.874 | 1.740 | 2.651 | −0.165 |
| | (1.534) | (1.402) | (2.707) | (0.282) |
| Implied coefficient on log(1 + rate) | 0.89 | 1.69 | 2.45 | −0.18 |
| | (1.49) | (1.17) | (2.43) | (0.31) |
| Pre-trend test, p (k = −2) | 0.493 | 0.297 | 0.367 | 0.227 |
| Controls | never treated | not yet treated | untreated observations | never treated |
| Treated flows | 84,866 | 84,866 | 84,866 | 88,449 |
| Flows in the panel | 130,629 | 130,629 | 130,629 | 106,875 |
| Adoption cohorts | 2018–2022 | 2018–2022 | 2018–2022 | 2018–2022 |
| Standard errors | bootstrap (999) | bootstrap (999) | jackknife | clustered |

Notes: Event = first year with a DST or VAT (destination x category). Effects in % of the treated flows' counterfactual trade. k = years relative to adoption, reference k = −1. Treated flows are compared only with control flows of the same origin and category. Standard errors clustered by destination in parentheses (multiplier bootstrap for (1)–(2), delete-one-destination jackknife for (3), clustered standard errors for (4)). * p<0.10, ** p<0.05, *** p<0.01. Column (4) is the Poisson estimator of Fig. 2 (PPML with origin × category × year and destination × category fixed effects, never-treated controls), with effects 100 × (exp(b) − 1).

## Table S46. Event study around DST + VAT adoption: digital product trade (headquarters attribution)

| | (1) Callaway–Sant'Anna | (2) de Chaisemartin–D'Haultfœuille | (3) Borusyak–Jaravel–Spiess | (4) Poisson (Wooldridge) |
|---|---|---|---|---|
| k = −2 | −0.36 | −8.37 | −3.72 | 2.39 |
| | (16.33) | (10.10) | (4.77) | (1.99) |
| k = −1 (reference) | 0 | 0 | 0 | 0 |
| k = 0 (adoption year) | 0.49 | 8.25 | 13.86 | −4.85*** |
| | (11.81) | (8.05) | (16.07) | (1.30) |
| k = +1 | 0.15 | 14.50 | 19.53 | −5.68 |
| | (18.24) | (12.68) | (21.57) | (4.34) |
| k = +2 | 3.36 | 17.89 | 19.03 | −0.65 |
| | (21.20) | (18.12) | (26.99) | (6.55) |
| Average post-adoption effect (%) | 1.33 | 13.55 | 17.48 | −3.75 |
| | (16.87) | (12.81) | (21.45) | (3.52) |
| % per percentage point of the rate | 0.106 | 1.077 | 1.389 | −0.292 |
| | (1.334) | (0.999) | (1.705) | (0.274) |
| Implied coefficient on log(1 + rate) | 0.11 | 1.08 | 1.37 | −0.32 |
| | (1.50) | (0.92) | (1.60) | (0.31) |
| Pre-trend test, p (k = −2) | 0.982 | 0.407 | 0.436 | 0.224 |
| Controls | never treated | not yet treated | untreated observations | never treated |
| Treated flows | 84,866 | 84,866 | 84,866 | 88,449 |
| Flows in the panel | 130,629 | 130,629 | 130,629 | 106,875 |
| Adoption cohorts | 2018–2022 | 2018–2022 | 2018–2022 | 2018–2022 |
| Standard errors | bootstrap (999) | bootstrap (999) | jackknife | clustered |

Notes: Event = first year with a DST or VAT (destination x category). Effects in % of the treated flows' counterfactual trade. k = years relative to adoption, reference k = −1. Treated flows are compared only with control flows of the same origin and category. Standard errors clustered by destination in parentheses (multiplier bootstrap

for (1)–(2), delete-one-destination jackknife for (3), clustered standard errors for (4)). * p<0.10, ** p<0.05, *** p<0.01. Column (4) is the Poisson estimator of Fig. 2 (PPML with origin × category × year and destination × category fixed effects, never-treated controls), with effects 100 × (exp(b) − 1).

## Table S47. Event study around VAT adoption: digital product trade (subsidiary attribution)

| | (1) Callaway–Sant'Anna | (2) de Chaisemartin–D'Haultfœuille | (3) Borusyak–Jaravel–Spiess | (4) Poisson (Wooldridge) |
|---|---|---|---|---|
| k = −2 | −19.02 | −22.02 | −10.63 | −3.21 |
| | (21.77) | (19.94) | (11.60) | (2.43) |
| k = −1 (reference) | 0 | 0 | 0 | 0 |
| k = 0 (adoption year) | −1.93 | 2.22 | 14.65 | −9.08*** |
| | (6.92) | (3.51) | (14.67) | (1.90) |
| k = +1 | 8.22 | 16.26 | 29.51 | −3.33 |
| | (15.44) | (12.46) | (28.08) | (4.99) |
| k = +2 | 38.17 | 47.68 | 60.80 | 7.81 |
| | (38.97) | (39.97) | (62.09) | (7.33) |
| Average post-adoption effect (%) | 14.82 | 22.05 | 34.99 | −1.78 |
| | (18.74) | (17.57) | (34.52) | (3.91) |
| % per percentage point of the rate | 1.180 | 1.756 | 2.786 | −0.139 |
| | (1.469) | (1.366) | (2.749) | (0.305) |
| Implied coefficient on log(1 + rate) | 1.18 | 1.70 | 2.56 | −0.15 |
| | (1.37) | (1.15) | (2.44) | (0.33) |
| Pre-trend test, p (k = −2) | 0.382 | 0.269 | 0.359 | 0.193 |
| Controls | never treated | not yet treated | untreated observations | never treated |
| Treated flows | 82,535 | 82,535 | 82,535 | 86,118 |
| Flows in the panel | 130,740 | 130,740 | 130,740 | 110,205 |
| Adoption cohorts | 2018–2022 | 2018–2022 | 2018–2022 | 2018–2022 |
| Standard errors | bootstrap (999) | bootstrap (999) | jackknife | clustered |

Notes: Event = first year with VAT on digital services (destination). Effects in % of the treated flows' counterfactual trade. k = years relative to adoption, reference k = −1. Treated flows are compared only with control flows of the same origin and category. Standard errors clustered by destination in parentheses (multiplier bootstrap for (1)–(2), delete-one-destination jackknife for (3), clustered standard errors for (4)). * p<0.10, ** p<0.05, *** p<0.01. Column (4) is the Poisson estimator of Fig. 2 (PPML with origin × category × year and destination × category fixed effects, never-treated controls), with effects 100 × (exp(b) − 1).

**Table S48. Event study around VAT adoption: digital product trade (headquarters attribution)**

| | (1) Callaway–Sant'Anna | (2) de Chaisemartin–D'Haultfœuille | (3) Borusyak–Jaravel–Spiess | (4) Poisson (Wooldridge) |
|---|---|---|---|---|
| k = −2 | −4.42 | −9.06 | −3.89 | 1.63 |
| | (13.25) | (9.71) | (4.76) | (2.13) |
| k = −1 (reference) | 0 | 0 | 0 | 0 |
| k = 0 (adoption year) | 3.48 | 8.74 | 15.16 | −5.07*** |
| | (10.52) | (7.89) | (16.14) | (1.48) |
| k = +1 | 5.10 | 15.31 | 21.39 | −5.16 |
| | (16.62) | (12.65) | (21.93) | (4.89) |
| k = +2 | 8.47 | 17.66 | 21.09 | −0.88 |
| | (20.06) | (17.98) | (27.37) | (6.80) |
| Average post-adoption effect (%) | 5.68 | 13.90 | 19.21 | −3.72 |
| | (15.56) | (12.72) | (21.73) | (3.81) |
| % per percentage point of the rate | 0.453 | 1.107 | 1.530 | −0.290 |
| | (1.230) | (0.989) | (1.730) | (0.296) |
| Implied coefficient on log(1 + rate) | 0.47 | 1.11 | 1.50 | −0.32 |
| | (1.32) | (0.92) | (1.60) | (0.33) |
| Pre-trend test, p (k = −2) | 0.738 | 0.351 | 0.414 | 0.441 |
| Controls | never treated | not yet treated | untreated observations | never treated |
| Treated flows | 82,535 | 82,535 | 82,535 | 86,118 |
| Flows in the panel | 130,740 | 130,740 | 130,740 | 110,205 |
| Adoption cohorts | 2018–2022 | 2018–2022 | 2018–2022 | 2018–2022 |
| Standard errors | bootstrap (999) | bootstrap (999) | jackknife | clustered |

Notes: Event = first year with VAT on digital services (destination). Effects in % of the treated flows' counterfactual trade. k = years relative to adoption, reference k = −1. Treated flows are compared only with control flows of the same origin and category. Standard errors clustered by destination in parentheses (multiplier bootstrap for (1)–(2), delete-one-destination jackknife for (3), clustered standard errors for (4)). * p<0.10, ** p<0.05, *** p<0.01. Column (4) is the Poisson estimator of Fig. 2 (PPML with origin × category × year and destination × category fixed effects, never-treated controls), with effects 100 × (exp(b) − 1).

## S3.5 Event study with bilateral digitally delivered services from BaTIS (Table S49)

Table S49 re-estimates the Poisson event study of Table S42, column (1), with the bilateral digitally delivered services trade from BaTIS of Section S2.5, to check whether the decline after adoption depends on how our bilateral flows are constructed. The treated units are the 34 destination–sector units of the 12 destinations that adopt a DST in 2018–2021. Relative to the year before adoption, trade is 5.5% lower a year after adoption and 5.5% lower two years after adoption, whereas the decline of 1.2% in the adoption year is not statistically significant. The average decline over the three years is 4.1% (p = 0.023). The decline is smaller than in our data, 7.7% and 9.7% in the first two years, but it is still present two years after adoption. Trade two years before adoption is 3.8% above the reference year, as in our data, so this pattern is not specific to the construction of our trade data. Without the delivered shares the declines are 2.6%, 6.8% and 6.6%. The short-run decline after adoption thus does not depend on how our bilateral flows are constructed, although the broader BaTIS sectors, which also contain services that DSTs do not tax, make it smaller.

**Table S49. Bilateral digitally delivered services trade from BaTIS: Poisson event study around DST adoption**

| | Our data (Table S42, column 1) | BaTIS, digitally delivered | BaTIS, digitally deliverable | BaTIS, digitally delivered, without SH |
|---|---|---|---|---|
| k = −2 | 10.4*** | 3.8*** | 4.1*** | 3.8*** |
| | [4, 17] | [1, 6] | [2, 7] | [1, 6] |
| k = 0 | −7.7*** | −1.2 | −2.6** | −1.2 |
| | [−11, −4] | [−4, 2] | [−5, 0] | [−4, 2] |
| k = +1 | −9.7** | −5.5*** | −6.8*** | −5.5*** |
| | [−17, −2] | [−8, −2] | [−10, −4] | [−8, −2] |
| k = +2 | −1.0 | −5.5** | −6.6** | −5.5** |
| | [−9, 8] | [−10, 0] | [−11, −2] | [−10, 0] |
| Average, k = 0 to +2 | −6.2** | −4.1** | −5.3*** | −4.1** |
| | [−12, 0] | [−7, −1] | [−8, −2] | [−7, −1] |
| Treated units | destination × category | destination × sector | destination × sector | destination × sector |
| Observations | 250,460 | 293,313 | 313,401 | 237,216 |

Notes: PPML with an indicator for each adoption cohort (2018–2021) and event year, never-treated controls, and origin × category (sector) × year and destination × category (sector) fixed effects. Effects in %, 100 × (exp(b) − 1), relative to the year before adoption, cohorts weighted by their number of treated units. 95% confidence intervals in brackets. Standard errors clustered by destination. * p<0.10, ** p<0.05, *** p<0.01.

### S3.6 Alternative definitions of the DST (Table S50)

Jurisdictions differ in which taxes on digital services they levy, so the results may depend on the taxes we treat as a DST. Table S50 repeats Table 1 and the Poisson event study of Table S42, column (1), under two alternative codings. The statutory definition keeps only the taxes that Hebous et al. (2026) classify as DSTs, the DSTs and the equalisation levies of India, and covers 11 jurisdictions. The broad definition adds the levies on streaming services of Panel B of Table S1 and covers 21 jurisdictions, against 17 in our main definition. The coefficient of column (3) is −0.180 under the statutory definition and −0.173 under the broad definition, against −0.174 in Table 1, and the estimates within destinations remain close to zero. In the event study, trade is 11.7% and 7.2% lower in the adoption year under the two codings, against 7.7% in the main definition, and back at its pre-adoption level two years after adoption. The statutory definition has only 8 destinations in the event study, which widens its confidence intervals. The results thus do not depend on which taxes we treat as a DST.

## Table S50. The DST coefficient and the event study under alternative definitions of the DST

| | Main definition | Statutory DSTs only | With streaming levies |
|---|---|---|---|
| Panel A. Table 1, DST rate (pp) | | | |
| Column (1) | 0.023 | 0.040 | 0.020 |
| | (0.063) | (0.066) | (0.062) |
| Column (2) | −0.132* | −0.130 | −0.132* |
| | (0.073) | (0.080) | (0.072) |
| Column (3) | −0.174*** | −0.180*** | −0.173*** |
| | (0.055) | (0.061) | (0.055) |
| Column (4) | 0.003 | 0.003 | 0.000 |
| | (0.042) | (0.045) | (0.041) |
| Column (5) | −0.017 | −0.019 | −0.018 |
| | (0.044) | (0.048) | (0.043) |
| Panel B. Poisson event study, % change | | | |
| k = −2 | 10.4*** | 11.1*** | 10.3*** |
| | [4, 17] | [4, 18] | [4, 17] |
| k = 0 | −7.7*** | −11.7*** | −7.2*** |
| | [−11, −4] | [−17, −6] | [−11, −3] |
| k = +1 | −9.7** | −10.8* | −9.2** |
| | [−17, −2] | [−21, 1] | [−17, −1] |
| k = +2 | −1.0 | 3.3 | −0.6 |
| | [−9, 8] | [−8, 16] | [−9, 8] |
| Average, k = 0 to +2 | −6.2** | −6.6* | −5.8* |
| | [−12, 0] | [−14, 1] | [−11, 0] |
| Jurisdictions with a positive rate | 17 | 11 | 21 |
| DST destinations in the estimation sample | 14 | 9 | 18 |
| Destinations in the event study | 12 | 8 | 14 |

Notes: Main definition: DSTs and other gross-basis taxes on the digital services of non-resident suppliers (Panel A of Table S1). Statutory: DSTs and equalisation levies only, following Hebous et al. (2026). With streaming levies: the main definition plus the levies of Panel B of Table S1. Panel A as Table 1 (PPML, observations used by every column, standard errors clustered by country pair in parentheses). Panel B as Table S42, column (1) (PPML, adoption cohorts 2018–2021, effects in % relative to the year before adoption, standard errors clustered by destination, 95% confidence intervals in brackets). * p<0.10, ** p<0.05, *** p<0.01.

## S4. Scenarios: robustness and additional calculations

The scenarios of Sections 4.3 and 4.4 of the main text rest on several choices, and this section shows how the results change when each of them is varied. In the main scenarios, baseline trade is the trade predicted by column (3) of Table 1 at the observed DST, and trade in a scenario is this baseline multiplied by exp(b × Δw), where Δw is the change in the fiscal wedge in percentage points and b = −0.020 is the Poisson event study effect in the adoption year (−7.7%) divided by the average DST of the treated flows (4.0%). The same coefficient applies to tariff equivalents. Revenues are the predicted flows multiplied by the relevant rates, before administrative and compliance costs. In 2024 this gives revenues of US$2.7, 6.9 and 4.3 billion in the DST baseline, the tariff-inclusive regime and the tariff-only regime, and trade that is US$9.3 billion lower in the tariff-inclusive regime and US$3.8 billion lower in the tariff-only regime than in the baseline. All values below refer to 2024 and are in US$ billion.

### S4.1 What drives the tariff-equivalent revenue ranking (Table S51)

Section 4.2 of the main text shows that the largest tariff-equivalent revenues would accrue to Brazil and Argentina, while the largest DST revenues accrue to large European markets. A high tariff-equivalent revenue can reflect a large value of digital imports or a high tariff on the physical analogues of these imports, and the two explanations have different implications for policy. Supplementary Table S51 therefore decomposes the revenue of each destination in the accounting benchmark, computed on the trade of the regression sample, into the value of its digital imports and the average tariff equivalent on them. We find that the ranking is driven mainly by the tariff rates. Brazil, first by revenue, is ninth by digital imports, but its average tariff equivalent of 6.5% is about nine times the sample average. Argentina, second by revenue, is eighteenth by imports with an average rate of 7.1%. India, sixth, combines a small import value with a rate of 6.1%, and Kenya enters the top ten with small imports and a rate of 10%. In a log decomposition relative to the average destination, the rate accounts for about 80% of the difference for Brazil and for all of it for Argentina. The revenue that tariff equivalents would raise thus depends more on the tariff schedules of the importing countries than on the size of their digital markets.

**Table S51. Tariff-equivalent revenue by destination: import value and average rate (2024, accounting benchmark on the trade of the regression sample)**

| Destination | Revenue (US$ bn) | Rank | Digital imports (US$ bn) | Rank | Average tariff equivalent (%) | Rank |
|---|---|---|---|---|---|---|
| BRA | 1.156 | 1 | 17.7 | 9 | 6.55 | 10 |
| ARG | 0.703 | 2 | 9.8 | 18 | 7.14 | 8 |
| MEX | 0.387 | 3 | 13.8 | 13 | 2.80 | 19 |
| IDN | 0.282 | 4 | 14.9 | 11 | 1.89 | 23 |
| RUS | 0.230 | 5 | 14.6 | 12 | 1.57 | 24 |
| IND | 0.204 | 6 | 3.3 | 39 | 6.11 | 11 |
| KEN | 0.164 | 7 | 1.6 | 43 | 10.14 | 5 |
| VNM | 0.163 | 8 | 6.0 | 31 | 2.72 | 20 |
| PAK | 0.150 | 9 | 3.2 | 40 | 4.75 | 14 |
| MYS | 0.149 | 10 | 6.9 | 25 | 2.16 | 22 |
| ECU | 0.131 | 11 | 1.4 | 44 | 9.17 | 7 |
| SAU | 0.099 | 12 | 8.0 | 23 | 1.25 | 26 |

Notes: Twelve largest recipients of 64 destinations. The import-weighted average tariff equivalent is 0.74%.

### S4.2 Including VAT on digital services (Table S52)

VAT is the main tax on cross-border digital supplies, whereas the main text compares only DSTs and tariff equivalents. Supplementary Table S52 shows how large the two instruments are relative to VAT and whether including VAT changes the ranking of the regimes. Applying the destination VAT rate on digital services to 2024 trade gives about US$101 billion, against US$2.9 billion of DST and US$4.6 billion of tariff equivalents in the accounting benchmark on the same trade, and 91% of the observations face a positive VAT rate. These amounts use the trade of the regression sample, US$652 billion in 2024, and are smaller than in Figure 3, which applies the rates to all estimated cross-border trade (US$1.05 trillion). This is an upper bound, because business-to-business supplies are generally reverse-charged and their VAT is deductible, and our trade data do not separate business from consumer sales. In the scenarios, VAT is held at its observed level in every regime, so its revenue changes only through the trade response. Total revenue including VAT is then US$113.0 billion in the tariff-inclusive regime, US$111.4 billion in the tariff-only regime and US$109.8 billion in the DST baseline. The ranking of the regimes is the same as without VAT, because the small trade response of the figures reduces the VAT base only slightly. With the coefficient of column (3) of Table 1 the ranking reverses, as the larger trade contraction of the tariff-inclusive regime reduces the VAT base by more than its additional DST and tariff revenue. DSTs and tariff equivalents thus change the total taxation of digital trade only at the margin.

**Table S52. Scenario revenues in 2024 including VAT on digital services (US$ billion, calibration of the figures)**

| Regime | Trade | DST and tariff revenue | VAT revenue | Total including VAT |
|---|---|---|---|---|
| DST baseline | 652.22 | 2.74 | 107.05 | 109.80 |
| Tariff-inclusive (DST + tariff) | 642.91 | 6.95 | 106.06 | 113.01 |
| Tariff only (no DST) | 648.46 | 4.33 | 107.04 | 111.37 |

Notes: VAT revenue is predicted trade multiplied by the destination VAT rate on digital services. VAT is not part of the fiscal wedge of the scenarios and is the same in all regimes. Gross amounts before administrative and compliance costs. Business-to-business supplies are not separated, so VAT revenue is an upper bound.

### S4.3 Concordance between digital products and HS6 analogues (Tables S53–S54)

The tariff equivalents rest on a concordance between digital categories and physical goods, and some of the analogues are weak, such as cinematographic film for video on demand or printed books for online education. Supplementary Table S53 therefore repeats the accounting and scenario calculations with other choices among the analogues. It replaces the average tariff of the analogues of each category and origin–destination pair by the lowest, the median or the highest tariff, it excludes the weakest analogues (cinematographic film for video on demand, printed books for online education and printed matter for online dating), and it applies tariff equivalents to one broad group of products at a time. Excluding the weakest analogues leaves the accounting tariff revenue on the trade of the regression sample practically unchanged, US$4.5 billion against US$4.6 billion with the average. Using the lowest or the highest tariff among the analogues gives US$1.2 billion or US$10.8 billion, respectively, and the tariff-inclusive regime then reduces trade by US$3 billion or US$19 billion. The tariff-inclusive regime raises the most revenue in every case. The tariff-only regime raises more than the DST baseline unless the tariff equivalents are low, as with the lowest analogue or when only one group of products is taxed. Platform services account for almost two thirds of the tariff-equivalent revenue, content products for one third and cloud services for 4%. Supplementary Table S54 reports the trade-weighted lowest, median, mean and highest tariff of each sector. The main conclusions of the scenarios thus do not depend on the weakest analogues.

## Table S53. Tariff-equivalent scenarios under alternative HS6 concordances (2024, US$ billion, calibration of the figures)

| HS6 concordance | Average tariff equivalent (%) | Accounting tariff revenue | Revenue, tariff only | Revenue, DST + tariff | Δ trade, DST + tariff | Δ trade, tariff only |
|---|---|---|---|---|---|---|
| Mean of analogues (paper) | 0.71 | 4.61 | 4.30 | 6.92 | −9.23 | −3.68 |
| Minimum analogue | 0.18 | 1.18 | 1.16 | 3.87 | −2.53 | 3.12 |
| Median analogue | 0.45 | 2.91 | 2.87 | 5.49 | −6.27 | −0.72 |
| Maximum analogue | 1.66 | 10.83 | 8.39 | 10.93 | −19.14 | −13.68 |
| Weakest analogues excluded | 0.69 | 4.53 | 4.22 | 6.84 | −9.07 | −3.52 |
| Content only (platform and cloud untaxed) | 0.24 | 1.55 | 1.43 | 4.13 | −3.19 | 2.45 |
| Platform services only | 0.44 | 2.89 | 2.69 | 5.35 | −5.65 | −0.05 |
| Cloud and hosting only | 0.03 | 0.17 | 0.18 | 2.93 | −0.40 | 5.28 |

Notes: Lowest, median and highest tariff among the HS6 analogues of a category for each origin, destination and year. The average tariff equivalent is weighted by 2024 trade. The DST baseline raises US$2.74 billion in all rows.

## Table S54. Trade-weighted tariff equivalents by digital sector across HS6 analogues (2024, %)

| Sector | Group | Lowest | Median | Mean (paper) | Highest | Weakest excluded | HS6 analogues |
|---|---|---|---|---|---|---|---|
| Web Hosting | cloud and hosting | 0.17 | 0.22 | 0.22 | 0.27 | 0.22 | 847149, 847150, 847170 |
| File Hosting Service | cloud and hosting | 0.16 | 0.20 | 0.20 | 0.24 | 0.20 | 847149, 847150, 847170 |
| Cloud Computing | cloud and hosting | 0.12 | 0.15 | 0.15 | 0.19 | 0.15 | 847149, 847150, 847170 |
| Video on Demand | content | 0.82 | 1.58 | 1.85 | 3.26 | 1.44 | 370610, 370690, 852329, 852349, 852351, 852359 |
| Gaming Networks | content | 1.58 | 1.58 | 1.58 | 1.58 | 1.58 | 950450 |
| PC and Console Games | content | 0.34 | 0.89 | 1.15 | 2.48 | 1.15 | 852349, 852351, 852359, 950430, 950440, 950450 |
| Mobile Games | content | 0.27 | 0.80 | 0.97 | 2.05 | 0.97 | 852349, 852351, 852359, 950430, 950440, 950450 |
| Digital Music Streaming & Downloads | content | 0.39 | 0.91 | 0.88 | 1.28 | 0.88 | 852329, 852349, 852351, 852359 |
| Cybersecurity | content | 0.44 | 0.81 | 0.85 | 1.32 | 0.85 | 852329, 852349, 852351, 852359, 852380 |
| Mobile Application | content | 0.49 | 0.75 | 0.77 | 1.06 | 0.77 | 852349, 852351, 852359 |
| Software | content | 0.39 | 0.73 | 0.73 | 1.05 | 0.73 | 852329, 852349, 852351, 852359, 852380 |
| Data Licensing | content | 0.35 | 0.66 | 0.64 | 0.87 | 0.64 | 852329, 852349, 852351, 852359, 852380 |
| Operating System | content | 0.34 | 0.51 | 0.52 | 0.71 | 0.52 | 852349, 852351, 852359 |
| Online Education | content | 0.05 | 0.24 | 0.34 | 0.87 | 0.54 | 490110, 490191, 490199, 852349, 852351, 852359 |
| eBooks | content | 0.01 | 0.03 | 0.16 | 0.60 | 0.16 | 490110, 490191, 490199, 490300, 490400, 490520, 491191, 491199 |
| Online Gambling | platform services | 0.88 | 1.04 | 1.04 | 1.19 | 1.04 | 950440, 950490 |
| Online Ride-Hailing | platform services | 0.27 | 0.63 | 1.02 | 2.80 | 1.02 | 847090, 847190, 851762, 852321, 852352 |
| Digital Advertising | platform services | 0.04 | 0.27 | 0.81 | 2.24 | 0.81 | 490110, 490199, 490210, 490290, 491110, 491191, 491199 |
| Online Food Ordering | platform services | 0.26 | 0.71 | 0.76 | 1.35 | 0.76 | 847021, 847050, 847090, 847190 |
| Payment Service | platform services | 0.08 | 0.47 | 0.68 | 1.73 | 0.68 | 847021, 847050, 847090, 847190, 852321, 852352 |
| Online Travel Market | platform services | 0.19 | 0.44 | 0.63 | 1.25 | 0.63 | 491110, 847090, 847190 |
| Online Marketplace (n.e.s.) | platform services | 0.01 | 0.34 | 0.59 | 2.24 | 0.59 | 490110, 490199, 491110, 847021, 847050, 847090, 847190 |
| Online Dating | platform services | 0.07 | 0.13 | 0.22 | 0.48 | 0.00 | 490110, 490210, 490290 |

Notes: Weighted by 2024 trade in each sector.

### S4.4 Headquarters attribution of the scenario geography (Table S55, Figures S1–S3)

The geographic results of Figures 4, 6 and 8 attribute trade to the country of the subsidiary that serves the destination. Large shares of digital sales are booked by subsidiaries in locations such as Ireland, so the origin side of these figures may reflect the attribution rule more than the location of the suppliers. Supplementary Table S55 and Supplementary Figures S1–S3 therefore repeat these results with trade attributed to the country where the parent company is located. The scenarios keep the calibration of the figures and use the column (3) model re-estimated with headquarters-attributed trade, whose own DST coefficient is −0.166 per percentage point against −0.174 under subsidiary attribution. The two rules differ in which sales count as cross-border trade. A sale to France served by an Irish subsidiary of a US parent is Ireland–France trade under subsidiary attribution and United States–France trade under headquarters attribution, whereas a sale served by a French subsidiary of the same parent is a domestic French sale under subsidiary attribution but United States–France trade under headquarters attribution. Cross-border trade is therefore larger under headquarters attribution, US$906 billion against US$652 billion in 2024 in the regression sample, and so are the totals in Supplementary Table S55, which use this sample.

The destination side, which shows who collects the revenue, changes little. The rank correlations across destinations are 0.96–1.00, and the largest recipients are the same, Canada, France, the United Kingdom, Italy and Turkey for the DST and Brazil, Argentina and Mexico for tariff equivalents. The origin side, which shows to whom payments are attributed, changes as expected. Under headquarters attribution Ireland and other subsidiary locations drop out of the largest payers, and payments concentrate further on the United States and China. Ireland is the largest attributed DST payer under subsidiary attribution (US$1.37 billion of US$2.92 billion), and the United States under headquarters attribution (US$3.34 billion of US$4.02 billion). The contrast in the main text between DST payments attributed to subsidiary locations and tariff payments attributed to the United States is thus largely a property of the attribution rule, and neither rule measures who ultimately bears the tax.

**Table S55. Geography of payments and receipts under subsidiary and headquarters attribution (2024, US$ billion)**

| Panel | Total: subsidiary | Total: headquarters | Five largest: subsidiary | Five largest: headquarters | Rank correlation | Top-5 overlap |
|---|---|---|---|---|---|---|
| Fig. 4(a) DST revenues by destination | 2.92 | 4.02 | CAN, FRA, ITA, TUR, GBR | GBR, FRA, CAN, ITA, TUR | 1.00 | 5/5 |
| Fig. 4(b) tariff-equivalent revenues by destination | 4.64 | 6.12 | BRA, ARG, MEX, IDN, RUS | BRA, ARG, MEX, IND, IDN | 0.96 | 4/5 |
| Fig. 4(c) DST payments by origin | 2.92 | 4.02 | IRL, USA, GBR, IND, CHN | USA, CHN, KOR, JPN, DEU | 0.97 | 2/5 |
| Fig. 4(d) tariff-equivalent payments by origin | 4.64 | 6.12 | USA, IRL, CHN, URY, GBR | USA, CHN, URY, JPN, NLD | 0.98 | 3/5 |
| Fig. 6(a) payments by origin, tariff only | 4.33 | 6.70 | USA, IRL, CHN, GBR, JPN | USA, CHN, JPN, KOR, CAN | 0.97 | 3/5 |
| Fig. 6(a) payments by origin, tariff only, DST destinations (Fig. 8) | 1.03 | 1.70 | USA, IRL, CHN, GBR, SGP | USA, CHN, CAN, SWE, JPN | 0.97 | 2/5 |
| Fig. 6(b) payments by origin, DST + tariff | 6.95 | 10.49 | USA, IRL, CHN, GBR, JPN | USA, CHN, JPN, KOR, CAN | 0.97 | 3/5 |
| Fig. 6(b) payments by origin, DST + tariff, DST destinations (Fig. 8) | 3.65 | 5.49 | USA, IRL, CHN, GBR, SGP | USA, CHN, CAN, KOR, NLD | 0.97 | 2/5 |
| Fig. 6(c) revenues by destination, tariff only | 4.33 | 6.70 | BRA, IND, MEX, ARG, RUS | BRA, IND, MEX, ARG, IDN | 0.97 | 4/5 |
| Fig. 6(c) revenues by destination, tariff only, DST destinations (Fig. 8) | 1.03 | 1.70 | IND, PAK, KEN, URY, COL | IND, PAK, KEN, TUR, URY | 0.98 | 4/5 |
| Fig. 6(d) revenues by destination, DST + tariff | 6.95 | 10.49 | BRA, IND, CAN, GBR, FRA | BRA, IND, GBR, FRA, ITA | 0.98 | 4/5 |
| Fig. 6(d) revenues by destination, DST + tariff, DST destinations (Fig. 8) | 3.65 | 5.49 | IND, CAN, GBR, FRA, ITA | IND, GBR, FRA, ITA, CAN | 0.97 | 5/5 |
| Fig. 6(e) export change by origin, tariff only | −3.76 | −6.32 | USA, CHN, IRL, GBR, JPN | USA, CHN, JPN, DEU, SWE | 0.97 | 3/5 |
| Fig. 6(f) export change by origin, DST + tariff | −9.30 | −14.37 | USA, IRL, CHN, GBR, JPN | USA, CHN, JPN, KOR, CAN | 0.97 | 3/5 |

Notes: Panels as in Figures 4 (accounting benchmark) and 6 and 8 (scenarios, calibration of the figures). The rank correlation is the Spearman correlation of the country totals under the two attribution rules. The overlap counts the

countries among the five largest under both. Export changes in panels (e) and (f) are negative. The five largest declines are listed. The scenarios under headquarters attribution use the column-(3) model re-estimated with headquarters-attributed trade. The table uses the estimation sample. Figure 4 of the main text and Figure S1 use all recorded cross-border flows, so their accounting totals differ somewhat from the first four rows.

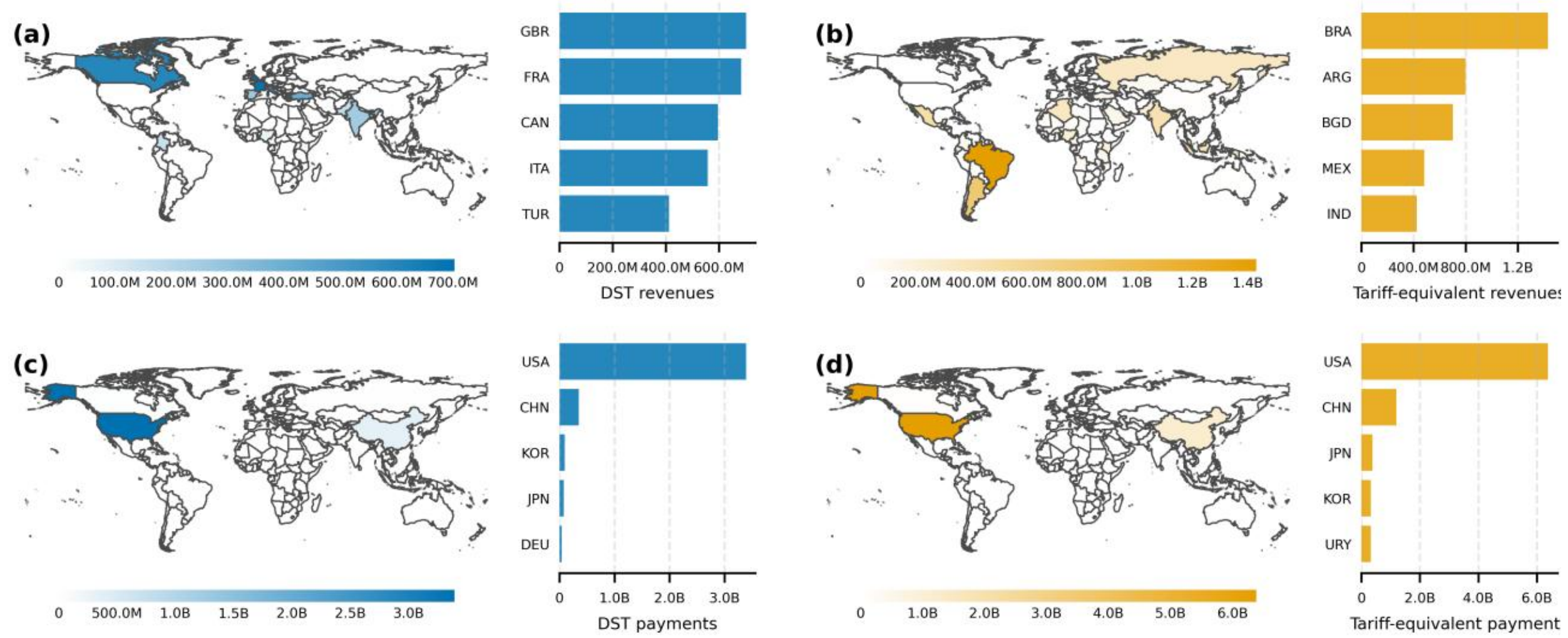

**Figure S1. Geographic distribution of DST and tariff-equivalent revenues and payments under headquarters attribution (2024).** As Figure 4 of the main text, with trade attributed to the country where the parent company is located. (a) DST revenues by destination. (b) Tariff-equivalent revenues by destination. (c) DST payments by origin. (d) Tariff-equivalent payments by origin. Accounting benchmark on all recorded cross-border flows, as in Figure 4.

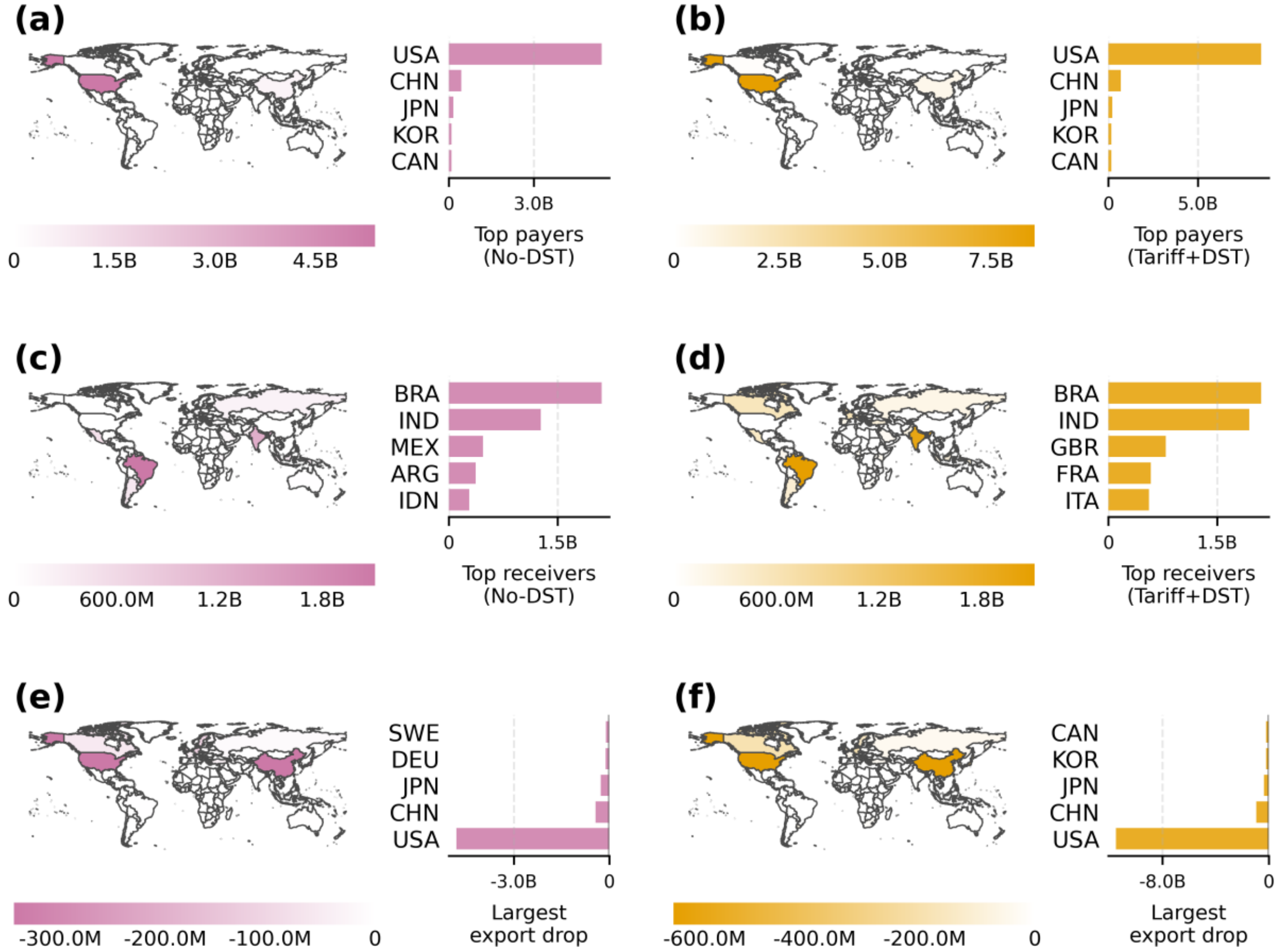

**Figure S2. Geographic allocation of payments, revenues and export changes in the scenarios under headquarters attribution (2024).** As Figure 6 of the main text. (a) And (b) Payments by origin under the tariff-only and the tariff-inclusive regime. (c) And (d) Revenues by destination. (e) And (f)

Export changes by origin. Calibration of the figures (−0.020 per percentage point) applied to the column-(3) model re-estimated with headquarters-attributed trade.

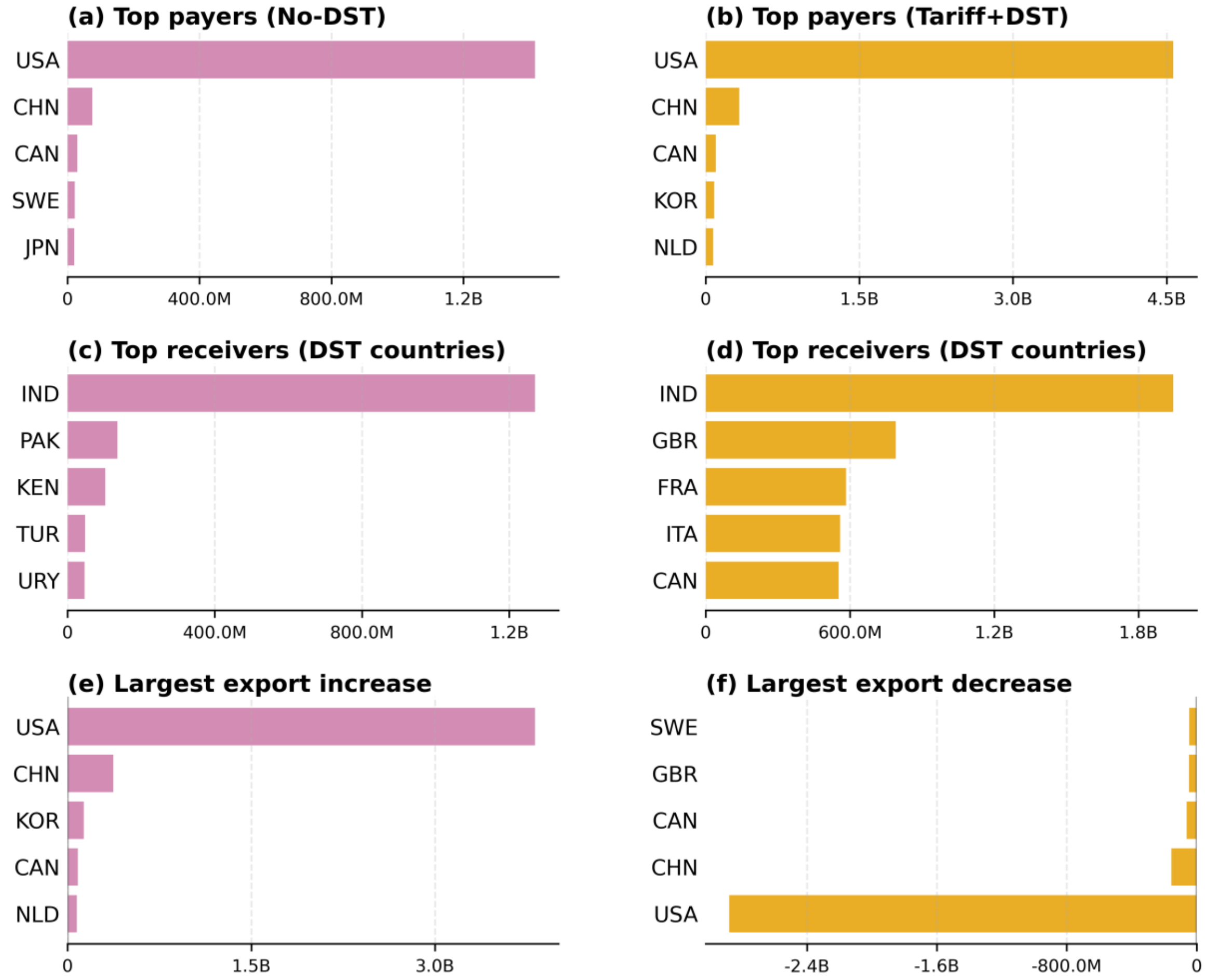

**Figure S3. Geographic allocation and trade adjustments within DST-adopting destinations under headquarters attribution (2024).** As Figure 8 of the main text. (a) And (b) Largest payers under the tariff-only and the tariff-inclusive regime. (c) And (d) Largest recipients. (e) And (f) Largest export changes. Calibration as in Figure S2.

### S4.5 Alternative values of the DST coefficient (Table S56)

The response of trade to the fiscal wedge is the main assumption of the scenarios. The figures use the Poisson event study effect in the adoption year, −0.020 per percentage point, because it rests on changes in the DST within destinations. Supplementary Table S56 repeats the scenarios with the coefficient of column (3) of Table 1 (−0.174 per percentage point), the largest response we consider, with the other event study coefficients of Section 4.1 (−1.64 to −3.58 on log(1 + DST)), with the cross-sectional coefficients of Table 1 in log(1 + DST) units (−13.5 and −17.8, Supplementary Table S4) and with no trade response. The baseline is the same in every row. The tariff-inclusive regime raises the most revenue for every coefficient, and its trade contraction ranges from US$7–15 billion at the event study coefficients (1.1–2.3% of covered trade) to US$49 billion with the coefficient of column (3) (7.6%). The ranking of the tariff-only regime and the DST baseline depends on the coefficient. With the event study coefficients, replacing the DST with tariff equivalents raises more revenue than the DST (US$3.9–4.5 billion against US$2.7 billion), whereas with the cross-sectional coefficients it raises less (US$1.8–2.2 billion), because the large response shrinks the taxed base. The main revenue result, that adding tariff equivalents to the DST raises the most revenue, thus holds for every coefficient.

**Table S56. Scenario outcomes in 2024 under alternative values of the DST coefficient (US$ billion)**

| Calibration | b | Δ trade, DST + tariff | Δ trade, tariff only | Revenue, DST baseline | Revenue, DST + tariff | Revenue, tariff only | Δ trade in DST destinations, tariff only |
|---|---|---|---|---|---|---|---|
| Event study, Poisson, per pp (figures) | −0.020 /pp | −9.30 | −3.76 | 2.74 | 6.95 | 4.33 | 3.42 |
| Table 1, col. (3), per pp (across destinations) | −0.174 /pp | −49.30 | 7.80 | 2.74 | 3.97 | 1.84 | 46.33 |
| No trade response | 0.00 | 0.00 | 0.00 | 2.74 | 7.81 | 5.07 | 0.00 |
| Event study, Poisson (Wooldridge), rel. to k=-1 | −1.64 | −7.42 | −3.00 | 2.74 | 7.13 | 4.48 | 2.76 |
| Event study, Callaway-Sant'Anna | −1.96 | −8.76 | −3.47 | 2.74 | 7.01 | 4.38 | 3.34 |
| Event study, Poisson (Wooldridge), standard | −2.81 | −12.16 | −4.53 | 2.74 | 6.71 | 4.13 | 4.93 |
| Event study, Borusyak-Jaravel-Spiess | −3.58 | −15.06 | −5.28 | 2.74 | 6.47 | 3.92 | 6.43 |
| Table 1, col. (2), log(1 + DST) | −13.50 | −41.59 | −0.63 | 2.74 | 4.48 | 2.25 | 31.93 |
| Table 1, col. (3), log(1 + DST) | −17.80 | −49.06 | 8.26 | 2.74 | 4.01 | 1.86 | 46.74 |

Notes: Δ trade is the change in 2024 digital trade in the regression sample relative to the DST baseline (US$652 billion). The event study coefficients are the average post-adoption effects of Section 4.1 converted to log(1 + DST) units: Poisson estimator of Wooldridge (2023) relative to k = −1 (−1.64) and in its standard form (−2.81), Callaway–Sant'Anna (−1.96) and Borusyak–Jaravel–Spiess (−3.58). The coefficient of the figures is the Poisson estimate for the adoption year per percentage point of the average DST of the treated flows. The first two rows enter the wedge linearly in percentage points, the others as log(1 + wedge). The last column restricts the sum to destinations with a DST in 2024 (Fig. 7).

### S4.6 A universal DST and digital supply of the HS6 analogues (Table S57)

Section 4.3 compares three regimes that correspond to the options available after the lapse of the WTO moratorium. Supplementary Table S57 adds two further scenarios, which show the revenue at stake under broader policy changes. The first applies a DST in every destination, at 3%, the most common DST rate and also the median DST rate, to all categories, keeping the existing DST rate where it is higher. This overstates the coverage of an actual DST, which applies only to in-scope activities of firms above revenue thresholds. The second replaces trade in the HS6 analogues by digital supply. With the calibration of the figures a universal 3% DST would raise US$18.9 billion, almost seven times the current DST receipts, and reduce trade by US$33 billion (5%). With the coefficient of column (3) it would raise US$12.9 billion, with a trade contraction of US$233 billion (36%). If trade in the HS6 analogues were entirely replaced by digital supply, the tariff revenue collected on these goods, US$1.7 billion on US$544 billion of trade in 2024 at an average applied tariff of 0.32%, would be lost while electronic transmissions remain duty-free, whereas the existing DSTs would raise US$1.9 billion on the same value. These amounts are of the same order as the US$4.6 billion of tariff equivalents and the US$2.9 billion of DST on the 2024 digital trade of the regression sample.

**Table S57. Scenarios with a universal DST and with digital supply of the HS6 analogues (2024, US$ billion)**

| Scenario | Figures (−0.020/pp): Δ trade | Column (3) (−0.174/pp): Δ trade | Figures (−0.020/pp): revenue | Column (3) (−0.174/pp): revenue |
|---|---|---|---|---|
| Baseline (DST) | 0.00 | 0.00 | 2.74 | 2.74 |
| DST in all destinations at 3% | −33.22 | −232.77 | 18.86 | 12.87 |
| Tariff only | −3.76 | 7.80 | 4.33 | 1.84 |
| DST + tariff | −9.30 | −49.30 | 6.95 | 3.97 |

Notes: Δ trade relative to the DST baseline. Revenue is DST plus tariff-equivalent revenue, excluding VAT.

### S4.7 A separate coefficient for the tariff equivalents (Tables S58–S60)

The figures apply the coefficient of the DST also to the tariff equivalents, because both raise the price of a cross-border digital transaction in proportion to its value. Tariffs and DSTs may nevertheless affect trade differently, as they differ in their legal payer, enforcement and pass-through. We therefore take the coefficient of the tariff equivalents from two other sources. First, Supplementary Table S58 estimates the tariff elasticity of goods trade in the HS6 analogues of Supplementary Table S2 with the specification of Table 1, with the same 112 origins and 68 destinations, the years 2016–2024, the gravity controls without the Digital STRI, which concerns digital services, and the five sets of fixed effects, with the HS6 product in place of the digital category. Trade is zero in 77% of the observations, and the zeros are kept. In column (3), a one-percentage-point higher applied tariff is associated with 9.0% lower trade (−0.095, SE 0.021), against 2.0% for the DST in the figures and 16.0% in column (3) of Table 1. Unlike the DST coefficient, the tariff coefficient remains negative and significant with destination and country-pair effects (−0.035 and −0.038), because applied tariffs differ across origins within a destination. Second, Supplementary Table S60 averages the product-level trade elasticities of Fontagné, Guimbard and Orefice (2022) over the analogues of each digital category. These elasticities are coefficients on log(1 + tariff), and their trade-weighted average is −11.5, against −10.1 in column (3) of Supplementary Table S58 in the same units.

Supplementary Table S59 applies these coefficients to the tariff equivalents and keeps the DST coefficient of the figures. Both imply a larger response than the event study coefficient, so the tariff-inclusive regime lowers 2024 trade by US$34–37 billion instead of US$9 billion and raises US$4.9–5.0 billion instead of US$6.9 billion. It remains the regime that raises the most revenue, whereas the tariff-only regime raises slightly less than the DST baseline (US$2.4–2.5 billion against US$2.7 billion), as with the coefficient of column (3) in Supplementary Table S56. These elasticities describe goods trade, whose demand and substitution possibilities differ from those of digital products, so the figures keep the common coefficient.

### Table S58. Goods trade in the HS6 analogues and the applied tariff (PPML, specification of Table 1)

| | (1) | (2) | (3) | (4) | (5) |
|---|---|---|---|---|---|
| Applied tariff (pp) | −0.145*** | −0.105*** | −0.095*** | −0.035** | −0.038*** |
| | (0.023) | (0.020) | (0.021) | (0.014) | (0.012) |
| log(1 + applied tariff) | −15.37*** | −11.18*** | −10.08*** | −3.66** | −4.05*** |
| | (2.39) | (2.14) | (2.26) | (1.55) | (1.33) |
| Year and HS6 product FE | No | Yes | Yes | Yes | Yes |
| Origin × HS6 product × year FE | No | No | Yes | Yes | Yes |
| Destination FE | No | No | No | Yes | Yes |
| Country-pair FE | No | No | No | No | Yes |
| Observations | 2,038,271 | 2,038,271 | 1,511,417 | 1,511,417 | 1,354,810 |

Notes: PPML. Dependent variable: bilateral imports of the HS6 analogues of Table S2 (excluding 847130, 847141 and 847149, as for the tariff equivalents), zeros included. Each row is a separate regression with the effectively applied tariff in percentage points or as log(1 + rate). All columns include log distance, contiguity, common colonizer, common official language and log GDP of origin and destination. Controls absorbed by the fixed effects are not estimated. Each column uses its own estimation sample. Standard errors clustered by origin–destination pair in parentheses. * p<0.10, ** p<0.05, *** p<0.01.

### Table S59. Scenarios with a separate coefficient for the tariff equivalents (2024, US$ billion)

| Coefficient of the tariff equivalents | Coefficient | Δ trade, DST + tariff | Δ trade, tariff only | Revenue, DST baseline | Revenue, DST + tariff | Revenue, tariff only |
|---|---|---|---|---|---|---|
| Same as the DST (figures) | -0.020 per pp (same as DST) | −9.30 | −3.76 | 2.74 | 6.95 | 4.33 |
| Goods trade, Table S58, column (3) | -0.0946 per pp | −33.62 | −28.43 | 2.74 | 4.99 | 2.52 |
| CEPII product-level elasticities (Table S60) | -11.52 on log(1 + tariff), by category | −36.91 | −31.77 | 2.74 | 4.89 | 2.44 |

Notes: The DST keeps the coefficient of the figures (−0.020 per percentage point) in every row. Trade changes are relative to the DST baseline. With the CEPII elasticities each digital category has its own coefficient on log(1 + tariff) (Table S60). The value shown is the trade-weighted average.

## Table S60. CEPII product-level trade elasticities of the HS6 analogues, by digital sector

| Sector | Analogues | Mean | Lowest | Highest | Replaced by HS4 average |
|---|---|---|---|---|---|
| Online Dating | 3 | −15.29 | −16.89 | −12.08 | 1 |
| Online Food Ordering | 4 | −14.60 | −15.98 | −10.46 | 3 |
| Payment Services | 6 | −14.16 | −19.22 | −7.38 | 3 |
| Online Marketplace (n.e.s.) | 7 | −13.71 | −16.30 | −9.23 | 3 |
| eBooks | 8 | −13.07 | −27.19 | −4.25 | 2 |
| Digital Advertising | 7 | −12.15 | −16.89 | −4.80 | 1 |
| Online Travel Market | 3 | −11.89 | −15.98 | −9.23 | 2 |
| Online Education | 6 | −11.51 | −16.30 | −5.92 | 3 |
| Online Ride-Hailing | 5 | −11.33 | −19.22 | −3.63 | 2 |
| Video on Demand | 6 | −11.07 | −15.63 | −5.92 | 3 |
| Web Hosting | 2 | −10.78 | −11.11 | −10.46 | 1 |
| File Hosting Service | 2 | −10.78 | −11.11 | −10.46 | 1 |
| Cloud Computing | 2 | −10.78 | −11.11 | −10.46 | 1 |
| Data Licensing | 5 | −9.09 | −10.29 | −5.92 | 3 |
| Cybersecurity | 5 | −9.09 | −10.29 | −5.92 | 3 |
| Software | 5 | −9.09 | −10.29 | −5.92 | 3 |
| Mobile Applications | 3 | −8.84 | −10.29 | −5.92 | 2 |
| Operating System | 3 | −8.84 | −10.29 | −5.92 | 2 |
| Digital Music Streaming | 4 | −8.79 | −10.29 | −5.92 | 2 |
| Mobile Games | 6 | −6.62 | −10.29 | −2.82 | 2 |
| PC and Console Games | 6 | −6.62 | −10.29 | −2.82 | 2 |
| Gaming Networks | 1 | −5.61 | −5.61 | −5.61 | 0 |
| Online Gambling | 2 | −3.98 | −4.79 | −3.17 | 0 |

Notes: Elasticities of trade with respect to 1 + tariff from ProTEE 0.1 (Fontagné, Guimbard and Orefice, 2022, HS revision 2007), averaged over the HS6 analogues of Table S2. Three analogues are matched to their HS 2007 codes (852349 to 852340, 950450 to 950410, 490520 to 490591). ProTEE replaces elasticities that are positive or not significant at the 1% level by the average of their HS4 heading. The last column counts these analogues. Software sub-categories use the analogues of Software.

### S4.8 Scenarios with BaTIS digitally delivered trade (Table S61)

The scenarios of Section 4.3 of the main text apply the fiscal wedges to our estimates of bilateral digital trade. Table S61 repeats them with the bilateral digitally delivered services trade from BaTIS of Section S2.5, to show how the revenue at stake and the trade changes depend on the trade base. The calculation is the same as for Figs. 5 and 7. Baseline trade is the trade predicted by column (3) of Table 1, re-estimated on the BaTIS panel with the EBOPS sector in place of the category and scaled in each year to observed trade, and the wedge of each regime changes trade by −0.020 per percentage point. The DST rate of a destination, sector and year and the tariff equivalent of an origin, destination, sector and year are the mean rates of the categories assigned to the sector, and the charges for the use of intellectual property (SH), which contain none of our categories, face neither. The BaTIS sectors also contain services that DSTs do not tax and that have no physical analogue, such as banking, telecommunications and consulting, so these rates overstate the taxed base, and the BaTIS amounts are an upper bound. Applied to our data, the same calculation reproduces Figs. 5 and 7.

BaTIS trade of the 68 destinations is US$3.0 trillion in 2024, 4.7 times our data. The revenues are larger accordingly, US$9.3 billion for the DST baseline, US$20.4 billion for the tariff-

inclusive regime and US$11.4 billion for the tariff-only regime, against US$2.7, 6.9 and 4.3 billion in Fig. 5. The ranking of the regimes is the same. The tariff-inclusive regime raises the most revenue and lowers trade the most, by US$24.5 billion (0.8%), and the tariff-only regime raises more than the DST baseline and lowers trade by US$5.7 billion (0.2%). Within the DST-adopting destinations the tariff-only regime again raises trade, by US$13.2 billion, as in Fig. 7. The proportional trade changes are smaller than in our data, partly because the charges for the use of intellectual property carry no tariff equivalent. The ranking of the regimes is thus the same with both trade bases, although the revenue totals are larger with BaTIS.

**Table S61. Scenarios with our data and with BaTIS digitally delivered trade (2024, US$ billion)**

| | **Our data (Figs. 5 and 7)** | **BaTIS, digitally delivered** |
|---|---|---|
| All destinations | | |
| Observed trade | 652.2 | 3,035.8 |
| Revenue, DST baseline | 2.7 | 9.3 |
| Revenue, DST + tariff equivalents | 6.9 | 20.4 |
| Revenue, tariff equivalents only | 4.3 | 11.4 |
| Δ trade, DST + tariff equivalents | −9.3 (−1.4%) | −24.5 (−0.8%) |
| Δ trade, tariff equivalents only | −3.8 (−0.6%) | −5.7 (−0.2%) |
| DST-adopting destinations | | |
| Observed trade | 159.3 | 721.2 |
| Revenue, DST baseline | 2.7 | 9.3 |
| Revenue, DST + tariff equivalents | 3.6 | 11.7 |
| Revenue, tariff equivalents only | 1.0 | 2.8 |
| Δ trade, DST + tariff equivalents | −2.1 (−1.1%) | −5.6 (−0.6%) |
| Δ trade, tariff equivalents only | +3.4 (+1.9%) | +13.2 (+1.4%) |

Notes: Scenario model of Figs. 5 and 7, with a response of −0.020 per percentage point of the wedge. Δ trade relative to the DST baseline. Revenue is DST plus tariff-equivalent revenue, excluding VAT. BaTIS: bilateral digitally delivered services trade of Section S2.5 (SG, SH, SI, SJ and SK), with the DST rates and tariff equivalents of our categories averaged within each sector. DST-adopting destinations have a positive DST rate in at least one year.

### S4.9 Decomposition of the tariff-only scenario (Table S62)

In the main text the tariff-only regime lowers global trade by US$3.8 billion (Figure 5) but raises trade within the DST-adopting destinations (Figure 7). Supplementary Table S62 reconciles the two results by decomposing the global change into DST and other destinations. Trade rises by about US$3.4 billion in DST destinations, where the tariff equivalent replaces a higher DST, and falls by US$7.2 billion elsewhere, where it is a new tax. The global decline thus comes from the destinations without a DST, and the two figures describe the same scenario.

## Table S62. Trade changes of the scenarios in DST and other destinations (2024, US$ billion)

| Δ trade vs DST baseline | All destinations | DST destinations | Other destinations |
|---|---|---|---|
| Tariff only | −3.76 | 3.42 | −7.18 |
| DST + tariff | −9.30 | −2.12 | −7.18 |

Notes: Calibration of the figures. DST destinations have a positive DST rate in at least one year.